\documentclass[acmsmall,screen,nonacm]{acmart}
\def\setappendix{} 
\setcopyright{acmlicensed}
\acmDOI{10.1145/3704853}
\acmYear{2025}
\acmJournal{PACMPL}
\acmVolume{9}
\acmNumber{POPL}
\acmArticle{17}
\acmMonth{1}
\received{2024-07-08}
\received[accepted]{2024-11-07} 
\newif\ifcomments
\ifdefined\setcomments
\commentstrue
\else
\commentsfalse
\fi

\newif\ifproofs
\ifdefined\setproofs
\proofstrue
\else
\proofsfalse
\fi

\newif\ifappendix
\ifdefined\setappendix
\appendixtrue
\else
\appendixfalse
\fi

\usepackage{cleveref}
\usepackage{paralist}
\usepackage{bm}
\usepackage{subcaption}
\usepackage{xparse}
\usepackage{enumitem}
\usepackage{multirow}
\usepackage{macrolist}
\usepackage{textcomp}

\usepackage{tikz}

\usetikzlibrary{arrows}
\usetikzlibrary{positioning}
\usetikzlibrary{shapes.geometric}

\tikzset{round/.style={circle,draw,minimum size=23pt,inner sep=0pt}}
\tikzset{value/.style={
		regular polygon,regular polygon sides=4,draw,
		minimum size=30pt,
		inner sep=0pt
}}
\tikzset{summary/.style={double}}

\crefname{equation}{eq.}{eqs.}
\crefname{figure}{fig.}{figs.}
\crefname{example}{ex.}{exs.}
\crefname{theorem}{thm.}{thms.}
\crefname{definition}{def.}{defs.}
\crefname{section}{sec.}{secs.}
\crefname{claim}{claim}{claims}
\crefname{remark}{remark}{remarks}

\newcommand{\sharon}[1]{\ifdim\lastskip>0pt\ignorespaces\fi}
\newcommand{\neta}[1]{\ifdim\lastskip>0pt\ignorespaces\fi}

\ifcomments
\renewcommand{\sharon}[1]{{\textcolor{purple}{SS: {\em #1}}}}
\renewcommand{\neta}[1]{{\textcolor{teal}{NE: \em #1}}}
\fi

\newcommand{\NameStyle}[1]{\ifmmode \text{#1}\kern-\scriptspace \else #1\fi }
\newcommand{\osc}{\NameStyle{OSC}}
\newcommand{\epr}{\NameStyle{EPR}}
\newcommand{\pa}{\NameStyle{PA}}

\newcommand{\parens}[1]{\left(#1\right)}
\newcommand{\angles}[1]{\left\langle#1\right\rangle}
\newcommand{\braces}[1]{\left\{#1\right\}}
\newcommand{\brackets}[1]{\left[#1\right]}
\newcommand{\abs}[1]{\left|#1\right|}

\newcommand{\para}[1]{\medskip \noindent \emph{#1}.}
\newcommand{\defeq}{\triangleq}

\newcommand{\foleq}{\mathrel{{\approx}}}
\newcommand{\liaeq}{\mathrel{{\approx}}}
\newcommand{\Land}{\bigwedge}
\newcommand{\Lor}{\bigvee}
\newcommand{\medLand}{\mathop{\text{$\textstyle\bigwedge$}}}
\newcommand{\sort}[1][s]{\bm{#1}}
\newcommand{\Sorts}{{\mathcal{S}}}
\DeclareMathOperator{\Formulas}{WFF}
\DeclareMathOperator{\QF}{QF}
\DeclareMathOperator{\FV}{FV}
\DeclareMathOperator{\Skolem}{Sk}

\newcommand{\LIA}{\NameStyle{LIA}}
\newcommand{\subLIA}{{_\LIA}}

\newcommand{\Domain}{\mathcal{D}}
\newcommand{\Interp}{{\mathcal{I}}}
\DeclareMathOperator{\Structures}{struct}

\newcommand{\Bounds}{\mathcal{B}}
\newcommand{\Explicate}{\mathcal{E}}
\newcommand{\ElemsOfNode}{\mathcal{E}}
\newcommand{\sinfty}{\sort[s^c]}
\DeclareMathOperator{\trans}{trans}
\DeclareMathOperator{\summary}{sum\vphantom{g}}
\DeclareMathOperator{\regular}{reg}

\newcommand{\Atoms}{\mathcal{A}}
\newcommand{\Chars}{\mathcal{C}}
\newcommand{\SymbolicCandidates}{\mathcal{F}}
\newcommand{\Reduced}[1]{#1^\star}
\newcommand{\oscReduced}{\ensuremath{\Reduced{\osc}}}

\newcommand{\N}{\mathbb{N}}

\newcommand{\SemanticsStyle}[1]{\mathbb{#1}}
\renewcommand{\S}{\SemanticsStyle{S}}

\NewDocumentCommand{\WFSemantics}{O{\Sigma}O{\prec}}{\SemanticsStyle{W}[{#1,#2}]}
\NewDocumentCommand{\BidiWFSemantics}{O{\prec}}{\SemanticsStyle{W}^{\updownarrow}[{#1}]}

\newcommand{\TheoryStyle}[1]{\mathtt{#1}}

\NewDocumentCommand{\PartialOrder}{O{\prec}}{\TheoryStyle{po}[{#1}]}
\NewDocumentCommand{\TotalOrder}{O{\prec}}{\TheoryStyle{to}[{#1}]}
\NewDocumentCommand{\WellOrder}{O{\prec}}{\TheoryStyle{WO}^{\downarrow}[{#1}]}
\NewDocumentCommand{\WFTheory}{O{\Sigma}O{\prec}}{\TheoryStyle{WF}^{\downarrow}[{#1,#2}]}
\NewDocumentCommand{\UpWFTheory}{O{\Sigma}O{\prec}}{\TheoryStyle{WF}^{\uparrow}[{#1,#2}]}
\NewDocumentCommand{\BidiWFTheory}{O{\Sigma}O{\prec}}{\TheoryStyle{WF}^{\updownarrow}[{#1,#2}]}
\NewDocumentCommand{\BidiWellOrder}{O{\prec}}{\TheoryStyle{WO}^{\updownarrow}[{#1}]}

\newcommand{\code}[1]{\texttt{\footnotesize #1}}
\newcommand{\codeSort}[1]{\code{\bfseries #1}}

\newcommand{\valueSort}{\codeSort{value}}
\newcommand{\roundSort}{\codeSort{round}}
\newcommand{\Safe}{\code{Safe}}
\newcommand{\Proposal}{\code{Proposal}}

\newcommand{\AutoExamples}{15}
\newcommand{\GuidedExamples}{13}
\newcommand{\AutoMinusGuidedExamples}{\the\numexpr \AutoExamples - \GuidedExamples \relax}
\newcommand{\PaxosExamples}{5}
\newcommand{\AllExamples}{\the\numexpr \AutoExamples + \PaxosExamples \relax}
\newcommand{\OscExamples}{4}

\AtEndPreamble{
	\newtheorem{claim}[theorem]{Claim}
	
}

\macronewlist{ProofsList}
\macrolistadd{ProofsList}{}
\newcommand{\Proofs}{\macrolistforeach{ProofsList}{\element}{\element}}

\ifproofs
\NewDocumentCommand{\DeferredProof}{mm+m}{\HereProof{#1}{#2}{#3}}
\else
\NewDocumentCommand{\DeferredProof}{mm+m}{\neta{proof deferred to \Cref{sec:proofs}}
\macrolistadd{ProofsList}{#2}
\newcommand{#2}{\begin{proof}[Proof of \Cref{#1}]#3
\end{proof}}
}
\fi

\newcommand{\PageLimit}{25}
\newcommand{\CheckPageLimit}{\vspace{0pt}\ifnum \PageLimit<\thepage\relax \ClassWarning{}{Over page limit}\fi } 
\begin{document}
\title{Axe 'Em: Eliminating Spurious States with Induction Axioms}

\author{Neta Elad}
\orcid{0000-0002-5503-5791}
\affiliation{\institution{Tel Aviv University}
	\city{Tel Aviv}
	\country{Israel}
}
\email{netaelad@mail.tau.ac.il}

\author{Sharon Shoham}
\orcid{0000-0002-7226-3526}
\affiliation{\institution{Tel Aviv University}
	\city{Tel Aviv}
	\country{Israel}
}
\email{sharon.shoham@cs.tau.ac.il} 
\begin{abstract}
\neta{this is a small comment to make sure I compiled without comments}First-order logic
has proved to be a versatile and expressive tool
as the basis of abstract modeling languages.
Used to verify complex systems with unbounded domains,
such as heap-manipulating programs and distributed protocols,
first-order logic,
and specifically uninterpreted functions and quantifiers,
strike a balance between expressiveness
and amenity to automation.
However, first-order logic semantics
may differ in important ways
from the intended semantics
of the modeled system,
due to the inability to distinguish
between finite and infinite
first-order structures,
for example,
or the undefinability of well-founded relations in first-order logic.
This semantic gap
may give rise
to spurious states and unreal behaviors,
which only exist as an artifact of the first-order abstraction
and impede the verification process.

In this paper we take a step towards
bridging this semantic gap.
We present an approach for soundly refining
the first-order abstraction
according to
either well-founded semantics
or finite-domain semantics,
utilizing induction axioms for an abstract order relation,
a common primitive in verification.
We first formalize 
sound axiom schemata
for each of the aforementioned semantics,
based on well-founded induction.
Second, we show how to use spurious
counter-models,
which are necessarily infinite,
to guide
the instantiation of these axiom schemata.
Finally, we present a
sound and complete
reduction
of well-founded semantics
and finite-domain semantics
to standard semantics
in the recently discovered
Ordered Self-Cycle (\osc{}) fragment of first-order logic,
and prove that satisfiability under these semantics
is decidable in \osc{}.

We implement a prototype tool to evaluate our approach,
and test it on various examples
where spurious models arise,
from the domains of distributed protocols
and
heap-manipulating programs.
Our tool quickly finds the necessary axioms
to refine the semantics,
and successfully completes the verification process,
eliminating the spurious system states
that blocked progress. \end{abstract}

 
\maketitle

\neta{Consistent style questions:
	\begin{itemize}
		\item Skolem, Skolemized, Skolemization - capital S? 
		\textbf{Just for Skolem (constants, functions, etc.), other forms not capitalized}
		\item \textbf{bound formula} / bounds formulas?
		\item Referring to previous paper: \textbf{past}/present tense?
		\item \textbf{First} / firstly, \textbf{second} / secondly
		\item finite models / structures, finite-structure (singular) semantics (hyphens in general); 
		upwards-well-foundedness?
		\item a FOL / an FOL, generally how to treat initialism
		\item when to refer to semantics as singular? 
		can we usually just use it as plural? 
		modulo (the) semantics?
		\item $:$ vs. \textbf{\textbackslash{}colon in math}
		\item resp. vs. respectively
		\item iff vs. $\iff$
		\item w.r.t.\
		\item \textbf{non-empty} vs. nonempty
		\item amount vs. number
		\item schemata vs. schemas
		\item proved / proven participle
		\item use sentences for formulas when appropriate
		\item paragraph names capitalized? \textbf{yes}
		\item which (extra info) vs. that (essential)
	\end{itemize}
}

\section{Introduction}
\label{sec:intro}
In recent years major strides have been made
in advancing the use of formal methods and logic
to verify correctness of systems
such as computer programs and distributed protocols.
The applicability and adoption of formal methods
is greatly affected by the trade-off between the
expressiveness of the logic used,
and the degree of automation it enjoys.

Recently, first-order logic (FOL) has proved to provide
a sweet spot in this space.
First-order solvers and theorem provers~\cite{z3,cvc5,vampire}
support various first-order theories,
as well as
uninterpreted functions and quantifiers,
allowing users to take advantage
of the expressive language of first-order logic.
Still, the combination of theories, uninterpreted functions
and quantifiers remains a significant challenge,
and as a result, many works based on first-order logic
either make extensive use of first-order theories
but
minimize use of
uninterpreted functions with quantifiers~\cite{chc-modulo-adt,smt-recursive-programs,chc-integers-arrays,chc-adt,reasoning-about-vectors},
or vice versa~\cite{paxos-made-epr,mcmillan-decidable-ivy,linked-lists-epr,heap-paths-epr,
	modularity-for-decidability,bounded-horizon,vericon-networks,duoai,distai}.

This paper falls under the latter category.
As an abstract modeling language,
uninterpreted first-order logic with quantifiers
is particularly suitable for specifying systems
that consist of unbounded domains,
such as programs
manipulating objects in the heap~\cite{linked-lists-epr,heap-paths-epr}
and distributed protocols~\cite{paxos-made-epr,bounded-horizon}.
The avoidance of theories
does mean that useful primitives
such as the natural numbers cannot be used.
Previous work
has showed that instead of such primitives,
a partial or total order can be used
to adequately model
numbered rounds of a distributed protocol~\cite{paxos-made-epr}
and reachable objects in the heap~\cite{linked-lists-epr},
among other things.

Unfortunately, FOL also has limitations
as an abstraction tool,
an important one being
the inability of FOL
to precisely capture the intended semantics
of the system it abstracts,
exposing a \emph{semantic gap}.
When FOL formulas are used to
encode the verification conditions
of a system, the semantic gap
may result in failure
to verify the correctness of the system
due to spurious states,
which do not exhibit the intended behavior of the system,
but satisfy the negated
verification condition formulas.
The semantics mismatch
and existence of spurious counter-models
leave the user
to deal with unresponsive tools,
manually attempting to amend their specification
towards their intended semantics.

Two prominent instances
where FOL semantics
introduce spurious behaviors
are when
finite-structure semantics
or well-founded semantics
are intended.
Since FOL in general does not distinguish between
finite and infinite models,
there is a semantic gap whenever the user wishes
to consider only finite objects,
but there exist infinite counter-models.
Similarly, since well-foundedness is not definable in FOL,
spurious non-well-founded counter-models may also arise
whenever an abstract order is used.
Notable examples where this might happen
are rounds in a distributed protocol,
that are meant to be natural numbers and thus well-founded,
or a linked list of nodes in the heap,
meant to be unbounded but finite.

Recent approaches for modeling and verifying systems 
using fragments of FOL that enjoy a finite-model property (e.g.,~\cite{paxos-made-epr})
bypass these issues.
In such fragments, whenever there exists a counter-model
there also exists a finite counter-model of bounded size,
that, due to its finiteness,
does not exhibit the semantics mismatch.
However, one major drawback of these approaches
is that the modeling requires manual creative effort
to ensure that the verification conditions of the system
fall into the desired fragment.
Moreover, the realm of what is expressible
within such fragments
is not completely understood,
and a user might spend a lot of effort trying
to come up with a suitable encoding to no avail.

In this paper,
we present a
complementary approach
for tackling the semantic gap
in systems modeled using abstract orders.
In our approach
the system and verification conditions
are encoded more naturally
in a less restrictive first-order logical language,
and automation is used
for soundly bridging the gap between FOL semantics
and well-founded semantics
or finite-domain semantics.
To this end, we first
formalize a framework
for augmenting verification conditions
with well-founded induction axioms
in order to refine the first-order abstraction.
We build on the connection
between well-foundedness
and well-founded induction
(also known as Noetherian induction),
and observe that for an ordered domain,
finiteness corresponds to bidirectional well-foundedness.
We develop a method for automatically generating
induction axioms,
by identifying behaviors
that violate the intended semantics
in spurious counter-models,
which are necessarily infinite.

A recent paper~\cite{infinite-needle} has shown
how to capture certain infinite counter-models
with a finite representation called symbolic structures,
allowing them to be found automatically
and presented to the user.
While examining the infinite counter-models
may give the user insight
on how to amend the first-order specification,
this process requires mental effort on part of the user,
and might not be trivial in some cases.
We show that
the abstraction can be refined
and the semantic gap closed
by identifying violations of the intended semantics
in the spurious symbolic models,
and automatically instantiating
axiom schemata
according to them.

Surprisingly,
we use
the set of axioms computed this way
to prove new decidability results
of satisfiability under well-founded semantics
and finite-structure semantics
for the Ordered Self-Cycle (\osc{})~\cite{infinite-needle}
fragment of FOL.
The \osc{} fragment extends
the many-sorted variant of the
Effectively PRopositional
fragment of FOL
(\epr{})~\cite{bsr-epr,epr-decidable},
itself a useful fragment,
capable of verifying systems from various domains,
such as heap-manipulating programs~\cite{linked-lists-epr,heap-paths-epr}
and distributed network protocols~\cite{paxos-made-epr}.
\osc{} relaxes some of the restrictions of \epr{}
on functions
for one totally-ordered sort,
which is a common construct in verification.
This increases the expressiveness of \osc{},
at the price of losing the finite-model property of \epr{}.
Namely, some \osc{} formulas
only admit infinite models,
thus, unlike \epr{},
well-founded semantics and finite-structure semantics
do not coincide with standard FOL semantics
in \osc{}.

Our decidability proofs show
that for \osc{},
the aforementioned set of axioms
provides a sound and complete
reduction
of well-founded semantics
and finite-structure semantics
to standard FOL satisfiability.
Together with decidability of
satisfiability of \osc{} under standard FOL semantics,
proved in~\cite{infinite-needle},
this shows
that the problems of
satisfiability by well-founded structures
and satisfiability by finite structures
are also decidable
in \osc{}.
Note that decidability under standard FOL semantics alone
does not readily imply decidability under the other semantics.
For example,
if the decision procedure of \osc{}
under standard FOL semantics determines
satisfiability by providing a model that is not finite
(or not well-founded),
it may or may not be the case that a model
of the intended semantics exists.
Thus, no information regarding
satisfiability under the other semantics
is obtained.

We evaluate our approach on
a suite of \AllExamples{} examples
from the domains of distributed protocols
and heap-manipulating programs
where natural FOL modeling results
in spurious infinite counter-models.
Using our automatically generated axioms,
we are able to exclude these spurious models,
leading ultimately to successful verification.

\paragraph{Contributions}
In summary, we present the following contributions:
\begin{itemize}
	\item
	We formalize a systematic approach
	for refining
	first-order specifications towards well-founded
	and finite-domain semantics,
	via sound induction axioms.
	
	\item
	We develop a method for automatically
	obtaining instances of induction axioms
	from spurious counter-models
	represented as symbolic structures.
	
	\item
We prove that for the
	Ordered Self-Cycle (\osc{})
	fragment of first-order logic
	we can compute a finite set of such axioms
that is both sound and complete,
	showing that
	satisfiability under well-founded
	or finite-structure semantics
	is decidable there.

	\item
	We implement the above ideas in a prototype tool,
and evaluate their applicability on \AllExamples{}
	examples,
	taken from two domains.
\end{itemize}

The rest of the paper is organized as follows.
\Cref{sec:overview} uses a simplified encoding
of the Paxos consensus protocol
to informally walk through the key ideas,
\Cref{sec:background} provides necessary background,
and
\Cref{sec:semantics,sec:sym-models-axioms,sec:decidable,sec:evaluation}
present the main contributions listed above.
Finally, \Cref{sec:related} discusses related work
and \Cref{sec:conclusion} concludes.
\ifappendix
We omit proofs of lesser salience
and defer them to
\Cref{sec:proofs}.
\else
We omit proofs of lesser salience.
See~\citep{axe-em-arxiv} for an extended version with detailed proofs.
\fi

 \section{Overview}
\label{sec:overview}
In this section we demonstrate
our approach using an example of
a simplified verification condition (VC)
from the safety proof of the Paxos distributed
consensus protocol~\cite{lamport-paxos}.
This example shows how using first-order logic (FOL)
to abstract the intended semantics of the system
may create semantic gaps,
which give rise to spurious counter-models,
and how our automatically generated axioms
are able to ignore and eliminate those counter-models.
As the example is in the \osc{} fragment,
a finite set of such axioms is sound and complete,
and necessarily allows us to either prove unsatisfiability
or produce a well-founded model.

\subsection{Motivation}
Paxos~\cite{lamport-paxos} is a well-known
and widely used
distributed consensus protocol.
In Paxos, an unbounded set of nodes
exchange messages during an unbounded series
of numbered rounds
until they are able to decide on a common value among themselves.
In recent works~\cite{paxos-made-epr,paxos-ic3po,infinite-needle}
the protocol was modeled
in uninterpreted FOL,
with the rounds abstracted
as a strict, total order.
In~\cite{paxos-made-epr}, manual and creative effort was made
in order to ensure that the formalization
of the protocol and its invariants
stay in the decidable fragment of \epr{},
which further guarantees that when a VC
cannot be verified,
a finite counter-model can be found.
In a finite counter-model,
a total order can always be embedded
in the standard order on the natural numbers. Thus, an artifact of the \epr{} modeling is
that the order abstraction tightly captures the intended semantics.

In~\cite{paxos-ic3po,infinite-needle}, however, a more natural
first-order modeling was used,
which
introduced infinite counter-models
where the rounds are not well-founded.
These infinite models are spurious,
and do not capture the intended semantics of the protocol,
due to a semantic gap between the total order abstraction
of the protocol rounds,
and their intended interpretation as a well-ordered set.
The semantic gap is exposed
due to the consideration of infinite models,
where a total order cannot necessarily be embedded
in the order on the natural numbers.
Unfortunately, infinite models cannot be avoided
in the general case since state-of-the-art solvers
adhere to the standard first-order semantics. This is not coincidental, as first-order logic
where the semantics is restricted to finite structures
does not have a complete proof system.
To overcome the semantic gap,
we need to refine our modeling
by incorporating additional axioms.
As well-foundedness is not finitely axiomatizable,
we cannot naturally include it in the modeling of Paxos
(as we do the strict, total order property),
but instead we can only add finitely many instances
of well-founded induction.
As we show in this paper,
this may still be enough
to bridge the semantic gap.

\subsection{Running Example: Paxos}
For our simplified presentation of Paxos
we will ignore the distributed network of nodes
that run the protocol,
and instead focus on the properties of the values
that are proposed in each round.
We want to verify that no two distinct values
may be proposed,
under the assumptions that a value may be proposed
only if it was deemed safe, or was proposed before
(i.e., in an earlier round),
and that there is only a single safe value.

Formally, we have
two sorts,
\valueSort{} and \roundSort{},
and a vocabulary
$\Sigma = \braces{
	{\prec}(\cdot, \cdot),
	\Safe(\cdot),
	\Proposal(\cdot, \cdot)
}$,
where $\prec$ is a binary relation over \roundSort{},
specifying a strict total order on rounds,
$\Safe$ is a unary relation over $\valueSort$,
encoding the safe values,
and $\Proposal$ is a binary relation
over $\roundSort \times \valueSort$,
stating which values are proposed in which rounds.
The complete formalization
of the assumptions and (negated) property
is given in \Cref{fig:paxos-simplified-vc}.
The conjunction of
\Cref{eq:paxos-simplified-vc:anti-reflexive,eq:paxos-simplified-vc:transitive,eq:paxos-simplified-vc:total,eq:paxos-simplified-vc:one-safe,eq:paxos-simplified-vc:invariant,eq:paxos-simplified-vc:safety}
is the negated VC of our running example.

\begin{figure}
	\begin{align}
		\label{eq:paxos-simplified-vc:anti-reflexive}
		&\forall x\colon \roundSort. x \not\prec x
		\\
		\label{eq:paxos-simplified-vc:transitive}
		&\forall x, y, z \colon \roundSort.
		\parens{ x \prec y \land y \prec z } \to x \prec z
		\\
		\label{eq:paxos-simplified-vc:total}
		&\forall x, y \colon \roundSort. x = y \lor x \prec y \lor y \prec x
		\\
		\label{eq:paxos-simplified-vc:one-safe}
		& \forall v, v' \colon \valueSort.
		\parens{ \Safe(v) \land \Safe(v') } \to v = v'
		\\
		\label{eq:paxos-simplified-vc:invariant}
		&\forall r \colon \roundSort, v \colon \valueSort.
		\Proposal(r, v) \to \parens{
			\Safe(v)
			\lor
			\exists r' \colon \roundSort. r' \prec r \land \Proposal(r', v)
		}
		\\
		\label{eq:paxos-simplified-vc:safety}
		& \exists r_1, r_2 \colon \roundSort, v_1, v_2 \colon \valueSort.
		v_1 \neq v_2 \land \Proposal(r_1, v_1) \land \Proposal(r_2, v_2)
	\end{align}
	
	\caption{FOL formalization of the simplified VC of Paxos (negated):
		\Cref{eq:paxos-simplified-vc:anti-reflexive,eq:paxos-simplified-vc:transitive,eq:paxos-simplified-vc:total}
		encode that $\prec$ is a strict, total order;
		\Cref{eq:paxos-simplified-vc:one-safe}
		encodes that there is at most one \Safe{} value;
		\Cref{eq:paxos-simplified-vc:invariant}
		encodes the \Proposal{} mechanism invariant;
		and \Cref{eq:paxos-simplified-vc:safety} encodes
		a violation of the safety property.
	}
	\Description{FOL formalization of the simplified VC of Paxos (negated)}
	\label{fig:paxos-simplified-vc}
\end{figure} 

\subsection{Infinite Counter-Model}
The negated VC corresponding to the conjunction of
\Cref{eq:paxos-simplified-vc:anti-reflexive,eq:paxos-simplified-vc:transitive,eq:paxos-simplified-vc:total,eq:paxos-simplified-vc:one-safe,eq:paxos-simplified-vc:invariant,eq:paxos-simplified-vc:safety}
is in the decidable \osc{} fragment,
and previous work~\cite{infinite-needle}
has provided an automated tool to find a
(potentially infinite) satisfying model,
which indicates a violation of the VC.
For convenience, we first skolemize the negated VC,
so that the infinite counter-model could be better understood.
We use the extended vocabulary
$\Sigma^\star = \Sigma \cup \braces{ r_1, r_2, v_1, v_2, w(\cdot) }$,
where $r_1, r_2$ are constants of the \roundSort{} sort,
specifying the rounds in which distinct values are proposed,
$v_1, v_2$ are constants of the \valueSort{} sort,
{specifying the distinct values,
and $w$ is a function from \roundSort{} to \roundSort{}\footnote{
	Formally, the Skolem function $w$ also depends on
	the \valueSort{} $v$.
	For simplicity, we restrict it to only depend on the \roundSort{}.
},
indicating for each round an earlier round
witnessing 
an earlier proposal.
\Cref{eq:paxos-simplified-vc:invariant,eq:paxos-simplified-vc:safety}
are replaced with the following two equations (resp.):
\begin{align}
	\label{eq:paxos-simplified-vc:invariant-skolem}
	&\forall r \colon \roundSort, v \colon \valueSort.
	\Proposal(r, v) \to
	\parens{
		\Safe(v)
		\lor
		\parens{ w(r) \prec r \land \Proposal(w(r), v) }
	}
	\tag{\ref{eq:paxos-simplified-vc:invariant}$^\star$}
	\\
	\label{eq:paxos-simplified-vc:safety-skolem}
	& v_1 \neq v_2 \land \Proposal(r_1, v_2) \land \Proposal(r_1, v_2)
	\tag{\ref{eq:paxos-simplified-vc:safety}$^\star$}
\end{align}

\begin{figure}
	\centering \begin{subfigure}[t]{.485\textwidth}\centering \scalebox{0.85}{
			\begin{tikzpicture}[->,>=stealth',shorten >=1pt,auto,scale=0.8]
				\node[round]
					(r1) at (1, 1)
					{ $r_1$ };
					
				\node
					(dots) at (3, 1)
					{ $\dots$ };
					
				\node[round]
					(before-r2) at (5, 1)
					{};
					
				\node[round]
					(r2) at (7, 1)
					{ $r_2$ };
					
				\node[value]
					(v1) at (1, 4)
					{ $v_1$ };
				\node
					() at (1, 4.75)
					{ \Safe{} };
					
				\node[value]
					(v2) at (5, 4)
					{ $v_2$ };

				\path 
					(r1) edge [loop left] node { $w$ } (r1)
					(r2) edge node { $w$ } (before-r2)
					(before-r2) edge node { $w$ } (dots)
				;
				
				\path
					(r1) edge node { \Proposal{} } (v1)
					(r2.north) edge [bend right] node [right] 
						{ \Proposal{} } (v2.east)
					(before-r2.north) edge node [right]
						{ \Proposal{} } (v2.south)
					(dots.north) edge [bend left] node [left]
						{ \Proposal{} } (v2.west)
				;
			\end{tikzpicture}
		}
		\caption{
			Model for the skolemized version 
			of \Cref{fig:paxos-simplified-vc}.
			There are infinitely many rounds,
			ordered from left to right.
			The \Safe{} value $v_1$ is proposed
			on round $r_1$,
			while the un-\Safe{} value $v_2$
			is proposed on all other rounds.
			Starting from $r_2$,
			there is an infinitely decreasing chain
			of rounds where $v_2$ is proposed,
			with each round having a witness
			for proposing $v_2$ through $w$.
		}
		\label{fig:paxos-simplified-vc:model}
	\end{subfigure}\hfill
	\begin{subfigure}[t]{.485\textwidth}\centering \scalebox{0.85}{
			\begin{tikzpicture}[->,>=stealth',shorten >=1pt,auto,scale=0.8]
				\node[round]
				(r1) at (1, 1)
				{ $r_1$ };
				
				\node[round,summary]
				(summary) at (4, 1)
				{};
				
				\node
				(summary-bound) at (4, 0)
				{ $i \leq 0$ };
				
				\node[value]
				(v1) at (1, 4)
				{ $v_1$ };
				
				\node
				() at (1, 4.75)
				{ \Safe{} };
				
				\node[value]
				(v2) at (5.5, 4)
				{ $v_2$ };
				
				\node[round]
				(r2) at (7, 1)
				{ $r_2$ };
				
				\path
				(r1) edge [loop left] node { $w$ } (r1)
				(summary)
				edge [loop left]
				node [above, xshift=-10pt, yshift=2pt] { $w \colon i - 1$ } (summary)
				
				(r2.north) edge [bend right] node [above] { $w\colon 0$ } (summary.north)
				;
				
				\path[->]
				(r1) edge node { \Proposal{} } (v1)
				(summary.north) edge [bend left] node { \Proposal{} } (v2.west)
				
				(r1) edge [bend right] node [below] { $\prec$ } (summary)
				(summary) edge [bend right] node [below] { $\prec$ } (r2)
				
				(r2.north) edge [bend right] node [right] { \Proposal{} } (v2.east)
				;
				
				\path[->](summary) edge
				[out=15,in=-15,loop] node
				[above, xshift=10pt, yshift=2pt] { $\prec \colon i_1 < i_2$ } (summary)
				;
			\end{tikzpicture}
		}
		\caption{
			Symbolic-structure representation of the model
			in \Cref{fig:paxos-simplified-vc:model}.
			The summary node (double circle in the middle)
			succinctly represents
			the infinitely many rounds
			where the un-\Safe{} value $v_2$ is proposed.
The infinitely decreasing chain of rounds
			is clearly visible as the summary node
			represents a copy of the non-positive integers
			(where $i \leq 0$),
			ordered between them according to the natural
			integer order ($i_1 < i_2$).
		}
		\label{fig:paxos-simplified-vc:symbolic-model}
	\end{subfigure}
	\caption{A violation of the VC 
		of \Cref{fig:paxos-simplified-vc}.
		Values are depicted as squares
		and rounds as circles.}
	\label{fig:paxos-simplified-vc:both-models}
	\Description{A violation of the VC of Paxos}
\end{figure} 
A satisfying infinite model is presented in
\Cref{fig:paxos-simplified-vc:model}.
In the model, two distinct values are proposed,
the \Safe{} $v_1$ on round $r_1$,
and the un-\Safe{} $v_2$ on all other rounds,
which form an infinitely decreasing chain
starting from $r_2$,
each comprising the witness of its successor.
Thus it is apparent that the domain of \roundSort{}
in the infinite counter-model
is not well-founded.

\subsection{Using Induction to Eliminate the Spurious Counter-Model}
Since the infinite counter-model does not adhere
to our intended, well-founded, semantics,
it does not in fact show that the VC is violated.
It is therefore necessary to
refine the first-order modeling
in order to hopefully better capture
the intended properties of the system.
Specifically, an induction schema of well-foundedness
can be used to capture
the intended semantics of the $\prec$ relation.
The induction schema we use,
formalized in \Cref{sec:semantics},
is based on well-founded induction:
given some property of elements in the well-founded set
(stated as a first-order formula),
if there exists any element satisfying that property,
then there exists a minimal element that does so.
In this example of Paxos,
consider the following two axioms,
which state that if there is a round where $v_1$ (resp. $v_2$)
is proposed,
then there exists a minimal proposal round:

\begin{align}
	\label{eq:paxos-simplified-vc:axiom-v}
	&
	\parens{ \exists r \colon \roundSort. \Proposal(r, v_1) }
	\to
	\parens{
		\exists r \colon \roundSort.
			\Proposal(r, v_1)
			\land
			\forall r' \colon \roundSort \prec r. \neg \Proposal(r', v_1)
	}
	\\
	\label{eq:paxos-simplified-vc:axiom-vprime}
	&
	\parens{ \exists r \colon \roundSort. \Proposal(r, v_2) }
	\to
	\parens{
		\exists r \colon \roundSort.
		\Proposal(r, v_2)
		\land
		\forall r' \colon \roundSort \prec r. \neg \Proposal(r', v_2)
	}
\end{align}

By instantiating the invariant
\Cref{eq:paxos-simplified-vc:invariant-skolem}
on the minimal rounds in which $v_1$ and $v_2$ are proposed, we get
that $v_1$ and $v_2$ must be safe,
and from \Cref{eq:paxos-simplified-vc:one-safe}
they must be equal --- contradicting the violation of safety.
Indeed, once these axioms are added to the negated VC, it becomes unsatisfiable
(even under standard FOL semantics),
and the safety property is verified.

\subsection{Using Symbolic Models to Pinpoint Semantics Violations}
The focus of this paper
is how to automatically find the right axioms for
eliminating spurious counter-models,
such as the one in \Cref{fig:paxos-simplified-vc:model},
instead of leaving this task to users.
Our key idea is to find a characterization
of the rounds along the decreasing sequence,
and apply induction to exclude
them.
Here we may leverage the representation
of the infinite model in
\Cref{fig:paxos-simplified-vc:model}
as a \emph{symbolic structure}, as shown in
\Cref{fig:paxos-simplified-vc:symbolic-model}.
This representation was proposed in~\cite{infinite-needle},
and is used by the FEST tool to
encode the infinite models it finds.

The symbolic model groups the elements
of the explicit model into nodes.
Each regular node represents a single element,
while each summary node, denoted by a double circle,
represents a non-empty set of elements specified by
a \emph{bound} formula (written underneath the node).
Specifically, for each integer $i$ that satisfies
the bound formula of the summary node $n$,
the set of elements represented by $n$
includes the element $\angles{n,i}$.
The interpretation of function and relation symbols for the
elements represented by summary nodes
is given symbolically, using \LIA{} terms and formulas.
For example, the interpretation of the Skolem function $w$
for the summary node in
\Cref{fig:paxos-simplified-vc:symbolic-model}
is $i - 1$,
which means that $w$ maps the element $\angles{n,i}$
to the element $\angles{n,i - 1}$.
Similarly, the interpretation of $\prec$
for the summary node is $i_1 < i_2$,
which means that elements $\angles{n,i_1}$ and $\angles{n,i_2}$
represented by $n$
are ordered according to their indices $i_1, i_2$.

Zooming in on the summary node (in the middle),
we can see that the combination of its bound formula ($i \leq 0$)
and the interpretation of $\prec$ for it ($i_1 < i_2$)
means that the set of explicit elements it represents
forms an infinitely decreasing chain.
A formula that captures these explicit elements
can therefore be used in an induction axiom to eliminate such a model.
In this case a suitable formula,
that results in the axiom in
\Cref{eq:paxos-simplified-vc:axiom-vprime},
is
$\psi_1(r) = \Proposal(r, v_2)$.
After eliminating this model, the symmetric case
arises,
where there is an infinitely decreasing chain
satisfying
$\psi_2(r) = \Proposal(r, v_1)$.
Instantiating the induction schema with this formula gives us the axiom in \Cref{eq:paxos-simplified-vc:axiom-v}.
These are the two axioms we have seen before,
which suffice for verifying the property.

We formalize and generalize this idea
of generating induction axioms from properties
of nodes in \Cref{sec:sym-models-axioms}.
Specifically, we show that a violation of the intended semantics
can always be witnessed by a single violating node.
Further, when the formula with which 
the induction schema is instantiated
uniquely identifies the violating node,
the spurious model is excluded by the generated axiom.

\subsection{Decidability of Well-Founded Satisfiability}
The simplified (negated) VC formula of Paxos belongs to
the \osc{} fragment of FOL,
introduced in~\cite{infinite-needle}.
For this fragment,
whose satisfiability was proved decidable in~\cite{infinite-needle},
we provide
in \Cref{sec:decidable}
a finite and complete set of induction axioms,
such that a formula is satisfiable by a well-founded model
iff it is satisfiable in conjunction with the axioms
under standard semantics.
In some sense,
this means that the set of axioms
excludes all non-well-founded models of formulas in \osc{},\footnote{
More precisely,
	if the formula with the axioms is satisfiable,
	it may still have non-well-founded models,
	but importantly, it will also have a well-founded model.
}
even though the property of well-foundedness
is not definable in first-order logic.

A finite and complete set of axioms
lets us reduce satisfiability
under well-founded semantics
to satisfiability under standard semantics.
This is already beneficial
because it means that we can harness
existing FOL solvers for checking satisfiability
under  well-founded semantics
in a sound and complete way.
Unfortunately, FOL satisfiability is undecidable in general,
hence the reduction alone does not imply decidability
of satisfiability under well-founded semantics.

However, standard satisfiability is decidable for the aforementioned class,
and the conjunction of any formula in the class
with the set of axioms
remains in \osc{},
thus giving us decidability for
satisfiability under well-founded semantics
in \osc{}.
Furthermore, by previous results~\cite{infinite-needle},
we have a finite search space of symbolic structures
in which to check satisfaction of \osc{} formulas,
and we show that given such a model that satisfies the induction axioms,
we can construct a well-founded one.
Thus, beyond determining satisfiability
under well-founded semantics,
we are able to provide a well-founded counter-model when it exists.

\subsection{From Well-Foundedness to Finite Semantics}
We consider now a variant of the VC of Paxos,
that is violated even under well-founded semantics,
and therefore remains satisfiable after adding
any (and all) well-founded induction axioms.
In this variant we have an additional
unary relation \code{ManuallySet},
and the proposal invariant
(\Cref{eq:paxos-simplified-vc:invariant-skolem})
is altered to
choose between two modes:
either proposing values that were deemed ``safe''
in a past round, as before, or
proposing values according to
``future'' manual setting:
\begin{equation}
	\label{eq:paxos-simplified-variant}
\forall v. \parens{
	\begin{array}{c}
		\parens{
			\forall r. \Proposal(r, v) \to \parens{ \Safe(v) \lor \parens{ w(r) \prec r \land \Proposal(w(r), v) }}
		}
		\\
		\Lor \\
			\parens{
					\forall r. \Proposal(r, v) \to \parens{ \code{ManuallySet}(v) \lor \parens{ r \prec w(r) \land \Proposal(w(r), v) }}
			}
	\end{array}
}.
\end{equation}
For this bidirectional variant of Paxos
we get an infinite (well-founded) counter-model
even when the complete set of axioms
that capture well-foundedness of $\prec$ is added.
The counter-model mirrors
the behavior of the model in
\Cref{fig:paxos-simplified-vc:symbolic-model},
with an infinitely increasing chain of rounds
where an un-\Safe{} value is proposed.
Indeed, the VC formula of this variant
can only be verified by assuming stronger semantics:
that the number of rounds is finite
(which is indeed the case when we consider only the rounds
that are active in every step of the protocol)\footnote{Every trace violating a safety property has a finite prefix
	which demonstrates the violation,
	and only finitely many rounds may be active in a finite prefix.
Therefore, in the context of safety verification it suffices
	to reason about states (structures)
	with a finite (but unbounded) set of rounds.
}.
Adopting this stronger semantics lets us
use induction on the inverse of the $\prec$ order,
which is also well-founded
when finitely many rounds are considered.
\neta{shorten?}

Using induction on the inverse of $\prec$
produces the following two axioms,
which can again be automatically generated
based on summary nodes
in spurious counter-models,
\begin{align}
	&
	\parens{ \exists r \colon \roundSort. \Proposal(r, v_1) }
	\to
	\parens{
		\exists r \colon \roundSort.
		\Proposal(r, v_1)
		\land
		\forall r' \colon \roundSort \bm{\succ} r. \neg \Proposal(r', v_1)
	}
	\\
	&
	\parens{ \exists r \colon \roundSort. \Proposal(r, v_2) }
	\to
	\parens{
		\exists r \colon \roundSort.
		\Proposal(r, v_2)
		\land
		\forall r' \colon \roundSort \bm{\succ} r. \neg \Proposal(r', v_2)
	},
\end{align}
and by conjoining them with
the negated VC formula of the Paxos variant
we indeed show unsatisfiability in finite-state semantics.

The idea of using induction based on $\prec$
and its inverse
can be generalized using the observation that
given a totally ordered domain,
considering a finite-state semantics is equivalent to assuming
well-foundedness of the order in both directions
(see \Cref{lem:order-and-finiteness}).
We can therefore soundly use axioms on both the relation and its inverse
when we are interested in a finite-state semantics.
(The actual result is more refined:
the axioms are sound also if the order is partial.
Totality is needed only for the completeness result.)

We show that for \osc{}
a finite set of such axioms is again complete
and lets us reduce finite satisfiability to satisfiability,
which is decidable in \osc{}.
The proof also gives us a finite-model property
for \emph{finite} satisfiability of \osc{}
(note that \osc{} itself
does not have a finite-model property:
it includes formulas that are only satisfied by infinite models~\cite{infinite-needle}).

\subsection{Applicability}
We implemented a prototype tool for automatically generating
induction axioms and checking satisfiability
using two approaches:
an enumeration of all possible axioms made of simple formulas,
and a symbolic-model-guided heuristic
that tries to first construct a symbolic model
and then use its summary nodes as hints to violations of well-foundedness.
Each of the two approaches is able to successfully eliminate
spurious infinite models from a various of examples,
where the model-guided approach is usually faster,
but since we only consider axioms based
on a single spurious symbolic model,
fails to find the correct axiom
in \AutoMinusGuidedExamples{} cases
where the simple enumeration does find it.
Most of the examples are taken from~\cite{infinite-needle},
which compiled examples from various sources~\cite{natural-proofs,paxos-ic3po,bounded-horizon},
with a few new examples modeling distributed algorithms~\cite{broadcast-echo,dijkstra-self-stabilizing}.

Using our tool
we can find the necessary axioms
for both variants of the
simplified VC of Paxos
in a couple of seconds,
and successfully verify the safety property
of the system.

 \section{Background}
\label{sec:background}
\paragraph{First-order logic}
We assume some familiarity with many-sorted
first-order logic (FOL) with equality.
A vocabulary $\Sigma$ consists
of a set of sorts $\Sorts$,
and sorted relation, function and constant
symbols.
Terms are constants $c$,
variables $x$ of some sort $\sort \in \Sorts$,
and well-sorted function applications
$f \parens{ t_1, \dots, t_k }$.
Atomic formulas are formed
from terms by equality,
denoted
$t_1 \foleq t_2$,
or by applying relations
$R \parens{ t_1, \dots, t_k }$.
Non-atomic formulas are built up
using the logical connectives
$\neg, \land, \lor, \to$
and quantifiers
$\forall, \exists$.
We denote the set of free variables
in a term $t$ (formula $\varphi$)
by $\FV(t)$ ($\FV(\varphi)$),
defined recursively over its structure.
We call a term $t$ with no free variables
($\FV(t) = \emptyset$)
a \emph{ground term},
and a formula where $\FV(\varphi) = \emptyset$
a \emph{sentence}.
We denote the set of all well-formed formulas
over $\Sigma$ by $\Formulas(\Sigma)$
\neta{propagate to use when currently saying $\varphi$ over $\Sigma$}
A structure $M$ over $\Sigma$ is a pair
$\parens{\Domain, \Interp}$,
where the domain $\Domain(\sort)$ maps sort $\sort$
to a non-empty set of elements,
and $\Interp(\alpha)$ gives the interpretation
of the symbol $\alpha$ over those sets.
We denote the class of all first-order structures
over $\Sigma$ by
$\Structures(\Sigma)$.
We define assignments and satisfaction in the usual way.
A formula $\varphi$ is satisfiable if
it has a satisfying structure
$M \models \varphi$, called a model.
We denote by $\top$ a logical tautology
and by $\bot$ a contradiction.
Given a formula $\varphi$ with free variables
$x_1, \dots, x_k$ we denote the universal closure
of $\varphi$ by 
$\varphi^\forall \defeq \forall x_1 \dots \forall x_k. \varphi$.

\paragraph{Orders}
Throughout the paper we use a binary relation
over a single sort $\prec \colon \sort \times \sort$
to denote a relation symbol  
axiomatized as
a strict partial or total order.
A \emph{strict partial order} 
is a transitive and
anti-reflexive
relation,
formulated in first-order logic
by
the conjunction of
$\forall x, y, z. \parens{ x \prec y \land y \prec z } \to x \prec z$
and $\forall x. x \not\prec x$,
and denoted by $\PartialOrder$.
A strict total order is a strict partial order
where every two distinct elements are ordered,
formulated by
$
\TotalOrder \defeq \PartialOrder \land
\parens{ \forall x, y. x \foleq y \lor x \prec y \lor y \prec x }
$.
As all orders we consider throughout the paper are strict,
we occasionally omit the word.
We use $\succ$ as a shorthand to mean
the inverse of $\prec$.

\paragraph{Linear Integer Arithmetic}
The interpreted theory of Linear Integer Arithmetic (\LIA{})
is defined over a vocabulary
that includes an integers sort,
and standard constant, function and relation symbols
($0, 1, +, -, <$).
Conventionally, the vocabulary is assumed to also include
infinitely many constants for all integers,
and shorthand relations such as $>, \geq, \leq$,
and we adopt these conventions.
We denote by $\QF_\subLIA$
the set of all quantifier-free \LIA{} formulas.
We denote by $\models_\subLIA$ satisfiability
in the standard model of the integers,
and note that checking satisfiability of (quantified) \LIA{}
formulas is decidable~\cite{lia-decidable}
and supported by existing tools~\cite{z3,cvc5}.
Throughout the paper we distinguish FOL variables,
denoted by $x, y$
with or without subscripts,
and \LIA{} variables,
denoted by $i$ and subscripts.
We denote integer values by $z$ and subscripts.

 \section{Semantics Refinement via Axioms}
\label{sec:semantics}
In this section we consider first-order logic (FOL)
as an abstract modeling language.
This abstraction creates a semantic gap
between first-order semantics and the intended semantics.
We discuss how this gap can be curtailed
by axioms that
soundly refine the abstraction
and bridge over the mismatch between
satisfying first-order structures
and the intended structures,
which in our case are either
well-founded structures
(\Cref{sec:semantics:wf})
or finite-domain structures
(\Cref{sec:semantics:fin}).
The content provided in this section
summarizes known results
and formulates them using the terminology
that we will use in the rest of the paper.

\neta{add reminders that we think of $\varphi$ as a VC}

\subsection{Intended Semantics}

We start by formulating the concept
of intended semantics.

\begin{definition}[Intended Semantics]
	Given a vocabulary $\Sigma$,
	an \emph{intended semantics} of $\Sigma$ is given by
	a class of structures $\S \subseteq \Structures(\Sigma)$.
	We say that a formula $\varphi \in \Formulas(\Sigma)$
	is \emph{satisfiable modulo $\S$}
	if there exists a model
	$M \in \S$
	such that
	$M \models \varphi$.
\end{definition}

If an intended semantics $\S$ is definable in FOL,
i.e., there exists a theory (set of sentences)
$T$ over $\Sigma$ such that
$M \models T$ iff $M \in \S$,
then satisfiability of $\varphi$ modulo $\S$
can be reduced to (standard) satisfiability
of $T \cup \{\varphi\}$.
However, this is often not the case,
and yet, we can sometimes still capture
the intended semantics via a set of FOL formulas,
in a weaker sense.

For example, consider the vocabulary
$\Sigma = \braces{
	0, 1, {+}, {<}
}$
and the standard model of the natural numbers
$M_\N$ as the intended semantics,
i.e., $\S = \braces{ M_\N }$.
There exists no set of FOL sentences
that defines $M_\N$
(any set of FOL formulas that is satisfied by $M_\N$
also admits non-standard models).
However, Presburger Arithmetic (\pa{})
provides a consistent and complete theory
for $\S$.
This means that for any first-order sentence
$\varphi$ over $\Sigma$,
exactly one of $\varphi$ and $\neg \varphi$
is implied by \pa{}.
Specifically, if $\varphi$ holds in the standard model
of the naturals ($M_\N \models \varphi$)
then it is implied by \pa{},
and if it does not,
then $\neg \varphi$ is implied by \pa{}.
Thus, even though the semantics induced by \pa{}
also includes non-standard models,
we can still use it to check first-order properties
of the standard model of the natural numbers
because
\[
M_\N \models \varphi \iff \pa \models \varphi
\iff \pa \not\models \neg \varphi \iff \pa \cup \{\varphi \} \text{ is satisfiable},
\]
and this is the property that may be generalized
to other intended semantics.

Alas, completeness in the sense that
for every $\varphi$,
exactly one of $\varphi$ and $\neg \varphi$
is implied by the theory,
does not generalize
to intended semantics
that include more than one structure, and specifically,
structures that satisfy different first-order sentences
(e.g., an intended semantics of finite structures).
For such intended semantics $\S$,
we cannot hope for a consistent and complete theory
that agrees with $\S$ on all first-order sentences
(even if we do not impose any computability requirements)
since the structures in $\S$ already disagree
among themselves.

Fortunately, completeness is also
not a necessary condition for our purposes.
What makes \pa{} capture
the standard model of the natural numbers is
the fact that the properties that distinguish
the standard model from non-standard ones
cannot be expressed in first-order logic.
Thus, if some non-standard model satisfies (or violates)
a first-order property, so does the standard model.
This quality of \pa{} can be stated more generally as follows.

Given an intended semantics $\S$,
we strive for a theory $T$
such that for every formula
$\varphi \in \Formulas(\Sigma)$,
\[
\forall M \in \S. M \models \varphi \iff T \models \varphi,
\]
or, equivalently,
\[
\exists M \in \S. M \models \varphi \iff T \cup \braces{ \varphi } \text{ is satisfiable}.
\]
Such $T$ allows
us to capture satisfiability modulo $\S$ of $\varphi$
with standard first-order satisfiability
(of $T \cup \braces{\varphi}$),
where in our setting
$\varphi$ encodes the negated VC of some system.
We call $T$ an elementary theory of $\S$
if the above holds,
and as we have discussed,
\pa{} is an elementary theory of $M_\N$ in this sense.
Another relevant example
is the intended semantics of well-orders
for a vocabulary containing only a binary relation $\prec$,
which also has an elementary theory~\cite{elementary-well-ordering}.
In the sequel we formally define this semantics
for arbitrary vocabularies,
as well as finite-domain semantics,
and discuss sound axiom schemata for both of them.

\subsection{Theory of Well-Founded Relations}
\label{sec:semantics:wf}
Given a vocabulary $\Sigma$
that includes a binary relation symbol
$\prec \colon \sort \times \sort$,
we are often interested in the class
of all structures over $\Sigma$
in which $\prec$ is interpreted as a well-founded relation.
That is, the class of all structures $M$,
where every non-empty subset of the domain
$S \subseteq \Domain^M(\sort)$ has a \emph{minimal}
element w.r.t. $\prec$.
For example,
this is the case when $\prec$ is meant
to abstract the order on the natural numbers.
We formally define well-founded relations
and well-founded semantics as follows:

\begin{definition}[Well-Founded Relation]
	A relation $R \subseteq U \times U$ is \emph{well-founded}
	if for any
	$\emptyset \neq S \subseteq U$,
	there exists $m \in S$ such that for all $s \in S$,
	$(s, m) \notin R$.
\end{definition}

\begin{definition}[Well-Founded Semantics]
	Given a vocabulary $\Sigma$
	with a binary relation symbol
	$\prec \colon \sort \times \sort$,
the \emph{well-founded semantics of $\parens{\Sigma, \prec}$}
	is the class of structures over $\Sigma$
	where $\prec$ is interpreted as a well-founded relation
	over the domain of sort $\sort$.
	We omit $\parens{\Sigma, \prec}$
	when clear from context.
\end{definition}

Though well-founded semantics 
is not definable in FOL
(for proof, see, e.g.,~\cite{open-logic-definability}),
we can use
a sound axiom schema
for well-founded semantics,
based on well-founded induction.
Namely, given a structure $M$ of $\Sigma$,
we can assume that if there exists an element
$e \in \Domain^M$
that satisfies some property $\psi$
(i.e., $M, \brackets{ x \mapsto e } \models \psi$),
then there is one that is minimal
w.r.t.\ the interpretation of $\prec$.
Applying the well-founded induction principle
is captured by the following axiom schema.
Note that $\psi$ can have arbitrarily many
free variables beyond $x$,
which the axiom schema universally quantifies over.
This can be understood as capturing in one axiom multiple inductions over a parametric property.

\begin{definition}[Well-Founded Induction]
	\label{def:wf-theory}
	Given a vocabulary $\Sigma$
	with a binary relation $\prec$,
	for any formula $\psi \in \Formulas(\Sigma)$,
	the following sentence is a \emph{well-founded induction axiom}:
	\[
	\xi_{\min}[\psi] \defeq
	\parens{ \parens{ \exists x. \psi }
	\to
	\parens{ \exists x_{\min}.
		\psi \brackets{ x_{\min} / x }
		\land
		\forall y. y \prec x_{\min} \to \neg \psi \brackets{ y / x }
	} }^\forall.
	\]
We denote the set of
	well-founded induction axioms over $\Sigma$
	by
$\WFTheory \defeq \braces{ \xi_{\min}[\psi]  \mid \psi \in \Formulas(\Sigma) }$.
\end{definition}

An alternative axiom schema,
$
\parens{
	\forall x. \parens{ \forall y. y \prec x \to \psi \brackets{ y / x } }
	\to
	\psi
}
\to
\forall x. \psi
$,
may be more familiar to the reader
(e.g., from the standard formulation
of the strong induction principle over natural numbers).
These two axiom schemata are equivalent:
the formula $\xi_{\min}[\neg \psi]$
is logically equivalent
to the aforementioned formula.
The axiom $\xi_{\min}[\neg \psi]$ can be thought of
as a contrapositive formulation of the induction principle for proving $\forall x.\psi$,
stating that if some element violates $\psi$,
then there exists a minimal element that does so.

We now show that the
$\WFTheory$ theory
is appropriate for sound refinement of formulas
according to well-founded semantics.
Formally:

\begin{claim}[Soundness]
	\label{claim:wf-axioms-sound}
Given a vocabulary $\Sigma$
	with a binary relation $\prec$,
	for any formula $\varphi$ over $\Sigma$,
if there exists a model $M \models \varphi$
	where $\prec^M$ is well-founded,
then $\WFTheory \cup \braces{\varphi}$
	is satisfiable.
\end{claim}

\DeferredProof{claim:wf-axioms-sound}{\WfAxiomsSoundProof}{Follows trivially from the definition of well-foundedness:
    Let $M \models \varphi$ where $\prec^M$ is well-founded, 
    and let $\xi_{\min}[\psi]  \in \WFTheory$. 
    We show that $M \models \xi_{\min}[\psi]$. 
    Let $x_1,\ldots,x_k$ be the free variables of $\psi$, apart from $x$ 
    (which may or may not be free in $\psi$),
    and let $e_1,\ldots,e_k$ be elements 
    from the domains of the corresponding sorts in $M$.
    
    Assume that 
    $M, \brackets{ x_1\mapsto e_1,\ldots,x_k\mapsto e_k } \models \exists x. \psi$. 
    We need to show that 
    \[
    M, \brackets{ x_1\mapsto e_1,\ldots,x_k\mapsto e_k } \models
    \exists x_{\min}.
		\psi \brackets{ x_{\min} / x }
		\land
		\forall y. y \prec x_{\min} \to \neg \psi \brackets{ y / x }.
	\]
    To this end, consider the set 
    \[
    \braces{
    	e \in \Domain^M(\sort) 
    	\mid 
    	M, \brackets{x \mapsto e,x_1\mapsto e_1,\ldots,x_k\mapsto e_k } \models \psi 
    }
   	\]
    This set is not empty 
    (since $M, \brackets { x_1\mapsto e_1,\ldots,x_k\mapsto e_k } \models \exists x. \psi$).
    Therefore, by well-foundedness of $\prec^M$ 
    it has a minimal element $e' \in \Domain^M(\sort)$. 
It follows that 
    \[
    M,[x_{\min} \mapsto e', x_1\mapsto e_1,\ldots,x_k\mapsto e_k] \models
		\psi \brackets{ x_{\min} / x }
		\land
		\forall y. y \prec x_{\min} \to \neg \psi \brackets{ y / x },
	\]
	which completes the proof.
}

\ifproofs
As the proof shows,
\else
In fact, 
\fi
a stronger property holds,
that if there exists
a well-founded model $M \models \varphi$,
then that same model
$M \models \WFTheory \cup \braces{\varphi}$,
but this stronger property is unnecessary
for our needs.

\subsection{Theory of Ordered Finite Domains}
\label{sec:semantics:fin}

Unlike well-founded semantics,
there is no immediately handy axiom schema
for finite-domain semantics.
There is, however, a connection between
\emph{ordered finite domains}
and \emph{bidirectional} well-foundedness,
which we explore in this section.
We start by defining
the semantics of finite domains, and
complementing our notion
of a well-founded relation
with the definitions of upwards well-founded
and bidirectionally well-founded relations.

\begin{definition}[Finite-Domain Semantics]
	\label{def:fin-semantics}
	Given a vocabulary $\Sigma$ and a sort $\sort$,
	the \emph{finite-domain semantics of $\parens{\Sigma, \sort}$}
	is the class of structures over $\Sigma$
	where the domain of $\sort$ is finite.
	We omit $\parens{\Sigma, \sort}$
	when clear from context.
\end{definition}

\begin{definition} A relation $R \subseteq U \times U$ is \emph{upwards well-founded}
if $R^{-1}$ is well-founded.
$R$ is \emph{bidirectionally well-founded}
if it is both well-founded and upwards well-founded.
\end{definition}

We first observe that finite partially ordered sets
must be bidirectionally well-founded.
Further, for totally ordered sets,
the properties of finiteness
and bidirectional well-foundedness
coincide.
Formally:

\begin{lemma}
	\label{lem:order-and-finiteness}
	Let $R$ be a strict partial order for a set $U$.
	Then, if $U$ is finite then
	$R$ is bidirectionally well-founded.
	Further, if $R$ is a strict total order,
	then $U$ is finite
	iff
	$R$ is bidirectionally well-founded.
\end{lemma}

\DeferredProof{lem:order-and-finiteness}{\OrderAndFinitenessProof}{Let $R$ be a partial order,
	and let us assume towards contradiction that there exists
	some infinitely increasing (or decreasing) sequence
	$a_0, a_1, \dots$ in $R$
	($\forall j. \parens{ a_j, a_{j + 1} } \in R$).
	Since $R$ is anti-reflexive and transitive,
	no element of $U$ may appear twice in the sequence,
	and thus $U$ is infinite --- contradiction.

	Now let $R$ be a total order,
	if $U$ finite then $R$ is bidirectionally well-founded.
	Let us assume then towards contradiction
	that $R$ is bidirectionally well-founded
	but $U$ is infinite.

	We show that we can construct an infinitely increasing sequence
	in $U$. 	
	Let us denote $U_0 = U$, $a_j = \min U_j$
	and $U_{j + 1} = U_j \setminus \braces{a_j}$.
	Observe that for any $j$, $U_j \supset U_{j + 1}$.
	Since $R$ is well-founded,
	for every non-empty $U_j$ there exists a minimal element $a_j$,
	and since $U$ is infinite, every $U_j$ is non-empty.
	Thus the sequence $a_0, a_1, \dots$ is well-defined.
	Since $R$ is a total order, $a_j \prec a_{j + 1}$
	(as $a_j, a_{j + 1} \in U_j$, and $a_j$ is minimal in $U_j$),
	and the sequence $a_0, a_1, \dots$ is an infinitely increasing sequence
	in $U$ --- in contradiction to the fact that $R$ is upwards well-founded.
}

Note that for the equivalence of well-foundedness
	and finiteness
	the requirement of a total order is necessary.
	(As a counterexample consider an infinite set $U$ and
	an empty relation $R$ over $R$.
	$R$ is bidirectionally well-founded and a partial order.)
However, for a sound refinement w.r.t.\ finite-domain semantics,
the one-sided implication guaranteed by a partial order is sufficient.

\sharon{fix here}
Building on the one-sided connection established
in \Cref{lem:order-and-finiteness},
we define an axiom schema
for bidirectionally well-founded induction,
and show that is a sound refinement
for finite-domain semantics.

\begin{definition}
	Given a vocabulary $\Sigma$
	with a binary relation $\prec$,
	for any formula $\psi \in \Formulas(\Sigma)$,
	the following sentence is an
	\emph{upwards well-founded induction axiom}:
	\[
	\xi_{\max}[\psi] \defeq
	\parens{
	\parens{ \exists x. \psi }
	\to
	\parens{
	\parens{
		\exists x_{\max}.
		\psi \brackets{ x_{\max} / x }
		\land
		\forall y. x_{\max} \prec y \to \neg \psi \brackets{ y / x }
	}
	}
	}^\forall.
	\]
	We denote the set of all
	bidirectionally well-founded induction axioms
	over $\Sigma$
	by
$
	\BidiWFTheory \defeq \WFTheory \cup \braces{
		\xi_{\max}[\psi] \mid \psi \in \Formulas(\Sigma)
	}
	$.
\end{definition}

\begin{claim}[Soundness]
	\label{claim:fin-sound-axioms}
	Given a vocabulary $\Sigma$
	with a binary relation
	$\prec \colon \sort \times \sort$,
	for any formula $\varphi$
such that $\varphi \models \PartialOrder$,
	if there exists a model $M \models \varphi$
	where $\Domain^M(\sort)$ is finite,
	then $\BidiWFTheory \cup \braces{\varphi}$
	is satisfiable.
\end{claim}

\DeferredProof{claim:fin-sound-axioms}{\FinSoundAxiomsProof}{Let $\varphi$ be a formula such that
	$\varphi \models \PartialOrder$,
	and let $M$ be a model $M \models \varphi$
	where $\Domain^M(\sort)$ is finite.
	Thus $M \models \PartialOrder$,
	and $\prec^M$ is a strict partial order for
	$\Domain^M(\sort)$.
	By \Cref{lem:order-and-finiteness},
	$\prec^M$ is bidirectionally well-founded,
	and similarly to the proof of
	\Cref{claim:wf-axioms-sound},
	$M \models \BidiWFTheory$.
}

\section{Generating Induction Axioms from Symbolic Counter-Models}
\label{sec:sym-models-axioms}
So far we have presented general axiom schemata
for capturing well-founded or finite-domain semantics.
In this section, we present an approach
for instantiating the axiom schemata
based on spurious counter-models, i.e.,
infinite models that do not adhere
to the intended semantics.
Previous work~\cite{infinite-needle}
introduced \emph{symbolic structures}
as a finite representation of (potentially) infinite structures.
Symbolic structures use \emph{summary nodes}
to capture repeating patterns in the infinite structure.
These repeating patterns 
sometimes cause violations of the intended semantics.
In such cases, we may exclude the violations 
by instantiating the axiom schemata
with formulas that are true 
in the corresponding summary nodes.

We start by recalling the notion of symbolic structures.
We then define the formulas 
that are true in a summary node,
and show how to instantiate induction axioms
from them.

\subsection{Symbolic Structures}
\begin{definition}[\cite{infinite-needle}]
A \emph{symbolic structure}
for a vocabulary $\Sigma$
with a set of sorts $\Sorts$
is a triple $S = (\Domain, \Bounds, \Interp)$,
where
\begin{itemize}
\item $\Domain$ is the \emph{domain} of $S$,
mapping each sort $\sort \in \Sorts$ to
a finite, non-empty, set of nodes $\Domain(\sort)$,
such that $\Domain(\sort) \cap \Domain(\sort') = \emptyset$
whenever $\sort \neq \sort'$.
For each sort $\sort \in \Sorts$,
the domain $\Domain(\sort)$ is partitioned into
``summary nodes'', denoted $\Domain^{\summary}_{\sort}$,
and ``regular nodes'', denoted $\Domain^{\regular}_{\sort}$,
such that
$\Domain(\sort) = 
\Domain^{\summary}_{\sort} \uplus \Domain^{\regular}_{\sort}$.

\item $\Bounds$ is the assignment of \emph{bound formulas}
to the nodes,
mapping for each sort $\sort$,
each summary node $n \in \Domain^{\summary}_{\sort}$
to a satisfiable \LIA{} formula $\Bounds(n) \in \QF_\LIA$
with at most one free variable,
$\FV \parens{ \Bounds(n) } \subseteq \braces{ i }$,
and each regular node $n \in \Domain^{\regular}_{\sort}$
to the bound formula $i \liaeq 0$.

\item
$\Interp$ is the \emph{interpretation} of the symbols in $\Sigma$.
Each constant symbol $c$ of sort $\sort$
is mapped by $\Interp$ to a pair $\angles{ n, z }$
where $n \in \Domain(\sort)$ 
and $z$ is an integer such that
$[i \mapsto z] \models_\subLIA \Bounds(n)$.
Each function symbol
$f \colon \sort_1 \times \dots \times \sort_k \to \sort$
is mapped by $\Interp$ to a function that maps 
tuples of nodes
$\parens{ n_1, \dots, n_k } \in \Domain(\sort_1) \times \dots \times \Domain(\sort_k)$
to a pair $\angles{ n, t }$
where $n \in \Domain(\sort)$
and $t$ is a LIA term
such that $\FV(t) \subseteq \braces{ i_1,\ldots, i_k }$
and
$\Land_{j=1}^k \Bounds(n_j)[i_j/i] \models_\subLIA \Bounds(n)[t/i]$.
A relation symbol
$R \colon \sort_1 \times \dots \times \sort_k$ is
mapped by $\Interp$
to a function that maps tuples of nodes
$\parens{ n_1, \dots, n_k } \in \Domain(\sort_1) \times \dots \times \Domain(\sort_k)$
to a LIA formula
$\psi \in  \QF_\LIA$
such that
$\FV(\psi) \subseteq \braces{ i_1, \dots, i_k }$.
For ease of read,
in case of unary function and relation symbols,
we denote the free variable in $t$, resp., $\psi$,
by $i$ instead of $i_1$.
\end{itemize}
\end{definition}

Intuitively, a symbolic structure $S$ is
a finite representation of a (possibly) infinite explicit
first-order structure, which we denote $\Explicate(S)$.
Each node $n$ in the domain of sort $\sort$ in $S$
represents possibly infinitely many elements
in the domain of sort $\sort$ in $\Explicate(S)$,
given by all pairs $\angles{ n, z }$ such that $z$
is an integer value that satisfies $n$'s bound formula $\Bounds(n)$. Note that the set of such elements $\angles{ n, z }$
is never empty since bound formulas
are required to be satisfiable.
In particular, regular nodes always represent a single element,
$\angles{n,0}$.
The interpretation function of $\Explicate(S)$
is induced by the interpretation function of $S$,
where the free variables $i_j$ that appear
in the interpretations of function and relation symbols in $S$
represent the elements of the nodes $n_j$
to which the function or relation is applied.
Here again, the additional requirements
on the terms and formulas
ensure that the induced interpretation is well-defined.

Formally, the \emph{explication} of a symbolic structure
$S = (\Domain, \Bounds, \Interp)$,
denoted $\Explicate(S)$,
is the structure $M = (\Domain^M,\Interp^M)$
defined as follows.
For a node $n$ of sort $\sort$, its set of elements is
$\ElemsOfNode(n) = \braces{
	\angles{ n, z } \mid z \in \Bounds(n)
}$,
where
with abuse of notation we write $\Bounds(n)$
for the set
$\braces{z \mid [i \mapsto z] \models_\subLIA \Bounds(n) }$.
The domain of sort $\sort \in \Sorts$ in $M$
is then the set
$\Domain^M(\sort)= \bigcup_{n \in \Domain(\sort)} \ElemsOfNode(n)$.
For a constant symbol $c$,
$\Interp^M(c) = \Interp(c)$.
For a function symbol $f$ of arity $k$,
if $\Interp(f)\parens{ n_1, \dots, n_k } = \angles{ n, t }$,
then for every $z_1,\ldots,z_k$ such that
$z_j \in \Bounds(n_j)$ for every $1 \leq j \leq k$,
\[
\Interp^M(f) \parens{\angles{ n_1, z_1 }, \dots, \angles{ n_k, z_k }}
= \angles{ n, \bar v_\subLIA(t) },
\]
where $v_\subLIA$ is an assignment that maps
$i_j$ to $z_j$ for every $1 \leq j \leq k$,
and $\bar v_\subLIA$ extends $v_\subLIA$
to all LIA terms in the standard way.
That is, the function maps the elements
$\angles{ n_1, z_1 }, \dots, \angles{ n_k, z_k }$
to an element contributed by node $n$
that is determined by evaluating the term $t$
on the specific elements $z_j$ of nodes $n_j$.
For a relation symbol $R$ of arity $k$,
if $\Interp(R)\parens{ n_1, \dots, n_k } = \psi$,
then
\[
\Interp^M(R) =
	\braces{
		\parens{ \angles{ n_1, z_1 }, \dots, \angles{ n_k, z_k } }
		\mid
		\parens{
			\forall 1 \leq j \leq k. \ z_j \in \Bounds(n_j)
		}
		\land
		[i_1 \mapsto z_1,\ldots, i_k \mapsto z_k] \models_\subLIA \psi
	}.
\]
That is, the tuples of elements in the relation are determined
by evaluating the LIA formula $\Interp(R) \parens{ n_1, \dots, n_k }$
on the specific elements $z_i$ of nodes $n_i$.

\begin{example}
	\label{ex:paxos-symbolic-model-formal}
Consider the symbolic model $S$ in
\Cref{fig:paxos-simplified-vc:symbolic-model}.
Formally, it has the domain
$\Domain^S(\valueSort) = \braces{ n_1, n_2 }$,
$\Domain^S(\roundSort) = \braces{ m_1, m_2, m }$,
where $m$ is a summary node with bound formula
$i \leq 0$, and the rest are regular nodes.
The constants $v_1, v_2, r_1, r_2$
are interpreted to the regular nodes $n_1, n_2, m_1, m_2$
respectively.
As the bound formula for $m$ is $\Bounds(m) = i \leq 0$,
the explicit elements of it are
$\Explicate(m) = \braces{ \angles{m, 0}, \angles{m, -1}, \dots }$,
and
the explicit domain of $\roundSort$ is
\[
\Domain^{\Explicate(S)}(\roundSort) = \braces{
	\angles{m_1, 0}, \angles{m_2, 0}, \angles{m, 0}, \angles{m, -1}, \dots
}.
\]
The explicit domain of the $\valueSort$ sort is
$
\Domain^{\Explicate(S)}(\valueSort) =
\braces{ \angles{n_1, 0}, \angles{n_2, 0} }
$.
The interpretation of \Proposal{} is
$\Interp^S(\Proposal)(m_1, n_1) = \top$,
$\Interp^S(\Proposal)(m_2, n_2) = \top$,
$\Interp^S(\Proposal)(m, n_2) = \top$,
and $\bot$ for all other pairs of nodes.
In the explication $\Explicate(S)$
the interpretation is
$
\Interp^{\Explicate(S)}(\Proposal) = \braces{
	\parens{ \angles{ m_1, 0 }, \angles{ n_1, 0 } },
	\parens{ \angles{ m_2, 0 }, \angles{ n_2, 0 } },
	\parens{ \angles{ m, 0 }, \angles{ n_2, 0 } },
	\parens{ \angles{ m, 0 }, \angles{ n_2, -1 } },
	\dots
}$.
The unary function $w$ is defined for $m$ as
$\Interp^S(w)(m) = \angles{ m, i - 1 }$.
Thus for every
$z \in \Bounds(m)$,
$
\Interp^{\Explicate(S)}(w) \parens{ \angles{ m, z } } = \angles{ m, z - 1 }
$.

\end{example}

\paragraph{Notation}
For a (symbolic or explicit) structure $S$
with interpretation function $\Interp^S$,
and a constant, function or relation symbol $\alpha$,
we often write $\alpha^S$ to denote $\Interp^S(\alpha)$.

\subsection{Extracting Induction Axioms from Summary Nodes}
Next we explain how
we use properties of summary nodes
in symbolic structures that violate the intended semantics
to instantiate induction axioms.
In particular, we are interested
in properties
that exclude spurious models,
refining a formula, e.g., a negated VC,
according to the intended semantics.

\subsubsection{Node Properties}
Symbolic structures are merely representations
of standard first-order structures.
As such, satisfaction of FOL formulas
in a symbolic structure $S$ is defined
by means of its explication $\Explicate(S)$.
Similarly, we can use FOL formulas
to reason about nodes in a symbolic structure.
To this end, we define properties of a node
as properties that are shared by all of its elements.
As we will see, such properties can sometimes be used
to eliminate spurious counter-models.
	
\begin{definition}[Properties of Nodes]
Let $S = (\Domain, \Bounds, \Interp)$
be a symbolic structure for $\Sigma$,
and $\psi \in \Formulas(\Sigma)$
such that $\FV(\psi) \subseteq \braces{x}$
where $x$ is a variable of sort $\sort$.
We say that a node $n \in \Domain(\sort)$
satisfies $\psi$,
or that $\psi$ is a \emph{property} of $n$,
denoted $S,[x \mapsto n] \models \psi$,
if for every $\angles{n,z} \in \ElemsOfNode(n)$,
we have that
$\Explicate(S), \brackets{x \mapsto \angles{n,z}} \models \psi$.
\end{definition}

Given a symbolic model $S$ and a node $n$,
checking if $S, [x \mapsto n] \models \psi$
can be done using the model-checking procedure
of~\cite{infinite-needle}.
For each node
$n \in \Domain(\sort)$
and formula $\psi$ with at most a single variable
$x$ of sort $\sort$,
their procedure constructs a \LIA{} formula,
denoted $\trans^S_{[ x \mapsto n]}(\psi)$,
with at most a single free integer variable $i$,
such that
$
\brackets{ i \mapsto z } \models_\subLIA \trans^S_{[x \mapsto n]}(\psi)
$
iff
$
\Explicate(S), \brackets{ x \mapsto \angles{n, z} } \models \psi
$.
Thus, determining if
$S, [x \mapsto n] \models \psi$
amounts to checking if the formula
$
\Bounds(n) \to \trans^S_{\brackets{ x \mapsto n }}(\psi)
$
is valid (in \LIA{}).

\begin{example}
	Continuing with the notations of
	\Cref{ex:paxos-symbolic-model-formal},
	let us consider the formula
	$\psi = w(x) \prec x$
	as a potential property for the summary node $m$.
	Formally,
	the interpretation of $w$ for $m$
	is defined as
	$w^S(m) = \angles{ m, i - 1 }$,
	and the interpretation of $\prec$ for $m$
	is defined as
	${\prec^S}(m, m) = i_1 < i_2$.
The resulting \LIA{} formula from the transformation is
	$
	\trans^S_{[x \mapsto m]}(\psi) = i - 1 < i
	$,
	and we need to check the validity of
	the formula
	$
	i \leq 0 \to i - 1 < i
	$
	(as $i \leq 0$ is the bound formula of $m$).
The formula is valid,
    which means that $w(x) \prec x$
    is a property of $m$.
\end{example}

\subsubsection{Induction Axioms Based on Node Properties}
Summary nodes in symbolic structures can help us
identify violations of an intended semantics
that can be eliminated by induction.
Specifically, in the case of a well-founded
or finite-domain semantics,
such violations always involve infinitely many elements,
spanning at least one summary node.
Indeed, we observe that under certain conditions,
such violations can always be witnessed by a single summary node
that violates well-foundedness or upwards-well-foundedness,
as defined next.

Throughout this section we fix
a vocabulary $\Sigma$
with a binary relation $\prec \colon \sort \times \sort$.
We consider the well-founded semantics
of $\parens{ \Sigma, \prec }$
and the finite-domain semantics of
$\parens{ \Sigma, \sort }$.

\begin{definition}[Violating Nodes]
Let $S$ be a symbolic structure for $\Sigma$.
We say that a node $n \in \Domain^S(\sort)$
 \emph{violates well-foundedness of $\prec$}
if the relation
\[
R = \braces{
	\parens{ z_1,z_2} \in \Bounds(n) \times \Bounds(n)
	\mid
	\brackets{ i_1 \mapsto z_1, i_2 \mapsto z_2 } \models_\subLIA {\prec^S}(n, n)
}
\]
on the set $\Bounds(n)$
is not well-founded.
We say that a node $n \in \Domain^S(\sort)$
\emph{violates upwards-well-foundedness of $\prec$}
if $R$ is not upwards well-founded.
\end{definition}

That is, a node $n$ violates well-foundedness
(or upwards well-foundedness) of $\prec$
if the relation induced by ${\prec^S}(n,n)$,
which is the
restriction of the interpretation of $\prec$ in $\Explicate(s)$
to the set of elements induced by $n$,
is not well-founded (resp., upwards well-founded).
Note that the set of elements induced
by a violating node is necessarily infinite.
Thus, a violating node is always a summary node.
In general, it may be difficult to determine
if a node violates well-foundedness
(or upwards well-foundedness) of $\prec$.
However, when symbolic structures are defined
using simple \LIA{} terms and formulas
(as is the case in~\cite{infinite-needle}),
it may be easy.
As a simple example,
consider the symbolic model $S$
from \Cref{fig:paxos-simplified-vc:symbolic-model}
as formally defined in
\Cref{ex:paxos-symbolic-model-formal}.
The bound formula of $m$
is $i \leq 0$,
and $\prec$ is interpreted as
${\prec^S}(m, m) = i_1 < i_2$,
which clearly violates well-foundedness.
Similarly, if $m$ had a bound formula
$i \geq k$
(for some integer $k$),
it would have violated upward well-foundedness.

\begin{lemma}
	\label{lem:violating-wf-exists}
Let $S$ be a symbolic structure for $\Sigma$
where $\prec^{\Explicate(S)}$ is a strict total order on
    $\Domain^{\Explicate(S)}(\sort)$.
	If $\prec^{\Explicate(S)}$ is not well-founded,
	then there exists a node
	$n \in \Domain^S(\sort)$
    that violates well-foundedness of $\prec$.
\end{lemma}

\DeferredProof{lem:violating-wf-exists}{\ViolatingWfExistsProof}{Let us assume to the contrary that no such node exist,
	and show in contradiction that every subset of the domain
	has a minimal element.
	Let $K \subseteq \Domain^{\Explicate(S)}(\sort)$.
	We can partition $K$ as
	\[
	K = \biguplus_{n \in \Domain^S(\sort)} U_n
	\]
	where $U_n \subseteq \Explicate(n)$.
	By our assumption, for every $n$,
	$\Bounds(n)$ is well-founded
	with the interpretation of
	${\prec^S}(n, n)$,
	and thus $\Explicate(n)$ is well-founded under
	$\prec^{\Explicate(S)}$.
	Let $e_n$ be a minimal element of
	$U_n \subseteq \Explicate(n)$ for every $n$.
Note that since $\prec^{\Explicate(S)}$ is a total order,
	each minimal element $e_n$
	is also a minimum of its respective $U_n$
	(i.e., all other elements of $U_n$ are greater than it).
$\Domain^S(\sort)$ is finite,
	and thus we have a finite set of minima
	$\braces{ e_n \mid n \in \Domain^S(\sort) }$.
Since $\prec$ is interpreted as a total order,
	there exists a minimum $m$ among
	$\braces{ e_n \mid n \in \Domain^S(\sort) }$,
	and by transitivity, $m$ is a minimum of $K = \biguplus U_n$,
	and in particular a minimal element thereof.
}

\begin{lemma}
	\label{lem:violating-fin-exists}
Let $S$ be a symbolic structure for $\Sigma$
	where $\prec^{\Explicate(S)}$ is a strict total order on
    $\Domain^{\Explicate(S)}(\sort)$.
	If $\Domain^{\Explicate(S)}(\sort)$ is infinite,
	then there exists a node
	$n \in \Domain^S(\sort)$
	that violates well-foundedness of $\prec$
	or upwards well-foundedness of $\prec$ (or both).
\end{lemma}

\DeferredProof{lem:violating-fin-exists}{\ViolatingFinExistsProof}{Since the domain of $\Explicate(s)$ is infinite,
a node $n$ such that $\Bounds(n)$ is infinite must exist.
Because $\prec^{\Explicate(S)}$ is a total order,
so is its restriction to $\Bounds(n)$,
which is exactly
$R = \bigl\{
	\parens{ z_1,z_2} \in \Bounds(n) \times \Bounds(n)
	\mid
	[i_1 \mapsto z_1, i_2 \mapsto z_2] \models_\subLIA {\prec^S}(n, n)
\bigr\}$.
Since $\Bounds(n)$ is infinite, by
\Cref{lem:order-and-finiteness},
at least one of $R$ or $R^{-1}$ is not well-founded.
}

	The previous lemmas establish that when
	well-founded semantics or
	finite-domain semantics are violated,
	there always exists a violating node
	(assuming the domain is totally ordered).
This motivates us to use properties
	of summary nodes for instantiations of induction axioms
	meant to exclude spurious counter-models.
	The idea is to exclude a spurious counter-model
	by excluding a (violating) node in it.
	If the node to be excluded has property $\psi$,
	then one way to exclude it is to say that $\psi$
	cannot be satisfied (or, in fact, $\neg \psi$ must hold).
	Our induction axioms $\xi_{\min}[\psi]$ (resp., $\xi_{\max}[\psi]$)
	can be viewed as facilitating an inductive proof of $\neg \psi$
	by asserting that if $\psi$ holds for some element,
	then there is a minimal (resp., maximal) element for which it holds.
	
\begin{example}
		\label{ex:paxos-properties-for-axiomns}
		In the case of the VC of
		\Cref{fig:paxos-simplified-vc},
		an example of a spurious counter-models
		is given in
		\Cref{fig:paxos-simplified-vc:symbolic-model}.
		In this counter-model the summary node
		violates well-foundedness of $\prec$,
		and we construct the induction axiom
		\Cref{eq:paxos-simplified-vc:axiom-vprime}
according to the property
		$\Proposal(r, v_2)$
		it satisfies.
The axiom eliminates the spurious model since
		it forces the existence of a minimal element
		for which $\Proposal(r, v_2)$
holds,
		and the minimal element leads to a contradiction:
by the invariant \Cref{eq:paxos-simplified-vc:invariant-skolem},
		the minimal element must be \Safe{},
		and by \Cref{eq:paxos-simplified-vc:one-safe}
		both \Safe{} values $v_1, v_2$ must be the same.
		Similarly, the axiom 
		\Cref{eq:paxos-simplified-vc:axiom-v}
		eliminates the model for the symmetric case
		where $v_1$ is the un-\Safe{} value,
		
\end{example}

	Recall that when we are interested in satisfiability modulo
	well-founded, resp., finite-domain, semantics,
	\Cref{claim:wf-axioms-sound}, resp.,
	\Cref{claim:fin-sound-axioms},
	ensure that
	it is always sound to add the axioms
	$\xi_{\min}[\psi]$ or $\xi_{\max}[\psi]$,
	regardless of whether $\psi$ is satisfied by a violating node.
Thus, for soundness of the approach,
    we do not even have to verify that a node is violating
    and/or that $\psi$ is its property.
Next, however, we show that under certain conditions,
	we can actually guarantee some form of progress by
	such instantiations of induction.

\begin{claim}
	\label{claim:axiom-eliminate-violating-node}
Let $S$ be a symbolic structure for $\Sigma$,
    $n \in \Domain^S(\sort)$,
and $\psi$ a formula over $\Sigma$
with a free variable $x$ of sort $\sort$ such that $S,[x \mapsto n] \models \psi$
	and for every $n' \neq n \in \Domain^S(\sort)$,
    $S, [x\mapsto n'] \models \neg\psi$.
    Then if $n$ violates well-foundedness of $\prec$,
    then
	$\Explicate(S) \not\models \xi_{\min}[\psi]$.
    If $n$ violates upwards well-foundedness of $\prec$,
    then
	$\Explicate(S) \not\models \xi_{\max}[\psi]$.
\end{claim}

\begin{proof}
	Let us consider the set of explicit elements
	that satisfy $\psi$,
	$
	K = \bigl\{
		\angles{m, z} \in \Domain^{\Explicate(S)}(\sort)
		\mid
		\Explicate(S), \brackets{ x \mapsto \angles{m, z} } \models \psi
	\bigr\}
	$.
	Since
	$S, [x \mapsto n] \models \psi$ ,
	we get $\Explicate(n) \subseteq K$,
	and since for every $n' \neq n$,
	$n' \models \neg \psi$
	we get $K \subseteq \Explicate(n)$.
	Since $\Bounds(n)$ is not well-founded,
	and $K = \Explicate(n)$,
	we have that $K \neq \emptyset$ and has no minimal element.
Therefore,
	$\Explicate(S) \not\models \xi_{\min}[\psi]$.
    The other case is symmetric.
\end{proof}

Thus, if $\psi$ is a \emph{unique} property
of a violating node $n$
(in the sense that any other node
satisfies the negation of $\psi$),
we can use $\xi_{\min}[\psi]$ or $\xi_{\max}[\psi]$
to exclude the spurious counter-model $\Explicate(S)$
of some formula $\varphi$
by conjoining the formula with the axiom.
Harking back to
\Cref{ex:paxos-properties-for-axiomns},
though the properties 
we instantiated the axioms with are not unique
(but nonetheless suffice to exclude the models),
there are unique properties of the violating node,
for example
$\Proposal(r, v_2) \land r \neq r_2$
for \Cref{fig:paxos-simplified-vc:symbolic-model},
which are in fact guaranteed to exclude the models.

We can in fact relax the uniqueness requirement
and generalize \Cref{claim:axiom-eliminate-violating-node}
by considering properties of a node $n$
that also depend on other elements in the model.
Such properties are captured by formulas $\psi$
with free variables $y_1,\ldots,y_k$ in addition to $x$.
The idea is that such $\psi$ is a property of $n$
and elements $\parens{e_1,\ldots,e_k}$ in the model
if for every element $\angles{n,z}$ in $\Explicate(n)$,
we have that
$\Explicate(S),
\brackets{ x \mapsto \angles{n,z}, y_1 \mapsto e_1, \ldots, y_k \mapsto e_k }
\models \psi$.
That is, all explicit elements of $n$ satisfy $\psi$
when combined with $\parens{e_1,\ldots,e_k}$.
(Note that this requirement can also be checked
using the model checking procedure of~\cite{infinite-needle}.)
Such properties $\psi$ can also be used to
instantiate induction axioms in order to exclude $n$.
Namely, we can attempt to exclude $n$
by showing (by induction) that $\psi$
cannot in fact hold for $n$ and $\parens{e_1,\ldots,e_k}$.
In this case,
$\xi_{\min}[\psi]$ (resp., $\xi_{\max}[\psi]$)
asserts that \emph{for every} tuple of elements
$\parens{e_1,\ldots,e_k}$,
if some element paired with
$\parens{e_1,\ldots,e_k}$
satisfies $\psi$,
then there exists a minimal (resp., maximal)
element that satisfies $\psi$
when paired with $\parens{e_1,\ldots,e_k}$.

For instance in
\Cref{fig:paxos-simplified-vc:symbolic-model}
we can exclude the violating node $m$
with a well-founded induction axiom
obtained from
$\psi = \Proposal(r, v)$,
which is a property of node $m$
and the element corresponding to $v_2$.
The resulting axiom,
which universally quantifies over $v$, subsumes the two axioms
\Cref{eq:paxos-simplified-vc:axiom-v,eq:paxos-simplified-vc:axiom-vprime}
and thus also excludes the symmetric spurious model
where $v_1$ is un-\Safe{}.

When considering a relaxed notion of node properties
as above we can similarly relax the uniqueness requirement
that ensures progress to only require that $\neg \psi$ holds for
the other nodes with the same tuple $\parens{e_1,\ldots,e_k}$ with which $n$ satisfies $\psi$:
\begin{claim}\label{claim:axiom-eliminate-violating-node-gen}
Let $S$ be a symbolic structure for $\Sigma$,
    $n \in \Domain^S(\sort)$,
and $\psi \in \Formulas(\Sigma)$
    such that $\FV(\psi) \subseteq \braces{ x, y_1, \dots, y_k }$,
    where $x, y_1, \dots, y_k$ are of sorts
    $\sort, \sort_1,\ldots,\sort_k$
    (not necessarily distinct),
    respectively.
    Suppose that there exist
    $(e_1,\ldots,e_k) \in \Domain^{\Explicate(S)}(\sort_1) \times \cdots \times \Domain^{\Explicate(S)}(\sort_k)$
    such that
    for all
	$z \in \Bounds(n)$,
	$\Explicate(S),
		\brackets{
			x \mapsto \angles{n, z},
			y_1 \mapsto e_1, \dots, y_k \mapsto e_k
}
	\models \psi
	$
	and for any other $n' \neq n \in \Domain^S(\sort)$,
	for all $z' \in \Bounds(n')$,
	$\Explicate(S),
	\brackets{
		x \mapsto \angles{n', z'},
		y_1 \mapsto e_1, \dots, y_k \mapsto e_k
	}
	\models \neg \psi
	$.
    Then if $n$ violates well-foundedness of $\prec$,
    then
	$\Explicate(S) \not\models \xi_{\min}[\psi]$.
    If $n$ violates upwards well-foundedness of $\prec$,
    then
	$\Explicate(S) \not\models \xi_{\max}[\psi]$.
\end{claim}

\begin{proof}
	Given a tuple of elements
    $(e_1,\ldots,e_k) \in \Domain^{\Explicate(S)}(\sort_1) \times \cdots \times \Domain^{\Explicate(S)}(\sort_k)$,
    we consider the set of tuples of explicit elements
	that satisfy $\psi$ with $(e_1,\ldots,e_k)$:
	\[
	K_{e_1,\ldots,e_k} \defeq
	\braces{
		\angles{m, z}
		\mid
		m \in \Domain^S(\sort),
        z \in \Bounds(m),
        \Explicate(S),
			\brackets{
				x \mapsto \angles{m, z},
				y_1 \mapsto e_1, \dots, y_k \mapsto e_k
			}
			\models \psi
	}.
	\]
As given, there exist $e_1, \dots, e_k$
	such that
	$K_{e_1, \dots, e_k} = \Explicate(n)$.
	If $\Bounds(n)$ is not well-founded,
	$K_{e_1, \dots, e_k} \neq \emptyset$ has no minimal element.
Therefore,
	$
	\Explicate(S) \not\models \xi_{\min}[\psi]
	$.
	The case for when $\Bounds(n)$ is not upwards well-founded
	is symmetrical.
\end{proof}

As demonstrated earlier, uniqueness of $\psi$,
in the sense of \Cref{claim:axiom-eliminate-violating-node}
or even the weaker sense of
\Cref{claim:axiom-eliminate-violating-node-gen},
is a sufficient
but not a necessary condition
for the axiom $\xi_{\min}[\psi]$
(resp., $\xi_{\max}[\psi]$)
to exclude $\Explicate(S)$.
Indeed, in our evaluation we use
easily computable properties of summary nodes,
in the form of conjunctions of literals
that are satisfied by the node,
without requiring that they are unique properties.

The astute reader might notice
that even when using unique properties
to instantiate induction axioms
that exclude spurious counter-models,
convergence is not implied.
First, there may be infinitely many different
spurious counter-models for $\varphi$
(in either of the semantics).
Furthermore, it is possible
that a spurious counter-model
still exists but cannot
be represented by a symbolic structure;
and even if a symbolic counter-model exists,
we cannot guarantee that a unique property exists
for a violating node in it
(a violating node, on the other hand, is guaranteed
to exist when the domain is totally ordered,
by \Cref{lem:violating-wf-exists,lem:violating-fin-exists}).

Despite these sources of incompleteness,
our evaluation shows that the approach can be effective,
even when the axioms are instantiated
with rather simple properties of summary nodes.
Moreover,
as we will see in the next section,
if the formula $\varphi$ at hand is in \osc{},
there does exist a finite set of axioms,
each derived from a unique property
of a potential
summary node
in a spurious symbolic counter-model,
such that adding these axioms essentially
excludes all spurious counter-models\footnote{In fact, even for an \osc{} formula,
	the axioms do not exclude all spurious counter-models;
	they only ensure that whenever
	a spurious counter-model exists,
	so does a model that obeys the intended semantics.
}.

 \section{Decidability of \osc{} under Well-Founded and Finite Semantics}
\label{sec:decidable}
In this section we consider the
Ordered Self-Cycle (\osc{}) fragment of FOL,
introduced in~\cite{infinite-needle}.
\osc{} extends the decidable EPR fragment
by allowing a designated sort, totally ordered by $\prec$,
to have cyclic functions (under certain conditions),
leading to cases where the only satisfying models
have infinite domains.
We show a reduction from
satisfiability modulo well-founded semantics
and finite-domain semantics
to standard FOL semantics for \osc{}.
Further, relying on the decidability
of satisfiability in \osc{},
we show that satisfiability modulo those
intended semantics is also decidable in \osc{}.
Beyond solving the decision problem,
our proofs
provide
algorithms that given a formula in \osc{},
either produce a satisfying model with the intended semantics,
or determine no such model exist.
Formally we prove the following:

\begin{theorem}[Decidability of Satisfiability modulo Well-Foundedness]
	\label{thm:wf-decidability}
	There exists an algorithm
	that given a formula $\varphi \in \osc$,
	either determines that $\varphi$ has no model
    where $\prec$ is interpreted as a well-founded relation,
	or returns a symbolic model $S$
	such that $\Explicate(S) \models \varphi$
	and $\prec^{\Explicate(S)}$ is well-founded.
\end{theorem}

\begin{theorem}[Decidability of Finite-Model Satisfiability]
	\label{thm:fin-decidability}
	There exists an algorithm
	that given a formula $\varphi \in \osc$,
	either determines that $\varphi$ has no finite model,
	or returns a finite model of $\varphi$.
\end{theorem}

By \Cref{thm:wf-decidability},
we can decidably verify the safety property of
\Cref{fig:paxos-simplified-vc}
under well-founded semantics,
and find a counter-model for the variant that uses
\Cref{eq:paxos-simplified-variant},
where well-founded semantics are insufficient.
By \Cref{thm:fin-decidability},
we can decidably verify the second example
under
finite-model semantics,
as both the example and its variant are in \osc{}.

We start by giving some background on
the \epr{} and \osc{} fragments
and their properties.

\subsection{The \osc{} Fragment}

\paragraph{EPR}
The decidable Effectively PRopositional (\epr{})\footnote{Note that, following previous work,
	we denote by \epr{} a strict extension
	of the original \epr{} fragment
	(which is single-sorted and does not allow
	any function symbols nor $\forall \exists$ quantifier alternations).
} fragment
of many-sorted FOL contains formulas
whose \emph{quantifier alternation graph} has no cycles.
Given a formula $\varphi$
in negation normal form\footnote{The requirement of negation normal form
	does not restrict the formulas in
	a meaningful way,
	as the negation normal form of a formula $\varphi$
	can be computed automatically from $\varphi$,
	and, in particular,
	without renaming quantified variables.
},
its quantifier alternation graph
is a directed graph whose set of vertices
is the set of sorts appearing in $\varphi$,
and whose set of edges
consists of all edges
$\sort_1 \to \sort_2$
such that one of the following holds: \begin{inparaenum}[(i)]\item $\sort_2$ is the output sort of some function $f$
	appearing in $\varphi$
	and $\sort_1$ is the sort of one of the arguments of $f$;
	or
	\item an $\exists y \colon \sort_2$ quantifier
	appears in the scope of some
	$\forall x \colon \sort_1$ quantifier
	in $\varphi$.
\end{inparaenum}

The acyclicity of the quantifier alternation graph
ensures that the vocabulary
of the skolemization of $\varphi$
generates only finitely many ground terms,
and thus \epr{} enjoys a finite-model property,
which follows from Herbrand's theorem.
In particular, satisfiability is decidable for formulas in \epr{}.

\paragraph{\osc{}}
The decidable Ordered Self-Cycle (\osc{}) fragment
is an extension of EPR,
which relaxes the acyclicity requirement in certain cases.
In \osc{}, one designated sort (denoted $\sinfty$)
is allowed to have self-loops
in the quantifier alternation graph.
Formally,
\osc{}
assumes vocabularies that include
the sort $\sinfty$ and a binary relation symbol
$\prec \colon \sinfty \times \sinfty$.
Formulas in \osc{}
are of the form
$\varphi \land \TotalOrder$,
where:
\begin{asparaenum}[(1)]
	\item
	$\varphi$ is in negation normal form;
	\item
	the only logical variable of sort $\sinfty$
	that appears in $\varphi$ is $x$;
\item
	in the quantifier alternation graph of $\varphi$
	the only cycles are self-cycles at $\sinfty$,
	and the only outgoing edges
	from $\sinfty$ are to $\sinfty$;
\item
	functions and relations in $\varphi$ other than $\prec$
	have at most one argument of sort $\sinfty$;
	and
\item
	the only non-ground term of sort $\sinfty$
	that appears as an argument to functions in $\varphi$
	is $x$.
\end{asparaenum}

\medskip
In~\cite{infinite-needle}, the satisfiability problem
of formulas in \osc{} is proved to be decidable,
in particular by showing that for every satisfiable formula
$\varphi \in \osc{}$
there exists a satisfying symbolic model with certain properties.
The proof works by first reducing formulas in \osc{}
to a simpler form and then showing
that satisfiability is decidable for those simpler-form formulas.
We prove
\Cref{thm:wf-decidability,thm:fin-decidability}
in the same manner,
by showing that analogous properties
hold for the simpler-form formulas.
For the first step,
we use the same reduction used in~\cite{infinite-needle},
which not only preserves satisfiability, but also
allows translation of models between formulas
and their reduced forms (in both directions).
Importantly, well-foundedness, resp., finiteness of the domain,
is maintained by the model translations.
Thus, the reduction preserves satisfiability
modulo the intended semantics we consider.
We start by defining the set of simpler-form
formulas in \osc:

\begin{definition}[\oscReduced]
	\label{def:osc-reduced}
	We denote the set of all \osc{} formulas
	where the only sort is $\sinfty$
	and all quantifiers are universal
	by \oscReduced{}.
\end{definition}

Note that since in \osc{},
the designated sort $\sinfty$
appears at most once as an argument
to function and relation symbols,
functions and relations in \oscReduced{}
are all unary.
Next we describe the reduction
from \osc{} to \oscReduced{}.

\paragraph{Reducing \osc{} Formulas to \oscReduced{}}
The reduction
transforms a given formula $\varphi \in \osc$
into a formula $\Reduced{\varphi} \in \oscReduced$
by performing the following steps.
First, $\varphi$ is skolemized such that
an existential quantifier
in the scope of universally quantified variables
$x_1, \dots, x_k$
is replaced by a Skolem function
whose arguments are
$x_1, \dots, x_k$.
Note that due to the single variable property of \osc{},
and the requirements of the quantifier alternation graph,
the designated $\sinfty$ sort never appears as an argument
of these Skolem functions.
This is a slight generalization
of the usual approach to skolemization
that does not require prenex normal form.

Next, since the quantifier alternation graph
of all sorts other than $\sinfty$ contains no cycles,
there are finitely many ground terms of those sorts,
and we can instantiate all remaining universal quantifiers of these sorts with these terms.
We replace each universal quantifier by a conjunction
over all instantiations produced
by the ground terms of the relevant sort,
and arrive at a formula
which contains only universal quantifiers
with the single variable $x \colon \sinfty$.

Finally, we remove all sorts $\sort \neq \sinfty$
from the formula,
``baking in'' the ground terms
of such sorts into functions and relations,
and adding constraints
that ensure that functions and relations
are the same if they are created by
different ground terms that are equal.
For example, given a predicate $P \colon \sort \times \sinfty$,
and ground terms $g, g' \colon \sort$,
the resulting formula will have two predicates,
$P_g \colon \sinfty, P_{g'} \colon \sinfty$.
Thus all functions and relations other than $\prec$
are unary.

The resulting formula,
$\Reduced{\varphi}$,
is a single-sort \osc{} formula,
with a single universally quantified variable
$x \colon \sinfty$.
Thus $\Reduced{\varphi} \in \oscReduced$.
The following claim formalizes
the correctness of the reduction:

\begin{claim}[\cite{infinite-needle}]
	\label{claim:osc-reduction}
	For every formula $\varphi \in \osc$,
	for every model
$M \models \varphi$, there exists
	$M' \models \Reduced{\varphi}$
	such that
	$\Domain^M(\sinfty) = \Domain^{M'}(\sinfty)$
	and
	${\prec^M} = {\prec^{M'}}$;
	and for every model $M' \models \Reduced{\varphi}$,
	there exists
	$M \models \varphi$
	such that
	$\Domain^M(\sinfty) = \Domain^{M'}(\sinfty)$,
	${\prec^M} = {\prec^{M'}}$,
and all sorts other than $\sinfty$
	have a finite domain.
\end{claim}

\Cref{claim:osc-reduction} ensures that
the reduction preserves
satisfiability modulo the intended semantics.
Further, by relying on the \epr{} properties
of all sorts other than $\sinfty$,
it guarantees that
the finite-domain semantics of the reduced formula
translate to finite-structure semantics of \osc{}.
Additionally,~\cite{infinite-needle} shows how $M'$
can be computed from $M$ and vice versa
(when both are given by symbolic structures).
Thus, to prove \Cref{thm:wf-decidability} and \Cref{thm:fin-decidability}
it suffices to consider formulas in \oscReduced{}.

Since formulas in $\oscReduced$
only include a single sort,
in the sequel we do not mention the sort explicitly
(neither in formulas nor in models).

\paragraph{Symbolic Models for $\oscReduced$}
Our proofs utilize properties of the symbolic models
computed by the decision procedure of~\cite{infinite-needle} for
satisfiability
of $\oscReduced$ in the standard semantics.
To understand these properties, we recall
the construction of a symbolic model $S$
from an arbitrary satisfying model $M$ for
$\varphi \in \oscReduced$ that the proof of~\cite{infinite-needle}
hinges on.
The construction partitions $M$'s domain
into equivalence classes according to a set of literals
of three kinds:
\begin{enumerate}
	\item
	\textbf{Element atoms}:
 $P(x),\neg P(x), x \bowtie g$
	
	\item
	\textbf{Image atoms}:
	$P(f(x)), \neg P(f(x)), f(x) \bowtie g$
	
	\item
	\textbf{Mixed atoms}:
	$x \bowtie f(x), f(x) \bowtie h(x)$
\end{enumerate}
Such literals are generated
for every relation symbol $P$,
ground term $g$
and function symbols $f,h$ that appear in $\varphi$,
where
${\bowtie} \in \braces{ \prec, \succ, \foleq }$.
The set of all possible atoms for a given formula $\varphi$ is denoted
by $\Atoms_\varphi$. Note that this set is finite.
The construction of $S$ in~\cite{infinite-needle} introduces
a node $n$ for every non-empty equivalence class
w.r.t.\ $\Atoms_\varphi$ of elements in $M$,
such that all the elements in $\Explicate(n)$ satisfy the same atoms.
Thus, every node in $S$ can be characterized by the atoms
satisfied by the elements in the equivalence class
that introduced it.
Formally:

\begin{definition}[Characterization]
A subset $C \subseteq \Atoms_\varphi$
is called a \emph{characterization} when it is satisfiable,
and for every atom $\alpha \in \Atoms_\varphi \setminus C$,
$C \cup \braces{ \alpha }$ is unsatisfiable.
We denote by $\Chars_\varphi$ the set of all
characterizations in $\varphi$,
and we partition this set in two:
$\Chars_\varphi = \Chars_\varphi^{\regular} \uplus \Chars_\varphi^{\summary}$,
such that any characterization $C \in \Chars_\varphi^{\summary}$
does not contain any $x \foleq g$ atoms,
and any $C \in \Chars_\varphi^{\regular}$
contains at least one such atom.
\end{definition}

Based on the construction of $S$ from $M$,
the proof in~\cite{infinite-needle} shows
there exists a computable function $\SymbolicCandidates$
such that
for every formula $\varphi \in \oscReduced$,
$\SymbolicCandidates(\varphi)$ is a finite set of symbolic structures,
and if $\varphi$ is satisfiable,
there exists a symbolic model
$S \in \SymbolicCandidates(\varphi)$
such that $\Explicate(S) \models \varphi$.
Further, every symbolic structure
$S \in \SymbolicCandidates(\varphi)$
has the following additional properties
(which will be useful for us).

\begin{lemma}[\cite{infinite-needle}]
	Given a formula $\varphi \in \oscReduced$,
	let $\ell$ be the number of function symbols appearing in $\varphi$,
	and let $S \in \SymbolicCandidates(\varphi)$.
	Then $S$ has the following properties:

	\begin{enumerate}[label={(Prop. \arabic*)},ref=Prop. \arabic*,leftmargin=*]
		\item
		\label{prop:osc-model:characterization}
		Each characterization $C \in \Chars_\varphi$
		introduces at most one node in $S$,
		and each node $n$ corresponds to a characterization $C$:
		$S, [x \mapsto n] \models \Land C$
		and for every $n' \neq n$,
		$S, [x \mapsto n'] \models \neg \Land C$.
        (That is, $\Land C$ is a unique property of $n$.)
		
		\item
		\label{prop:osc-model:regular}
		A node $n$ with characterization $C$ is a regular node
		iff
		there exists some ground term $g$ such that
		$x \foleq g \in C$ (i.e., iff $C \in \Chars_\varphi^{\regular}$).
		
		\item
		\label{prop:osc-model:fun-terms}
		For any regular node $n$
		and function symbol $f$,
		$f^S(n) = \angles{n', k}$
		for some $n'$ and $0 \leq k \leq \ell$.\footnote{There is an oversight in~\cite{infinite-needle},
			where function terms for regular nodes
			are always given as $0$.
			We confirmed with the authors that
a more intricate consideration is needed,
	        where the function terms for regular nodes
	        are integers between $0$ and $\ell$.
	        \neta{added:}
	        See amended proof in~\citep{infinite-needle-arxiv}.
}
		
		For any summary node $n$
		with characterization $C \in \Chars_\varphi^{\summary}$,
		and function symbol $f$,
		$f^S(n) = \angles{ n',  t }$
		such that if $n'$ is a regular node then $t = 0$,
		and otherwise, $t = i + k$
		for some $-\ell \leq k \leq \ell$.
		Moreover, $k < 0$ iff $x \succ f(x) \in C$
		and $k > 0$ iff $x \prec f(x) \in C$.

		\item
		\label{prop:osc-model:image}
		For every node $n$ with characterization $C$
		and function $f$ such that
		$f^S(n) = \angles{n', t}$
		for some node $n'$ with characterization $C'$,
		let $K \subseteq C'$ be the set of element atoms in $C'$
		and let $I \subseteq C$ be the set of
		$f$-image atoms in $C$ (where $f$-image atoms are image atoms involving $f(x)$),
then $\braces{ \alpha\brackets{ f(x) / x } \mid \alpha \in K } = I$.
		
		\item
		\label{prop:osc-model:order}
		The binary relation $\prec$
		is interpreted as a total order in $\Explicate(S)$,
		and for every summary node $n$,
		${\prec}^S(n, n) = i_1 < i_2$,
		and for every regular node $n$,
		${\prec}^S(n, n) = \bot$.
		
		\item
		\label{prop:osc-model:bounds}
		All bound formulas $\Bounds(n)$
		are $\top$ for summary nodes
		and $i \liaeq 0$ for regular nodes.
	\end{enumerate}
\end{lemma}

\subsection{Decidability of \osc{} Modulo Well-Founded Semantics}
\label{sec:decidable:well-founded}
Since $\varphi \models \TotalOrder[\prec]$
for every $\varphi \in \oscReduced$,
the induction schema from \Cref{def:wf-theory}
is sound for the well-founded semantics of \oscReduced{}.
To prove decidability,
we first show that in the case of \oscReduced{},
we can compute for each formula
a finite subset of the induction axioms from
\Cref{def:wf-theory} that allows us to reduce
satisfiability modulo well-founded semantics
to standard satisfiability
in a sound and complete way.

\begin{definition}[Well-Founded Axioms for \oscReduced{}]
	For a formula $\varphi \in \oscReduced$,
	we define
	\[
	\Xi_{\min}^\varphi \defeq \Land_{C \in \Chars_\varphi^{\summary}} \xi_{\min} [\medLand C].
	\]
\end{definition}

\begin{lemma}[Completeness]
	\label{lem:well-founded-completeness}
	Given a formula $\varphi \in \oscReduced$,
	if
	$\varphi \land \Xi_{\min}^\varphi$
	is satisfiable
	then $\varphi$ has a model where $\prec$ is interpreted as a well-founded relation.
\end{lemma}
Equivalently, if $\varphi$ is unsatisfiable
modulo well-founded semantics,
then $\varphi \land \Xi_{\min}^\varphi$ is unsatisfiable.
Intuitively, this means that the axioms
are sufficient to capture the intended semantics.
Note that this does not mean that
$\Xi_{\min}^\varphi$
excludes all non-well-founded models of $\varphi$;
it only means that if a non-well-founded model exists for
$\varphi \land \Xi_{\min}^\varphi$,
there must also exist one that is well-founded.
In particular, while \Cref{claim:wf-axioms-sound} ensures
that every well-founded model satisfies
$\Xi_{\min}^\varphi$,
in the converse direction it is possible that a model satisfies
$\Xi_{\min}^\varphi$ even though the model
is \emph{not} well-founded.
Still, in the latter case,
\Cref{lem:well-founded-completeness}
guarantees that a (possibly different)
well-founded model exists as well,
and in fact, the proof shows that it can be constructed.

The proof of
\Cref{lem:well-founded-completeness}
shows how to construct a (symbolic) well-founded model
for $\varphi$ from a symbolic model of
$\varphi \land \Xi_{\min}^\varphi$.
The construction consists of two steps.
First, we show that given a symbolic model of
$\varphi \land \Xi_{\min}^\varphi$
it is possible to amend its function interpretations
such that all elements interpreting ground terms
do not form infinitely decreasing chains,
\neta{added:}
roughly by using the minimal elements guaranteed by the axioms
to break such chains.
This is achieved by
\Cref{lem:well-founded-completeness:non-negative-terms}
which shows that all function interpretations can be amended
to only use terms of the form $i + k, k$
where $k \geq 0$.
Second, in \Cref{lem:well-founded-completeness:has-semi-bounded},
we show that we can obtain a well-founded model of $\varphi$
as a sub-structure of the model constructed by the first step
by amending the bound formulas to be $i \geq 0$
(or $i \liaeq 0$ for regular nodes).

\begin{lemma}
	\label{lem:well-founded-completeness:non-negative-terms}
	If $\varphi \land \Xi_{\min}^\varphi$ is satisfiable,
	then $\varphi$ has a model which is
the explication of
a symbolic model $S$,
where
	all bound formulas are
	$i \liaeq 0$ for regular nodes
	or $\top$ for summary nodes,
	all function terms are
	$k$ or $i + k$ for some $k \geq 0$,
	and for every summary node $n$,
	${\prec^S}(n, n) = i_1 < i_2$.
\end{lemma}

\begin{lemma}
	\label{lem:well-founded-completeness:has-semi-bounded}
 	If $\varphi$ has a model which is
the explication of
a symbolic model $S$
	with the properties listed in
	\Cref{lem:well-founded-completeness:non-negative-terms},
then there exists a symbolic model $S'$
	which is identical to $S$
	except that
	all bound formulas
	are either $i \liaeq 0$ or $i \geq 0$,
	and $\Explicate(S') \models \varphi$.
\end{lemma}

\begin{proof}[Proof of \Cref{lem:well-founded-completeness}]
Let us assume that $\varphi \land \Xi_{\min}^\varphi$ is satisfiable.
	From
	\Cref{lem:well-founded-completeness:non-negative-terms}
	there exists a symbolic model
	$S$
	with the properties listed above.
Thus, from
	\Cref{lem:well-founded-completeness:has-semi-bounded}
	there exists a symbolic model
	$S'$
	which is identical to $S$,
	except that all bound formulas are
	$i \liaeq 0$ or $i \geq 0$,
	and
	$\Explicate(S') \models \varphi$.
	
	The bound formula for every summary node $n$ in $S'$
	is $i \geq 0$,
	and since $S'$ is identical to $S$
	except for the bound formulas,
	for every summary node $n$,
	${\prec^{S'}}(n, n) = i_1 < i_2$.
	Thus $n$ does not violate well-foundedness of $\prec$,
	and of course, any regular node does not violate well-foundedness.
	Thus, from
	\Cref{lem:violating-wf-exists},
	$\Explicate(S')$ is well-founded.
\end{proof}

\begin{proof}[Proof of
	\Cref{lem:well-founded-completeness:non-negative-terms}
]
	Let us denote
	$\varphi' \defeq \varphi \land \Skolem \parens{ \Xi_{\min}^\varphi }$,
where
	$\Skolem \parens{ \Xi_{\min}^\varphi }$
	skolemizes each induction axiom
	$\xi_{\min}[\medLand C]$ in $\Xi_{\min}^\varphi$ separately,
	introducing a fresh Skolem constant
	for the minimal element of $C$.
	Note that $\varphi' \in \oscReduced$, and
    since $\varphi \land \Xi_{\min}^\varphi$
    is satisfiable and skolemization preserves satisfiability,
	$\varphi'$ is also satisfiable.
Since $\varphi' \in \oscReduced$ is satisfiable,
	there exists a symbolic model
$S' \in \SymbolicCandidates(\varphi')$
	such that $\Explicate(S') \models \varphi'$.
We construct a new symbolic model
$S$
	such that $\Explicate(S) \models \varphi$
	and $S$ is identical to $S'$
	except for function interpretations of summary nodes,
	and in particular all function terms in $S'$
	are either $k$ or $i + k$
	for some $k \geq 0$.

	Let $n$
    be a summary node in $S'$ with interpretation
	$f^{S'}(n) = \angles{n', i + k}$.
	If $k \geq 0$, we keep $f^{S}(n)$ unchanged.
    Otherwise, to define $f^{S}(n)$,
    consider the characterization
    $C \in \Chars_{\varphi'}$ of $n$ in $S'$
	(\ref{prop:osc-model:characterization}).
We denote its $\varphi$-atoms by
	$C' = C \cap \Atoms_\varphi$.
	From $\Skolem \parens{ \Xi^{\min}_\varphi }$,
	and in particular $\Skolem \parens{ \xi_{\min}[\medLand C'] }$,
	there exists a node $m$ in $S'$
to which the Skolem constant
    of the ``minimal'' element of $C'$ is interpreted.
    The node $m$ is a regular node (\ref{prop:osc-model:regular})
    and $S', \brackets{ x \mapsto m } \models \Land C'$.
Since $m$ is a regular node,
	we know that $f^{S'}(m) = \angles{m', k'}$
	for some $m'$ and $0 \leq k' \leq \ell$
	(\ref{prop:osc-model:fun-terms}).
We thus redefine in $S$,
	$f^{S}(n) = \angles{m', k'}$.
	
	Note that all the relation interpretations,
	in particular for $\prec$,
	and all bound formulas
	remain unchanged between $S'$ and $S$.
	Namely, all bound formulas for $S$ are
	$i \liaeq 0$ for regular nodes
	or $\top$ for summary nodes
	(\ref{prop:osc-model:bounds}),
	and for every summary node $n$,
	${\prec^S}(n, n) = i_1 < i_2$
	(\ref{prop:osc-model:order}).

	We now prove that after all functions $f, h, \dots$
	appearing in $\varphi$ have been redefined,
	all nodes satisfy the same atoms
	$\alpha \in \Atoms_\varphi$
    that they satisfied in $S'$.
	This is enough in order to show,
	by the same induction as in~\cite{infinite-needle},
	that $\Explicate(S) \models \varphi$,
	due to the single variable property of \osc{}.

	Let $n$ be a node with characterization $C$ in $S'$
	and let us denote $C' = C \cap \Atoms_\varphi$.
	From \ref{prop:osc-model:characterization},
	$S', \brackets{ x \mapsto n } \models \Land C \models \Land C'$,
	and we now prove that
	$S, \brackets{ x \mapsto n } \models \Land C'$.
	
	All element atoms in $C'$ are still satisfied by $n$,
	since we did not change the interpretation of
	any ground terms,
	unary relations,
	or the binary relation ${\prec}$.

    As for $f$-image atoms in $C'$,
    they may only be affected in nodes for which $f^S(n)$ was
    redefined.
    In that case, recall that $f^{S}(n)$ is redefined to be $\angles{m', k'}$, where
    $f^{S'}(m) = \angles{m', k'}$ for $m$ such that
    $S', \brackets{ x \mapsto m } \models \Land C'$.
Therefore, from \ref{prop:osc-model:image}
	the $f$-image atoms of $m$
	match the element atoms of $m'$,
	and transitively,
	the $f$-image atoms of $n$,
	which agree with $m$, match the element atoms of $m'$.

	Next we consider mixed atoms in $C'$.
	We start from mixed atoms of the form $x \bowtie f(x)$.
Again, these may only be affected if $f^S(n)$ is redefined.
    In that case, the explication of the ``minimal'' node $m$
    is also a minimum
    among the elements in $\Explicate(S')$ that satisfy $\Land C'$,
	since $\prec$ is interpreted as a total order
	(\ref{prop:osc-model:order}).
	Thus for all $z \in \Bounds(n)$
	$
	\angles{m, 0} \prec^{\Explicate(S')} \angles{n, z}
	$.
Since $f^{S'}(n) = \angles{n', i + k}$
	where $k < 0$,
	from \ref{prop:osc-model:fun-terms}
	we know that $n$ is marked with $x \succ f(x)$.
Since $x \succ f(x) \in C'$,  $S', \brackets{ x \mapsto m } \models \Land C'$,
    and $f^{S'}(m) = \angles{m', k'}$,
    we have that
	$\angles{m', k'}
	\prec^{\Explicate(S')} \angles{m, 0}
	\prec^{\Explicate(S')} \angles{n, z}$,
	for all $z \in \Bounds(n)$.
Thus, after defining $f^S(n) = \angles{m', k'}$,
    we still have $S, \brackets{ x \mapsto n } \models x \succ f(x)$.

	Lastly, we consider mixed atoms $f(x) \bowtie h(x)$.
    We explain the case of $f(x) \prec h(x)$.
    The case of $f(x) \foleq h(x)$ is similar.
	If $x \prec f(x) \prec h(x)$,
	we did not redefine $f$ and $h$,
	and thus  $f(x) \prec h(x)$
	is still satisfied by $n$.
	If $f(x) \prec x \prec h(x)$
	then we did not redefine $h(x)$,
	and thus $x \prec h(x)$ is still satisfied,
	and as shown above,
	$f(x) \prec x$ is also still satisfied.
	Since ${\prec}$ is interpreted as a transitive relation,
	and was not changed,
	we have transitively that
	$f(x) \prec h(x)$ is still satisfied by $n$.
	Finally, if $f(x) \prec h(x) \prec x$,
	then both $f(x)$ and $h(x)$ were redefined
	according to the same minimal node $m$.
	Let us denote
	$f^{S'}(m) = \angles{m_1, k_1}$
	and
	$h^{S'}(m) = \angles{m_2, k_2}$.
	Since
	$S', \brackets{ x \mapsto m} \models \Land C'$,
	and $f(x) \prec h(x) \in C'$,
	we have
	$\angles{m_1, k_1} \prec^{\Explicate(S')} \angles{m_2, k_2}$,
	and thus
	$S, \brackets{ x \mapsto n } \models f(x) \prec h(x)$
(again, note that $\prec$ is interpreted identically in $S'$ and $S$).
	
Combining the above,
	we get that $S$ is a well-defined
symbolic model
	with the properties listed in
	\Cref{lem:well-founded-completeness:non-negative-terms},
	such that
	$\Explicate(S) \models \varphi$.
\end{proof}

\begin{proof}[Proof of
	\Cref{lem:well-founded-completeness:has-semi-bounded}]
	Let $S$ be a symbolic structure
	with the properties listed in \Cref{lem:well-founded-completeness:non-negative-terms},
	such that $\Explicate(S) \models \varphi$.
We prove that a symbolic model $S'$
	which is identical to $S$,
	except that all bound formulas of summary nodes
	are set to $i \geq 0$,
	is well-defined and
	$\Explicate(S') \models \varphi$.

	Recall that for a symbolic model to be well-defined
	all function interpretations must adhere to the bound formulas.
	I.e., for every function $f$,
	with interpretation
	$f^{S'}(n) = \angles{n', t}$,
	for any $z \in \Bounds(n)$,
	$t[z / i] \in \Bounds(n')$.
	
If $n'$ is a regular node,
	we did not modify its bound formulas,
	and since $S$ is well-defined,
	$t = 0$.
Otherwise, we need to show that
	for any $z \geq 0$,
	$t[z / i] \geq 0$.
	Since $t$ is either $k$ or $i + k$
	for some $k \geq 0$,
	$S'$ is well-defined.

Note that for any node $n$,
	$\Bounds^{S'}(n) \subseteq \Bounds^S(n)$,
	and thus the structure $\Explicate(S')$
	is a sub-structure of $\Explicate(S)$.
	Since $\varphi$ is a universal formula,
	we have
	$\Explicate(S) \models \varphi
	\Rightarrow \Explicate(S') \models \varphi$.
\end{proof}

\Cref{lem:well-founded-completeness} together with
\Cref{claim:wf-axioms-sound}
imply that the finite set of axioms of
$\Xi_{\min}$ is sound and complete
for formulas in \osc{} with well-founded models:

\begin{corollary}[Soundness and Completeness]
	Given a formula $\varphi \in \oscReduced$,
	$\varphi \land \Xi_{\min}^\varphi$
	is satisfiable
	iff
	$\varphi$ has a model where $\prec$ is interpreted as a well-founded relation.
\end{corollary}

\Cref{thm:wf-decidability} then follows since satisfiability of $\osc$,
to which  $\varphi \land \Xi_{\min}^\varphi$ belongs,
is decidable~\cite{infinite-needle},
and our proof of \Cref{lem:well-founded-completeness}
provides a construction of a well-founded model from
a model of $\varphi \land \Xi_{\min}^\varphi$
that is returned by the decision procedure of~\cite{infinite-needle}.

\subsection{Decidability of \osc{} Modulo Finite-Structure Semantics}
We move on to prove decidability
of satisfiability of \oscReduced{}
modulo finite-structure semantics
(\Cref{thm:fin-decidability}).
As in the case of the well-founded semantics,
we do so by identifying a finite set
of axioms that allows us to reduce
the problem to standard satisfiability.
We leverage the relation between finiteness
of a totally-ordered domain
and bidirectional well-foundedness of the order,
and augment the axioms used in
\Cref{lem:well-founded-completeness}
with their upwards variants:

\begin{definition}[Upwards Well-Founded Axioms for \osc{}]
	For a formula $\varphi \in \oscReduced$,
	we define
	\[
	\Xi_{\max}^\varphi \defeq \Land_{C \in \Chars_\varphi^{\summary}} \xi_{\max}[\medLand C].
	\]
\end{definition}

\begin{lemma}[Completeness]
	\label{lem:finite-completeness}
	Given a formula $\varphi \in \oscReduced$,
	if
	$\varphi \land \Xi_{\min}^\varphi \land \Xi_{\max}^\varphi$
	is satisfiable
	then $\varphi$ has a finite model.
\end{lemma}

In order to prove this, we use the following two lemmas for
$\varphi \in \oscReduced{}$
(which similarly to the well-founded case,
show that we can construct a finite model for $\varphi$
by amending function interpretations and bound formulas
of a symbolic model for
$\varphi \land \Xi_{\min}^\varphi \land \Xi_{\max}^\varphi$):

\begin{lemma}
	\label{lem:finite-completeness:fin-terms}
If $\varphi \land \Xi_{\min}^\varphi \land \Xi_{\max}^\varphi$
	is satisfiable,
	then $\varphi$ has a model
	which is the explication of a symbolic model $S$,
	where all bound formulas are
	$i \liaeq 0$ for regular nodes
	or $\top$ for summary nodes,
	all function terms are
	$i$ or $k$
for some $0 \leq k \leq \ell$,
	and for every summary node $n$,
	${\prec^S}(n, n) = i_1 < i_2$.
\end{lemma}

\begin{lemma}
	\label{lem:finite-completeness:fin-bounds}
If $\varphi$ has a model
	which is the explication of a symbolic model $S$
	with the properties listed in
	\Cref{lem:finite-completeness:fin-terms},
then there exists a symbolic model $S'$
	which is identical to $S$
	except that
all bound formulas are
	either $i \liaeq 0$ for regular nodes or
	$0 \leq i \leq \ell$ for summary nodes,
	and $\Explicate(S') \models \varphi$.
\end{lemma}

\begin{proof}[Proof of \Cref{lem:finite-completeness}]
	Similarly to the proof of \Cref{lem:well-founded-completeness},
	by combining
	\Cref{lem:finite-completeness:fin-terms,lem:finite-completeness:fin-bounds}
	we get that there exists a symbolic model $S'$
such that $\Explicate(S') \models \varphi$
	and all bound formulas are either $i \liaeq 0$
	or $0 \leq i \leq \ell$.
	Thus, for all nodes $n$,
	$\Explicate(n)$ is finite,
	and immediately $\Explicate(S')$
	is finite.
\end{proof}

\begin{proof}[Proof of
	\Cref{lem:finite-completeness:fin-terms}
	]
	In the same vain as the proof
	of \Cref{lem:well-founded-completeness:non-negative-terms},
	we consider the formula
	$
	\varphi' = \varphi \land \Skolem \parens{
	 \Xi_{\min}^\varphi \land \Xi_{\max}^\varphi
	}
	$.
	Since $\varphi' \in \oscReduced$
	is satisfiable,
	there exists a symbolic model
	$S' \in \SymbolicCandidates(\varphi')$
    such that
	$\Explicate(S') \models \varphi'$,
	and we construct a new symbolic model
	$S$ as required by similarly fixing the function interpretations.
	
	For every definition
	$f^S(n) = \angles{ n', i + k }$
	where $k < 0$, we use the minimal regular node
	$m_{\min}$ as in
	\Cref{lem:well-founded-completeness:non-negative-terms}.
Similarly,
	for $f^S(n) = \angles{n', i + k}$
	where $k > 0$,
we fix the interpretation
	by using a maximal regular node $m_{\max}$,
	which must exist due to $\Xi_{\max}^\varphi$.

	Once all functions $f, h, \dots$
	appearing in $\varphi$
	have been redefined,
	all nodes in $S$ satisfy the same atoms
	$\alpha \in \Atoms_\varphi$
	in much the same way as in the previous proof.

	Similarly to the previous proof,
	if $f^S(n)$ was redefined based on $m_{\max}$,
    the characterization of $n$ necessarily contains
	$x \prec f(x)$
	(\ref{prop:osc-model:fun-terms}).
    Let $f^{S'}(m_{\max}) = \angles{ m', k' }$,
	then for all $z \in \Bounds(n)$,
	$
	\angles{ n, z } \prec^{\Explicate(S')} \angles{ m_{\max}, 0 }
	\prec^{\Explicate(S')} \angles{ m', k' }
	$.
	Thus $x \prec f(x)$ is still satisfied in $n$.

	All element atoms and image atoms
	are preserved in much the same way.
	The only other interesting case is the mixed atom
	$f(x) \prec h(x)$ when $x \succ f(x)$ and $x \prec h(x)$.

	Let $m_1, m_2$ be the $f$- and $h$-images
	of $m_{\min}, m_{\max}$ resp.,
	with terms $k_1, k_2$.
	Then
$
	\angles{ m_1, k_1 }
	\prec^{\Explicate(S')}
	\angles{ m_{\min}, 0 }
	\prec^{\Explicate(S')}
	\angles{ m_{\max}, 0}
	\prec^{\Explicate(S')}
	\angles{ m_2, k_2 }
	$
	and thus
	$
	S, \brackets{ x \mapsto n } \models f(x) \prec h(x)
	$.
\end{proof}

\begin{proof}[Proof of
	\Cref{lem:finite-completeness:fin-bounds}]
	Let $S$ be a symbolic model
	with properties listed in \Cref{lem:finite-completeness:fin-terms},
	such that $\Explicate(S) \models \varphi$.
We prove that a symbolic model
	$S'$ which is identical to $S$,
	except that all bound formulas (of summary nodes)
	are set to $0 \leq i \leq \ell$,
	is well-defined and
	$\Explicate(S') \models \varphi$.
	This follows the same logic as in the proof of
	\Cref{lem:well-founded-completeness:has-semi-bounded},
	as the function terms $i$, $0 \leq k \leq \ell$
	will always remain within the $\brackets{ 0, \ell }$
	bounds.
\end{proof}

As in \Cref{sec:decidable:well-founded},
\Cref{lem:finite-completeness} combined
with \Cref{claim:fin-sound-axioms} implies the following corollary,
and together with the decision procedure of~\cite{infinite-needle}
for standard satisfiability of \osc{}
and the finite-model construction presented
in the proof of \Cref{lem:finite-completeness},
completes the proof of \Cref{thm:fin-decidability}.

\begin{corollary}[Soundness and Completeness]
	Given a formula $\varphi \in \oscReduced$,
	$\varphi \land \Xi_{\min}^\varphi \land \Xi_{\max}^\varphi$
	is satisfiable
	iff
	$\varphi$ has a finite model.
\end{corollary}

\subsubsection{Small Model Property for Finite Satisfiability}
From the proofs above we can conclude that if
$\varphi \in \oscReduced$ has a finite model,
then there exists a symbolic model $S$
such that $\Explicate(S) \models \varphi$
and all bound formulas in $S$ are either $i \liaeq 0$
or $0 \leq i \leq \ell$,
where $\ell$ is the number of function symbols appearing in $\varphi$.
Next we show that we can in fact compute a bound on the
size of the finite model $\Explicate(S)$ in terms of syntactical parameters of $\varphi$
(the number of function symbols, unary relation symbols, and ground terms appearing in $\varphi$),
resulting in a \emph{small-model property}
for finite satisfiability of \osc{}.

Let $r$ be the number of regular nodes in $S$,
and let $s$ be the number of summary nodes in $S$,
then the size of $\Explicate(S)$
is $r + s \cdot \ell$.
To compute a bound on $r$ and $s$, let us first denote
$
\varphi' \defeq \varphi \land \Skolem \parens{ \Xi_{\min}^\varphi \land \Xi_{\max}^\varphi }
$.
From
\Cref{lem:finite-completeness:fin-terms,lem:finite-completeness:fin-bounds},
there exists a symbolic model
$S' \in \SymbolicCandidates(\varphi')$
which has the same symbolic domain as $S$.
Let $\ell'$ be the number of function symbols appearing in $\varphi'$,
$m'$ the number of unary relations in $\varphi'$,
and $g'$ the number of ground terms in $\varphi'$.
Then, according to~\cite{infinite-needle},
the number of regular nodes in $S'$,
i.e., $r$,
is bounded by $g'$,
and the number of summary nodes,
i.e., $s$,
is bounded by
$
b(\ell' + 1) \cdot 2^{m' \parens{ \ell' + 1 }} \cdot \parens{ g' + 1}^{\ell' + 1}
$
where $b(n)$ is the $n$'th ordered Bell number.

The number of functions and unary relations appearing in
$\varphi$ and $\varphi'$ is identical,
therefore $\ell' = \ell$, and $m' = m$
(the number of unary relation appearing in $\varphi$).
What remains is to express the number of ground terms $g'$,
which consist of the ground terms of $\varphi$,
whose number is denoted $g$,
together with Skolem constants introduced by
the induction axioms.
Each axiom $\xi_{\min}, \xi_{\max}$,
once skolemized,
introduces an additional constant,
and every characterization
$C \in \Chars_\varphi^{\summary}$
introduces two axioms.
From~\ref{prop:osc-model:characterization},
there are at most as many characterizations
as the bound of summary nodes in
$\SymbolicCandidates(\varphi)$.
Thus,
$
g' = g + 2 \cdot b(\ell + 1) \cdot 2^{m \parens{ \ell + 1 }} \cdot \parens{ g + 1}^{\ell + 1}
$.

\sharon{actually we can have a similar property for well-foundeness in terms of a small symbolic model property, but not sure if it's super interesting. Maybe here we can write this as a remark. }

\subsection{Complexity Analysis}
The decision procedures of
\Cref{thm:wf-decidability,thm:fin-decidability}
reduce satisfiability of \osc{}
under the intended semantics (well-founded or finite-domain semantics)
to satisfiability of \osc{} under standard FOL semantics, 
and then use a decision procedure for \osc{} under standard semantics.
The reduction is done by adding at most
$\abs{ \Chars_\varphi^{\summary} }$
induction axioms
for a given formula $\varphi$
under well-founded semantics,
or
$2 \cdot \abs{ \Chars_\varphi^{\summary} }$
under finite-domain semantics.
The number of summary node characterizations
$\abs{ \Chars_\varphi^{\summary} }$
is exponential in the number of atoms of $\varphi$,
$\abs{\Atoms_\varphi}$,
that is in turn polynomial in the number of
relation symbols, function symbols and ground terms,
and exponential in the maximum arity of function and relation symbols
(which is typically a small number).

The decision procedure of \osc{} under FOL semantics
from~\citep{infinite-needle}
reduces FOL satisfiability to LIA queries
that are linear in the size of the formula
and the maximum size of the symbolic structures that are considered,
which is at most $2^{\abs{\Atoms_\varphi}}$.
However, the optimal complexity of satisfiability in \osc{}
has not been investigated to the best of our knowledge.

The reduction-based decision procedures have the advantage that  
it is sound to use any off-the-shelf FOL solver once the reduction is applied. 
While existing solvers may not implement decision procedures for \osc{} and thus may diverge, 
if they succeed in proving (un)satisfiability under standard FOL semantics, 
their result is correct under the intended semantics. Moreover, the reductions can be applied lazily,
adding the induction axioms incrementally. 
When only a subset of the axioms is added, 
satisfiability under FOL semantics
may be spurious under the intended semantics, 
but unsatisfiability is sound at each step, 
i.e., it reflects the intended semantics. 
These properties, which we leverage in our evaluation, 
are especially useful in the context of verification, where we often seek to prove
unsatisfiability of a formula.
 \section{Implementation and Evaluation}
\label{sec:evaluation}
We implemented a prototype tool
for automatically generating and testing induction axioms
according to
user-specified intended semantics,
available at~\citep{axe-em-artifact-latest}.
Our tool supports two modes of operations:
a brute-force search,
and a symbolic-model-guided search,
which first uses FEST~\cite{infinite-needle}
to construct a single (spurious)
symbolic model,
and then examines its summary nodes
in order to generate
$\xi_{\min}, \xi_{\max}$
induction axioms
that might eliminate the model
({\`a} la \Cref{sec:sym-models-axioms}).
In the guided search mode,
the literals that are satisfied by the summary nodes
are used as building blocks for formulas,
whereas in the brute-force search all literals are used.
Our tool is implemented in Python~\cite{python},
receives as input
a negated VC as a
formula
in SMTLIB syntax~\cite{smtlib},
the intended semantics
(either well-founded, upwards well-founded or finite),
and the order relation symbol.
The tool either produces axioms
that ensure the formula is unsatisfiable
or fails.
Our tool uses the SMT solver Z3~\cite{z3}
to check if the proposed axioms do indeed ensure unsatisfiability.
Therefore there are two possible reasons for failure:
either the tool is unable to find suitable axioms
(i.e., the axioms it considers do not eliminate all spurious models)
or the underlying solver is unable to prove
unsatisfiability of the resulting formula
(e.g., due to timeouts).
In the former case,
it is possible to run FEST to
find an infinite counter-model,
after the axioms are added.
We leave using such models iteratively for future work.

\paragraph{Implementation}
Our tool uses several heuristics to improve performance
(in both modes of operation):
first, instead of considering
complete characterizations in axioms,
it uses simpler formulas, containing at most two literals;
second, for literals that contain terms of other sorts,
it succinctly captures many axioms
by replacing those terms with free variables,
that are then universally quantified;
third, in order not to slow down the underlying SMT solver
with large SMT queries,
our tool tries to use axioms in 4 phases:
\begin{inparaenum}[(i)]
	\item
	using each single-literal axiom;
\item
	using a conjunction of all single-literal axioms;
\item
	using each two-literal axiom; and
\item
	using a conjunction of all single-literal and two-literal axioms.
\end{inparaenum}
Additionally, the user can define derived relations
to capture more complex properties that are not
simply expressible with two literals.

\paragraph{Benchmark}
We evaluate the applicability of our approach
on a benchmark of \AllExamples{} examples,
from various domains.
The bulk of the examples is taken from~\cite{infinite-needle},
which compiled them
from previous work~\cite{natural-proofs,paxos-ic3po,bounded-horizon},
with several additional examples
modeling
Dijkstra's self-stabilizing
system~\cite{dijkstra-self-stabilizing},
and
the termination phase of
the Broadcast \& Echo protocol~\cite{broadcast-echo},
as described below.

\para{Examples from~\cite{infinite-needle}}
The benchmark suite of~\cite{infinite-needle}
consists of two parts:
distributed protocols and properties of linked lists.
We consider all the examples of distributed protocols,
and add the two examples of the simplified Paxos VC
from \Cref{sec:overview}.
Of the linked-list examples of~\cite{infinite-needle},
3 have finite counter-models,
thus not relevant for us
and excluded from the benchmark.
4 others that deal with a ternary
``segmented-reachability'' relation
are also excluded from the benchmark,
since it is not clear what
the ordered and well-founded semantics
of the ternary relation
should be.
The other linked-list examples are taken as-is,
with derived relations added for two of them.

\para{Dijkstra's Self-Stabilizing System}
In Dijkstra's self-stabilizing system~\cite{dijkstra-self-stabilizing},
an unbounded number of threads are arranged on a loop,
with a designated ``first'' thread.
Each thread is given a key,
with privilege determined
according to the key
(for the first thread, if its key is equal
to the key of the last thread,
for all other threads, if their key is distinct
from the key of their previous thread).
Once privileged, a thread may take a step
and modify its key
(for the first thread,
increment the key,
for others, copy the key of their predecessor).
We prove two properties of the algorithm:
one,
which we nickname ``sanity'',
that at any state of the system,
at least one thread is privileged;
second, that if the system reaches a safe state,
where exactly one thread is privileged,
it stays safe.

\para{Broadcast \& Echo Termination}
These two examples are VC's
for proving a safety property of the termination of the
Broadcast \& Echo protocol~\cite{broadcast-echo}
under the specific topologies of
nodes on a line and in a tree.
After the broadcast phase of the protocol,
each node in the network has a designated parent.
Before terminating, each node notifies its parent
that it received the broadcast.
The VC's check
that the root node cannot terminate
before all other nodes have terminated.
The encoding we use is inspired by~\cite{raz-squeezers}.

\newcommand{\EvalTableHeader}{\toprule
\multirow{2}{*}{Example}
& \multicolumn{1}{c}{Brute-Force}
& \multicolumn{1}{c}{Model-Guided}
\\
& \multicolumn{1}{c}{Time (s)}
& \multicolumn{1}{c}{Time (s)}
\\
\midrule }
\begin{table}
	\caption{Evaluation results for
		automatic induction-axiom generation.
		We report the mean and standard deviation
		of run time across 10 runs.
		``-'' denotes failure to 
		generate or verify induction axioms.
		We denote by ``$\star$''
		examples where derived relations are used.
		The times for the model-guided mode
		include the time to find the spurious model.
		\textbf{Bold} marks examples that are in \osc{}.
		The \PaxosExamples{} non-simplified Paxos examples,
		which were not solved,
		are excluded from the table.
	}
	\begin{scriptsize}
		\centering
		\begin{minipage}[t]{0.45\textwidth}
			\vspace{0pt}
			\centering
			\begin{tabular}{lrr}
				\EvalTableHeader{}\textbf{Echo Machine} & $0.72 \pm 0.61$ & $0.24 \pm 0.07$
				\\
				Ring Leader & $0.32 \pm 0.26$ & $0.15 \pm 0.09$
				\\
				Line Leader & $31.47 \pm 7.20$ & $23.53 \pm 0.62$
				\\
				\textbf{Paxos (simplified)} & $0.89 \pm 0.49$ & $0.29 \pm 0.06$
				\\
				\textbf{Paxos bidi (simplified)} & $1.65 \pm 0.01$ & $0.57 \pm 0.01$
				\\
				\midrule
				Self-Stabilizing (sanity) & $0.34 \pm 0.23$ & $1.42 \pm 0.23$
				\\
				Self-Stabilizing (safety) & $3.31 \pm 6.85$ & $13.97 \pm 9.42$
				\\
				B\&E Termination (line) & $0.76 \pm 0.47$ & $1.29 \pm 0.37$
				\\
				B\&E Termination (tree) & $0.64 \pm 0.49$ & $1.89 \pm 0.41$
				\\
				\bottomrule
			\end{tabular}
		\end{minipage}\begin{minipage}[t]{0.45\textwidth}
			\vspace{0pt}
			\centering
			\begin{tabular}{lrr}
				\EvalTableHeader{}List length & $23.24 \pm 17.97$ & -
				\\
				Doubly-linked list & $0.81 \pm 0.67$ & $0.60 \pm 0.29$
				\\
				Doubly-linked length$\star$ & $32.01 \pm 8.27$ & $21.84 \pm 0.55$
				\\
				Sorted list & $0.86 \pm 0.48$ & $0.29 \pm 0.12$
				\\
				Sorted list length$\star$ & $40.46 \pm 13.35$ & -
				\\
				\textbf{Sorted list max} & $0.50 \pm 0.34$ & $0.13 \pm 0.07$
				\\
				\bottomrule
			\end{tabular}
		\end{minipage}\end{scriptsize}

	\label{tab:evaluation-results}
\end{table}

\paragraph{Evaluation}
We run our tool on these examples
using a Macbook Pro with an Apple M1 Pro CPU
and 32GB RAM,
with Z3 version 4.12.2.
We do not include an empirical
comparison with other tools
since we are not aware of existing tools
\neta{changed:}
that check satisfiability
under well-founded or finite-domain semantics
for this kind of uninterpreted FOL encoding.
\Cref{tab:evaluation-results} summarizes
the results of the examples
our tool manages to solve (in either mode).
These include all examples
except for
\PaxosExamples{}
non-simplified Paxos variants from~\cite{infinite-needle}.
For each solved example, the table indicates
whether or not the example was verified with each approach,
reporting the average run time across 10 runs,
and whether or not derived relations were used
(denoted by ``$\star$'').
In the model-guided mode, the time includes the time for finding
an infinite~model.
\OscExamples{} of the examples are in \osc{}
(marked in \textbf{bold}).

Our tool is able to verify \AutoExamples{} out of \AllExamples{} examples
using the brute-force search,
and \GuidedExamples{} examples using the model-guided search,
which is generally faster.
In this prototype we implemented
the model-guided approach as a single iteration,
constructing a single symbolic model
and generating axioms from it.
We conjecture that using more iterations will bring
the model-guided approach on par with the brute-force search,
while staying faster in the common case.

The \PaxosExamples{} unsolved examples
consist of verification conditions of safety properties
for variants of Paxos (taken from~\cite{infinite-needle}),
which lie beyond \osc{}.
These examples are relatively large,
with complex quantifier alternations.
Though finding the correct induction axioms
is a necessary condition for proving these VC's,
it is not sufficient,
as finding the correct quantifier instantiations
for a proof is in and of itself hard.
In fact, we have checked manually the axiom generated
for the main variant of Paxos,
and indeed it proves the safety of the protocol,
but existing tools are unable to find the correct
quantifier instantiations,
and instead diverge.
When we tame the space of quantifier instantiations
by simplifying the VC's
to model only the core proposal mechanism,
as described in \Cref{sec:overview},
SMT solvers are able to find a proof
and verify safety under the intended semantics.

\neta{
	The axiom needed for Paxos
	is a standard well-founded induction axiom,
	with the following property
	$\code{isSafeAt}(x, v) \land x \succ b_0$,
	where $v, b_0$ are free variables (to be universally quantified).
	If we had $\code{isSafeAt}$ as a derived relation
	we could find this property,
	as this is essentially just two literals
	(though currently the implementation skips literals
	like $x \succ b_0$,
	since they're not needed in other examples).
	So maybe we could mention this,
	and say that this makes apparent that
	there is also a non-trivial issue
	with quantifier instantiations,
	on top of axiom generation.
}  \section{Related Work}
\label{sec:related}

\neta{works with other modeling
	have different things;
	separation logic}
	
\neta{something on other
	LFP semantics? matching $\mu$-logic}

\paragraph{Automating Induction}
In the context of least-fixpoint semantics
there has been a lot of work
on automating proofs by induction
in different logics,
such as separation logic~\cite{separation-logic-inductive-predicates,entailment-inductive-definitions}
or matching logic~\cite{unified-fixpoint-matching-logic}.
In this paper we focus on
verification conditions encoded in
first-order logic.
When using FOL to approximate
least-fixpoint semantics,
synthesis of inductive lemmas~\cite{fossil}
was shown to be a relatively-complete
method of proving validity of formulas.
There, lemmas that might or might not hold
are generated from partial finite models,
and then validated and used for furthering the proof.
Our method on the other hand produces
axioms which are always valid,
generated from semantics-violating infinite models.
A lot of previous work on automating induction
considered specific domains or sorts,
in particular
integer induction~\cite{vampire-integer-induction},
algebraic data types~\cite{automating-inductive-proofs},
recursive definitions~\cite{vampire-recursive-induction,natural-proofs},
or a combination thereof~\cite{vampire-reflection-induction,vampire-induction-generalization}.
In contrast, we consider uninterpreted FOL
with an order abstraction.
In the context of higher-order
proof assistants~\cite{coq,isabelle},
tactics have been developed to automate
proofs by induction,
and in the context of SMT solvers,
\citet{induction-smt-solver} proposed a
heuristic for when to use induction to
prove the validity of a given property.
Similarly,~\citet{dafny-module-based-induction}
showed how to use modules in Dafny~\cite{dafny}
to encode induction in order to prove inductive properties.
In our case, however,
induction is used as part of a larger
verification condition,
and the properties
proved by induction
are not themselves the goal.
\neta{think on this}

\paragraph{Refining First-Order Logic}
In~\cite{theory-refinement-for-verification}
first-order theories are abstracted and then refined
in a modular way to adjust the modeling precision
and speed up SMT solvers.
This is similar to our approach,
as theories can be understood as definable semantics,
though in our case the intended semantics are not definable.
\citet{finite-theory-axiomatizations}
use a finite representation of
a potentially infinite sets of saturated clauses
to find finite, local axiomatizations of theories.
This bears some resemblance
to our decidability results
that show that in \osc{}
well-founded semantics
and finite-structure semantics
can be captured by a finite set of axioms.

\paragraph{Decidability of Well-Founded and Finite-Model Semantics}
Several fragments of FOL
have a finite-model property,
meaning that every satisfiable formula
has a finite model,
most relevant to our work is \epr{}~\cite{bsr-epr}.
Verification of safety properties in such fragments,
e.g.,~\cite{paxos-made-epr,linked-lists-epr,heap-paths-epr},
is effectively verification modulo finite-domain semantics.
In contrast, we consider
formulas outside of such fragments,
where the different semantics do not coincide.
\neta{added:}
\citet{well-founded-lists,well-founded-reachability,well-founded-reachability-revisiting}
develop the concept of well-founded reachability,
that uses a known set of program variables,
denoted \emph{heads},
in order to break cycles and form
the foundation for reachability of heap objects.
Some of our examples verify general properties
of linked-lists,
not tied to a specific program context
where such heads can be provided.
Further, the decision procedures given by
\citet{well-founded-lists,well-founded-reachability,well-founded-reachability-revisiting}
\neta{changed:}
do not allow the combination of quantification and functions
present in our examples
(notably the literal $x \prec f(x)$
where $x$ is universally quantified,
which is allowed in \osc{}).
Many classic results like~\cite{shelah1977decidability,danielski-unfo}
show decidability of finite-structure semantics
in certain fragments of FOL.
Those fragments are incomparable to \osc{},
and well-founded semantics was not considered.
In~\cite{well-orders-theory-note,elementary-well-ordering}
well-founded semantics are shown to be decidable,
but under a limited vocabulary containing only the
well-founded relation symbol.
The seminal works on the specific infinite structures
of S1S and S2S~\cite{buchi-mso-s1s,buchi-deciable-mso,rabin-mso-s2s}
can also be viewed as
satisfiability modulo
the class of those specific structures.

 \section{Conclusion}
\label{sec:conclusion}
We have presented an approach
to soundly refine FOL semantics
in order to match the intended semantics
of the user,
specifically semantics of well-founded relations
or finite domains.
We have described axiom schemata
for well-founded semantics
and finite-domain semantics,
under an abstract order relation,
and we have shown how to use
spurious, semantic-violating counter-models
to guide the instantiations of these axioms.
We have developed a prototype tool,
and evaluated this approach
on examples from various domains,
where a natural first-order modeling
leads to a semantic gap.
In most cases our tool is able to quickly find
the necessary axioms
and allow the verification process
to successfully complete.
Beyond the empirical results
we have also proved
decidability results for
satisfiability under well-founded semantics
and finite-structure semantics
in the \osc{} fragment of FOL.

We consider this a first step towards
better aligning the intentions of the user
with the semantics of the specification language,
though there are challenges ahead.
One major hurdle we encountered
is that regardless of the induction axioms we add,
the formulas we consider can 
have complicated quantifier alternations,
making it difficult to prove their unsatisfiability.
Finding the quantifier instantiations
necessary for this purpose
is an orthogonal problem
of great interest.

\neta{not talk about quantifier alternation from axioms,
	quantifier instantiation orthogonal problem;
	not talk about future work of quant-inst;
	maybe different future work:
	implicit partial order, reachable segment
} \section*{Data-Availability Statement}
An extended version of this paper with detailed proofs
can be found at~\citep{axe-em-arxiv}.
An artifact for reproducing the results is available 
at~\citep{axe-em-artifact},
and the latest version of the tool can be obtained
at~\citep{axe-em-artifact-latest}.

\begin{acks}
We thank the anonymous reviewers and the artifact evaluation committee 
for comments which improved the paper.
We thank Raz Lotan, Eden Frenkel 
and Yotam Feldman for insightful discussions and comments.
The research leading to these results has received funding from the
European Research Council under the European Union's Horizon 2020 research and
innovation programme (grant agreement No [759102-SVIS]).
This research was partially supported by the Israeli Science Foundation (ISF) grant No.\ 2117/23.

\end{acks}
 \bibliography{includes/references.bib}


\begin{thebibliography}{60}


\ifx \showCODEN    \undefined \def \showCODEN     #1{\unskip}     \fi
\ifx \showDOI      \undefined \def \showDOI       #1{#1}\fi
\ifx \showISBNx    \undefined \def \showISBNx     #1{\unskip}     \fi
\ifx \showISBNxiii \undefined \def \showISBNxiii  #1{\unskip}     \fi
\ifx \showISSN     \undefined \def \showISSN      #1{\unskip}     \fi
\ifx \showLCCN     \undefined \def \showLCCN      #1{\unskip}     \fi
\ifx \shownote     \undefined \def \shownote      #1{#1}          \fi
\ifx \showarticletitle \undefined \def \showarticletitle #1{#1}   \fi
\ifx \showURL      \undefined \def \showURL       {\relax}        \fi
\providecommand\bibfield[2]{#2}
\providecommand\bibinfo[2]{#2}
\providecommand\natexlab[1]{#1}
\providecommand\showeprint[2][]{arXiv:#2}

\bibitem[Ball et~al\mbox{.}(2014)]%
        {vericon-networks}
\bibfield{author}{\bibinfo{person}{Thomas Ball}, \bibinfo{person}{Nikolaj~S.
  Bj{\o}rner}, \bibinfo{person}{Aaron Gember}, \bibinfo{person}{Shachar
  Itzhaky}, \bibinfo{person}{Aleksandr Karbyshev}, \bibinfo{person}{Mooly
  Sagiv}, \bibinfo{person}{Michael Schapira}, {and} \bibinfo{person}{Asaf
  Valadarsky}.} \bibinfo{year}{2014}\natexlab{}.
\newblock \showarticletitle{VeriCon: towards verifying controller programs in
  software-defined networks}. In \bibinfo{booktitle}{\emph{{ACM} {SIGPLAN}
  Conference on Programming Language Design and Implementation, {PLDI} '14,
  Edinburgh, United Kingdom - June 09 - 11, 2014}},
  \bibfield{editor}{\bibinfo{person}{Michael F.~P. O'Boyle} {and}
  \bibinfo{person}{Keshav Pingali}} (Eds.). \bibinfo{publisher}{{ACM}},
  \bibinfo{pages}{282--293}.
\newblock
\urldef\tempurl%
\url{https://doi.org/10.1145/2594291.2594317}
\showDOI{\tempurl}


\bibitem[Barbosa et~al\mbox{.}(2022)]%
        {cvc5}
\bibfield{author}{\bibinfo{person}{Haniel Barbosa}, \bibinfo{person}{Clark~W.
  Barrett}, \bibinfo{person}{Martin Brain}, \bibinfo{person}{Gereon Kremer},
  \bibinfo{person}{Hanna Lachnitt}, \bibinfo{person}{Makai Mann},
  \bibinfo{person}{Abdalrhman Mohamed}, \bibinfo{person}{Mudathir Mohamed},
  \bibinfo{person}{Aina Niemetz}, \bibinfo{person}{Andres N{\"{o}}tzli},
  \bibinfo{person}{Alex Ozdemir}, \bibinfo{person}{Mathias Preiner},
  \bibinfo{person}{Andrew Reynolds}, \bibinfo{person}{Ying Sheng},
  \bibinfo{person}{Cesare Tinelli}, {and} \bibinfo{person}{Yoni Zohar}.}
  \bibinfo{year}{2022}\natexlab{}.
\newblock \showarticletitle{cvc5: {A} Versatile and Industrial-Strength {SMT}
  Solver}. In \bibinfo{booktitle}{\emph{Tools and Algorithms for the
  Construction and Analysis of Systems - 28th International Conference, {TACAS}
  2022, Held as Part of the European Joint Conferences on Theory and Practice
  of Software, {ETAPS} 2022, Munich, Germany, April 2-7, 2022, Proceedings,
  Part {I}}} \emph{(\bibinfo{series}{Lecture Notes in Computer Science},
  Vol.~\bibinfo{volume}{13243})}, \bibfield{editor}{\bibinfo{person}{Dana
  Fisman} {and} \bibinfo{person}{Grigore Rosu}} (Eds.).
  \bibinfo{publisher}{Springer}, \bibinfo{pages}{415--442}.
\newblock
\urldef\tempurl%
\url{https://doi.org/10.1007/978-3-030-99524-9\_24}
\showDOI{\tempurl}


\bibitem[Barrett et~al\mbox{.}(2017)]%
        {smtlib}
\bibfield{author}{\bibinfo{person}{Clark Barrett}, \bibinfo{person}{Pascal
  Fontaine}, {and} \bibinfo{person}{Cesare Tinelli}.}
  \bibinfo{year}{2017}\natexlab{}.
\newblock \bibinfo{booktitle}{\emph{{The SMT-LIB Standard: Version 2.6}}}.
\newblock \bibinfo{type}{{T}echnical {R}eport}.
  \bibinfo{institution}{Department of Computer Science, The University of
  Iowa}.
\newblock
\newblock
\shownote{Available at {\tt www.SMT-LIB.org}}.


\bibitem[Brotherston et~al\mbox{.}(2014)]%
        {separation-logic-inductive-predicates}
\bibfield{author}{\bibinfo{person}{James Brotherston}, \bibinfo{person}{Carsten
  Fuhs}, \bibinfo{person}{Juan Antonio~Navarro P{\'{e}}rez}, {and}
  \bibinfo{person}{Nikos Gorogiannis}.} \bibinfo{year}{2014}\natexlab{}.
\newblock \showarticletitle{A decision procedure for satisfiability in
  separation logic with inductive predicates}. In
  \bibinfo{booktitle}{\emph{Joint Meeting of the Twenty-Third {EACSL} Annual
  Conference on Computer Science Logic {(CSL)} and the Twenty-Ninth Annual
  {ACM/IEEE} Symposium on Logic in Computer Science (LICS), {CSL-LICS} '14,
  Vienna, Austria, July 14 - 18, 2014}},
  \bibfield{editor}{\bibinfo{person}{Thomas~A. Henzinger} {and}
  \bibinfo{person}{Dale Miller}} (Eds.). \bibinfo{publisher}{{ACM}},
  \bibinfo{pages}{25:1--25:10}.
\newblock
\urldef\tempurl%
\url{https://doi.org/10.1145/2603088.2603091}
\showDOI{\tempurl}


\bibitem[B{\"u}chi(1990)]%
        {buchi-deciable-mso}
\bibfield{author}{\bibinfo{person}{J~Richard B{\"u}chi}.}
  \bibinfo{year}{1990}\natexlab{}.
\newblock \showarticletitle{On a decision method in restricted second order
  arithmetic}.
\newblock In \bibinfo{booktitle}{\emph{The collected works of J. Richard
  B{\"u}chi}}. \bibinfo{publisher}{Springer}, \bibinfo{pages}{425--435}.
\newblock
\urldef\tempurl%
\url{https://doi.org/10.1007/978-1-4613-8928-6\_23}
\showDOI{\tempurl}


\bibitem[B{\"{u}}chi and Landweber(1969)]%
        {buchi-mso-s1s}
\bibfield{author}{\bibinfo{person}{J.~Richard B{\"{u}}chi} {and}
  \bibinfo{person}{Lawrence~H. Landweber}.} \bibinfo{year}{1969}\natexlab{}.
\newblock \showarticletitle{Definability in the Monadic Second-Order Theory of
  Successor}.
\newblock \bibinfo{journal}{\emph{J. Symb. Log.}} \bibinfo{volume}{34},
  \bibinfo{number}{2} (\bibinfo{year}{1969}), \bibinfo{pages}{166--170}.
\newblock
\urldef\tempurl%
\url{https://doi.org/10.2307/2271090}
\showDOI{\tempurl}


\bibitem[Chen et~al\mbox{.}(2020)]%
        {unified-fixpoint-matching-logic}
\bibfield{author}{\bibinfo{person}{Xiaohong Chen}, \bibinfo{person}{Minh{-}Thai
  Trinh}, \bibinfo{person}{Nishant Rodrigues}, \bibinfo{person}{Lucas
  Pe{\~{n}}a}, {and} \bibinfo{person}{Grigore Rosu}.}
  \bibinfo{year}{2020}\natexlab{}.
\newblock \showarticletitle{Towards a unified proof framework for automated
  fixpoint reasoning using matching logic}.
\newblock \bibinfo{journal}{\emph{Proc. {ACM} Program. Lang.}}
  \bibinfo{volume}{4}, \bibinfo{number}{{OOPSLA}} (\bibinfo{year}{2020}),
  \bibinfo{pages}{161:1--161:29}.
\newblock
\urldef\tempurl%
\url{https://doi.org/10.1145/3428229}
\showDOI{\tempurl}


\bibitem[Claessen et~al\mbox{.}(2013)]%
        {automating-inductive-proofs}
\bibfield{author}{\bibinfo{person}{Koen Claessen}, \bibinfo{person}{Moa
  Johansson}, \bibinfo{person}{Dan Ros{\'{e}}n}, {and}
  \bibinfo{person}{Nicholas Smallbone}.} \bibinfo{year}{2013}\natexlab{}.
\newblock \showarticletitle{Automating Inductive Proofs Using Theory
  Exploration}. In \bibinfo{booktitle}{\emph{Automated Deduction - {CADE-24} -
  24th International Conference on Automated Deduction, Lake Placid, NY, USA,
  June 9-14, 2013. Proceedings}} \emph{(\bibinfo{series}{Lecture Notes in
  Computer Science}, Vol.~\bibinfo{volume}{7898})},
  \bibfield{editor}{\bibinfo{person}{Maria~Paola Bonacina}} (Ed.).
  \bibinfo{publisher}{Springer}, \bibinfo{pages}{392--406}.
\newblock
\urldef\tempurl%
\url{https://doi.org/10.1007/978-3-642-38574-2\_27}
\showDOI{\tempurl}


\bibitem[Cooper(1972)]%
        {lia-decidable}
\bibfield{author}{\bibinfo{person}{David~C Cooper}.}
  \bibinfo{year}{1972}\natexlab{}.
\newblock \showarticletitle{Theorem proving in arithmetic without
  multiplication}.
\newblock \bibinfo{journal}{\emph{Machine intelligence}} \bibinfo{volume}{7},
  \bibinfo{number}{91-99} (\bibinfo{year}{1972}), \bibinfo{pages}{300}.
\newblock
\urldef\tempurl%
\url{https://doi.org/10.1007/10930755_5}
\showDOI{\tempurl}


\bibitem[Danielski and Kieronski(2019)]%
        {danielski-unfo}
\bibfield{author}{\bibinfo{person}{Daniel Danielski} {and}
  \bibinfo{person}{Emanuel Kieronski}.} \bibinfo{year}{2019}\natexlab{}.
\newblock \showarticletitle{Finite Satisfiability of Unary Negation Fragment
  with Transitivity}. In \bibinfo{booktitle}{\emph{{MFCS}}}
  \emph{(\bibinfo{series}{LIPIcs}, Vol.~\bibinfo{volume}{138})}.
  \bibinfo{publisher}{Schloss Dagstuhl - Leibniz-Zentrum f{\"{u}}r Informatik},
  \bibinfo{pages}{17:1--17:15}.
\newblock
\urldef\tempurl%
\url{https://doi.org/10.4230/LIPIcs.MFCS.2019.17}
\showDOI{\tempurl}


\bibitem[de~Moura and Bj{\o}rner(2008)]%
        {z3}
\bibfield{author}{\bibinfo{person}{Leonardo~Mendon{\c{c}}a de Moura} {and}
  \bibinfo{person}{Nikolaj~S. Bj{\o}rner}.} \bibinfo{year}{2008}\natexlab{}.
\newblock \showarticletitle{{Z3:} An Efficient {SMT} Solver}. In
  \bibinfo{booktitle}{\emph{Tools and Algorithms for the Construction and
  Analysis of Systems, 14th International Conference, {TACAS} 2008, Held as
  Part of the Joint European Conferences on Theory and Practice of Software,
  {ETAPS} 2008, Budapest, Hungary, March 29-April 6, 2008. Proceedings}}
  \emph{(\bibinfo{series}{Lecture Notes in Computer Science},
  Vol.~\bibinfo{volume}{4963})}, \bibfield{editor}{\bibinfo{person}{C.~R.
  Ramakrishnan} {and} \bibinfo{person}{Jakob Rehof}} (Eds.).
  \bibinfo{publisher}{Springer}, \bibinfo{pages}{337--340}.
\newblock
\urldef\tempurl%
\url{https://doi.org/10.1007/978-3-540-78800-3\_24}
\showDOI{\tempurl}


\bibitem[Dijkstra(1974)]%
        {dijkstra-self-stabilizing}
\bibfield{author}{\bibinfo{person}{Edsger~W. Dijkstra}.}
  \bibinfo{year}{1974}\natexlab{}.
\newblock \showarticletitle{Self-stabilizing Systems in Spite of Distributed
  Control}.
\newblock \bibinfo{journal}{\emph{Commun. {ACM}}} \bibinfo{volume}{17},
  \bibinfo{number}{11} (\bibinfo{year}{1974}), \bibinfo{pages}{643--644}.
\newblock
\urldef\tempurl%
\url{https://doi.org/10.1145/361179.361202}
\showDOI{\tempurl}


\bibitem[Doner et~al\mbox{.}(1978)]%
        {elementary-well-ordering}
\bibfield{author}{\bibinfo{person}{John~E Doner}, \bibinfo{person}{Andrzej
  Mostowski}, {and} \bibinfo{person}{Alfred Tarski}.}
  \bibinfo{year}{1978}\natexlab{}.
\newblock \showarticletitle{The Elementary Theory of Well-Ordering—A
  Metamathematical Study—}.
\newblock In \bibinfo{booktitle}{\emph{Studies in Logic and the Foundations of
  Mathematics}}. Vol.~\bibinfo{volume}{96}. \bibinfo{publisher}{Elsevier},
  \bibinfo{pages}{1--54}.
\newblock


\bibitem[Elad et~al\mbox{.}(2023)]%
        {infinite-needle-arxiv}
\bibfield{author}{\bibinfo{person}{Neta Elad}, \bibinfo{person}{Oded Padon},
  {and} \bibinfo{person}{Sharon Shoham}.} \bibinfo{year}{2023}\natexlab{}.
\newblock \showarticletitle{An Infinite Needle in a Finite Haystack: Finding
  Infinite Counter-Models in Deductive Verification}.
\newblock \bibinfo{journal}{\emph{CoRR}}  \bibinfo{volume}{abs/2310.16762}
  (\bibinfo{year}{2023}).
\newblock
\urldef\tempurl%
\url{https://doi.org/10.48550/ARXIV.2310.16762}
\showDOI{\tempurl}
\showeprint[arXiv]{2310.16762}


\bibitem[Elad et~al\mbox{.}(2024)]%
        {infinite-needle}
\bibfield{author}{\bibinfo{person}{Neta Elad}, \bibinfo{person}{Oded Padon},
  {and} \bibinfo{person}{Sharon Shoham}.} \bibinfo{year}{2024}\natexlab{}.
\newblock \showarticletitle{An Infinite Needle in a Finite Haystack: Finding
  Infinite Counter-Models in Deductive Verification}.
\newblock \bibinfo{journal}{\emph{Proc. {ACM} Program. Lang.}}
  \bibinfo{volume}{8}, \bibinfo{number}{{POPL}} (\bibinfo{year}{2024}),
  \bibinfo{pages}{970--1000}.
\newblock
\urldef\tempurl%
\url{https://doi.org/10.1145/3632875}
\showDOI{\tempurl}


\bibitem[Elad and Shoham({[n.\,d.]})]%
        {axe-em-artifact-latest}
\bibfield{author}{\bibinfo{person}{Neta Elad} {and} \bibinfo{person}{Sharon
  Shoham}.} \bibinfo{year}{[n.\,d.]}\natexlab{}.
\newblock \bibinfo{title}{{Axe 'Em: Eliminating Spurious States with Induction
  Axioms (Artifact)}}.
\newblock
\newblock
\urldef\tempurl%
\url{https://doi.org/10.5281/zenodo.13912208}
\showDOI{\tempurl}


\bibitem[Elad and Shoham(2024a)]%
        {axe-em-arxiv}
\bibfield{author}{\bibinfo{person}{Neta Elad} {and} \bibinfo{person}{Sharon
  Shoham}.} \bibinfo{year}{2024}\natexlab{a}.
\newblock \showarticletitle{Axe 'Em: Eliminating Spurious States with Induction
  Axioms}.
\newblock  (\bibinfo{year}{2024}).
\newblock
\urldef\tempurl%
\url{https://doi.org/10.48550/ARXIV.2410.18671}
\showDOI{\tempurl}
\showeprint[arXiv]{2410.18671}


\bibitem[Elad and Shoham(2024b)]%
        {axe-em-artifact}
\bibfield{author}{\bibinfo{person}{Neta Elad} {and} \bibinfo{person}{Sharon
  Shoham}.} \bibinfo{year}{2024}\natexlab{b}.
\newblock \bibinfo{title}{{Axe 'Em: Eliminating Spurious States with Induction
  Axioms (Artifact)}}.
\newblock
\newblock
\urldef\tempurl%
\url{https://doi.org/10.5281/zenodo.13912279}
\showDOI{\tempurl}


\bibitem[Feldman et~al\mbox{.}(2017)]%
        {bounded-horizon}
\bibfield{author}{\bibinfo{person}{Yotam M.~Y. Feldman}, \bibinfo{person}{Oded
  Padon}, \bibinfo{person}{Neil Immerman}, \bibinfo{person}{Mooly Sagiv}, {and}
  \bibinfo{person}{Sharon Shoham}.} \bibinfo{year}{2017}\natexlab{}.
\newblock \showarticletitle{Bounded Quantifier Instantiation for Checking
  Inductive Invariants}. In \bibinfo{booktitle}{\emph{{TACAS} {(1)}}}
  \emph{(\bibinfo{series}{Lecture Notes in Computer Science},
  Vol.~\bibinfo{volume}{10205})}. \bibinfo{pages}{76--95}.
\newblock
\urldef\tempurl%
\url{https://doi.org/10.1007/978-3-662-54577-5\_5}
\showDOI{\tempurl}


\bibitem[Goel and Sakallah(2021)]%
        {paxos-ic3po}
\bibfield{author}{\bibinfo{person}{Aman Goel} {and} \bibinfo{person}{Karem~A.
  Sakallah}.} \bibinfo{year}{2021}\natexlab{}.
\newblock \showarticletitle{Towards an Automatic Proof of Lamport's Paxos}. In
  \bibinfo{booktitle}{\emph{{FMCAD}}}. \bibinfo{publisher}{{IEEE}},
  \bibinfo{pages}{112--122}.
\newblock
\urldef\tempurl%
\url{https://doi.org/10.34727/2021/isbn.978-3-85448-046-4\_20}
\showDOI{\tempurl}


\bibitem[Hajd{\'{u}} et~al\mbox{.}(2020)]%
        {vampire-induction-generalization}
\bibfield{author}{\bibinfo{person}{M{\'{a}}rton Hajd{\'{u}}},
  \bibinfo{person}{Petra Hozzov{\'{a}}}, \bibinfo{person}{Laura Kov{\'{a}}cs},
  \bibinfo{person}{Johannes Schoisswohl}, {and} \bibinfo{person}{Andrei
  Voronkov}.} \bibinfo{year}{2020}\natexlab{}.
\newblock \showarticletitle{Induction with Generalization in Superposition
  Reasoning}. In \bibinfo{booktitle}{\emph{Intelligent Computer Mathematics -
  13th International Conference, {CICM} 2020, Bertinoro, Italy, July 26-31,
  2020, Proceedings}} \emph{(\bibinfo{series}{Lecture Notes in Computer
  Science}, Vol.~\bibinfo{volume}{12236})},
  \bibfield{editor}{\bibinfo{person}{Christoph Benzm{\"{u}}ller} {and}
  \bibinfo{person}{Bruce~R. Miller}} (Eds.). \bibinfo{publisher}{Springer},
  \bibinfo{pages}{123--137}.
\newblock
\urldef\tempurl%
\url{https://doi.org/10.1007/978-3-030-53518-6\_8}
\showDOI{\tempurl}


\bibitem[Hajd{\'{u}} et~al\mbox{.}(2021)]%
        {vampire-recursive-induction}
\bibfield{author}{\bibinfo{person}{M{\'{a}}rton Hajd{\'{u}}},
  \bibinfo{person}{Petra Hozzov{\'{a}}}, \bibinfo{person}{Laura Kov{\'{a}}cs},
  {and} \bibinfo{person}{Andrei Voronkov}.} \bibinfo{year}{2021}\natexlab{}.
\newblock \showarticletitle{Induction with Recursive Definitions in
  Superposition}. In \bibinfo{booktitle}{\emph{{FMCAD}}}.
  \bibinfo{publisher}{{IEEE}}, \bibinfo{pages}{1--10}.
\newblock
\urldef\tempurl%
\url{https://doi.org/10.34727/2021/isbn.978-3-85448-046-4\_34}
\showDOI{\tempurl}


\bibitem[Ho and Pit{-}Claudel(2024)]%
        {dafny-module-based-induction}
\bibfield{author}{\bibinfo{person}{Son Ho} {and} \bibinfo{person}{Cl{\'{e}}ment
  Pit{-}Claudel}.} \bibinfo{year}{2024}\natexlab{}.
\newblock \showarticletitle{Incremental Proof Development in Dafny with
  Module-Based Induction}.
\newblock \bibinfo{journal}{\emph{CoRR}}  \bibinfo{volume}{abs/2401.16233}
  (\bibinfo{year}{2024}).
\newblock
\urldef\tempurl%
\url{https://doi.org/10.48550/ARXIV.2401.16233}
\showDOI{\tempurl}
\showeprint[arXiv]{2401.16233}


\bibitem[Horbach and Sofronie{-}Stokkermans(2013)]%
        {finite-theory-axiomatizations}
\bibfield{author}{\bibinfo{person}{Matthias Horbach} {and}
  \bibinfo{person}{Viorica Sofronie{-}Stokkermans}.}
  \bibinfo{year}{2013}\natexlab{}.
\newblock \showarticletitle{Obtaining Finite Local Theory Axiomatizations via
  Saturation}. In \bibinfo{booktitle}{\emph{Frontiers of Combining Systems -
  9th International Symposium, FroCoS 2013, Nancy, France, September 18-20,
  2013. Proceedings}} \emph{(\bibinfo{series}{Lecture Notes in Computer
  Science}, Vol.~\bibinfo{volume}{8152})},
  \bibfield{editor}{\bibinfo{person}{Pascal Fontaine},
  \bibinfo{person}{Christophe Ringeissen}, {and} \bibinfo{person}{Renate~A.
  Schmidt}} (Eds.). \bibinfo{publisher}{Springer}, \bibinfo{pages}{198--213}.
\newblock
\urldef\tempurl%
\url{https://doi.org/10.1007/978-3-642-40885-4\_14}
\showDOI{\tempurl}


\bibitem[Hozzov{\'{a}} et~al\mbox{.}(2021)]%
        {vampire-integer-induction}
\bibfield{author}{\bibinfo{person}{Petra Hozzov{\'{a}}}, \bibinfo{person}{Laura
  Kov{\'{a}}cs}, {and} \bibinfo{person}{Andrei Voronkov}.}
  \bibinfo{year}{2021}\natexlab{}.
\newblock \showarticletitle{Integer Induction in Saturation}. In
  \bibinfo{booktitle}{\emph{{CADE}}} \emph{(\bibinfo{series}{Lecture Notes in
  Computer Science}, Vol.~\bibinfo{volume}{12699})}.
  \bibinfo{publisher}{Springer}, \bibinfo{pages}{361--377}.
\newblock
\urldef\tempurl%
\url{https://doi.org/10.1007/978-3-030-79876-5\_21}
\showDOI{\tempurl}


\bibitem[Hyv{\"{a}}rinen et~al\mbox{.}(2017)]%
        {theory-refinement-for-verification}
\bibfield{author}{\bibinfo{person}{Antti E.~J. Hyv{\"{a}}rinen},
  \bibinfo{person}{Sepideh Asadi}, \bibinfo{person}{Karine Even{-}Mendoza},
  \bibinfo{person}{Grigory Fedyukovich}, \bibinfo{person}{Hana Chockler}, {and}
  \bibinfo{person}{Natasha Sharygina}.} \bibinfo{year}{2017}\natexlab{}.
\newblock \showarticletitle{Theory Refinement for Program Verification}. In
  \bibinfo{booktitle}{\emph{Theory and Applications of Satisfiability Testing -
  {SAT} 2017 - 20th International Conference, Melbourne, VIC, Australia, August
  28 - September 1, 2017, Proceedings}} \emph{(\bibinfo{series}{Lecture Notes
  in Computer Science}, Vol.~\bibinfo{volume}{10491})},
  \bibfield{editor}{\bibinfo{person}{Serge Gaspers} {and} \bibinfo{person}{Toby
  Walsh}} (Eds.). \bibinfo{publisher}{Springer}, \bibinfo{pages}{347--363}.
\newblock
\urldef\tempurl%
\url{https://doi.org/10.1007/978-3-319-66263-3\_22}
\showDOI{\tempurl}


\bibitem[Itzhaky et~al\mbox{.}(2014)]%
        {heap-paths-epr}
\bibfield{author}{\bibinfo{person}{Shachar Itzhaky}, \bibinfo{person}{Anindya
  Banerjee}, \bibinfo{person}{Neil Immerman}, \bibinfo{person}{Ori Lahav},
  \bibinfo{person}{Aleksandar Nanevski}, {and} \bibinfo{person}{Mooly Sagiv}.}
  \bibinfo{year}{2014}\natexlab{}.
\newblock \showarticletitle{Modular reasoning about heap paths via effectively
  propositional formulas}. In \bibinfo{booktitle}{\emph{The 41st Annual {ACM}
  {SIGPLAN-SIGACT} Symposium on Principles of Programming Languages, {POPL}
  '14, San Diego, CA, USA, January 20-21, 2014}},
  \bibfield{editor}{\bibinfo{person}{Suresh Jagannathan} {and}
  \bibinfo{person}{Peter Sewell}} (Eds.). \bibinfo{publisher}{{ACM}},
  \bibinfo{pages}{385--396}.
\newblock
\urldef\tempurl%
\url{https://doi.org/10.1145/2535838.2535854}
\showDOI{\tempurl}


\bibitem[Itzhaky et~al\mbox{.}(2013)]%
        {linked-lists-epr}
\bibfield{author}{\bibinfo{person}{Shachar Itzhaky}, \bibinfo{person}{Anindya
  Banerjee}, \bibinfo{person}{Neil Immerman}, \bibinfo{person}{Aleksandar
  Nanevski}, {and} \bibinfo{person}{Mooly Sagiv}.}
  \bibinfo{year}{2013}\natexlab{}.
\newblock \showarticletitle{Effectively-Propositional Reasoning about
  Reachability in Linked Data Structures}. In \bibinfo{booktitle}{\emph{{CAV}}}
  \emph{(\bibinfo{series}{Lecture Notes in Computer Science},
  Vol.~\bibinfo{volume}{8044})}. \bibinfo{publisher}{Springer},
  \bibinfo{pages}{756--772}.
\newblock
\urldef\tempurl%
\url{https://doi.org/10.1007/978-3-642-39799-8\_53}
\showDOI{\tempurl}


\bibitem[Jer{\'{a}}bek(2024)]%
        {well-orders-theory-note}
\bibfield{author}{\bibinfo{person}{Emil Jer{\'{a}}bek}.}
  \bibinfo{year}{2024}\natexlab{}.
\newblock \showarticletitle{A note on the theory of well orders}.
\newblock \bibinfo{journal}{\emph{CoRR}}  \bibinfo{volume}{abs/2405.05779}
  (\bibinfo{year}{2024}).
\newblock
\urldef\tempurl%
\url{https://doi.org/10.48550/ARXIV.2405.05779}
\showDOI{\tempurl}
\showeprint[arXiv]{2405.05779}


\bibitem[K. et~al\mbox{.}(2022)]%
        {chc-modulo-adt}
\bibfield{author}{\bibinfo{person}{Hari Govind~V. K.}, \bibinfo{person}{Sharon
  Shoham}, {and} \bibinfo{person}{Arie Gurfinkel}.}
  \bibinfo{year}{2022}\natexlab{}.
\newblock \showarticletitle{Solving constrained Horn clauses modulo algebraic
  data types and recursive functions}.
\newblock \bibinfo{journal}{\emph{Proc. {ACM} Program. Lang.}}
  \bibinfo{volume}{6}, \bibinfo{number}{{POPL}} (\bibinfo{year}{2022}),
  \bibinfo{pages}{1--29}.
\newblock
\urldef\tempurl%
\url{https://doi.org/10.1145/3498722}
\showDOI{\tempurl}


\bibitem[Katelaan et~al\mbox{.}(2019)]%
        {entailment-inductive-definitions}
\bibfield{author}{\bibinfo{person}{Jens Katelaan}, \bibinfo{person}{Christoph
  Matheja}, {and} \bibinfo{person}{Florian Zuleger}.}
  \bibinfo{year}{2019}\natexlab{}.
\newblock \showarticletitle{Effective Entailment Checking for Separation Logic
  with Inductive Definitions}. In \bibinfo{booktitle}{\emph{Tools and
  Algorithms for the Construction and Analysis of Systems - 25th International
  Conference, {TACAS} 2019, Held as Part of the European Joint Conferences on
  Theory and Practice of Software, {ETAPS} 2019, Prague, Czech Republic, April
  6-11, 2019, Proceedings, Part {II}}} \emph{(\bibinfo{series}{Lecture Notes in
  Computer Science}, Vol.~\bibinfo{volume}{11428})},
  \bibfield{editor}{\bibinfo{person}{Tom{\'{a}}s Vojnar} {and}
  \bibinfo{person}{Lijun Zhang}} (Eds.). \bibinfo{publisher}{Springer},
  \bibinfo{pages}{319--336}.
\newblock
\urldef\tempurl%
\url{https://doi.org/10.1007/978-3-030-17465-1\_18}
\showDOI{\tempurl}


\bibitem[Komuravelli et~al\mbox{.}(2015)]%
        {chc-integers-arrays}
\bibfield{author}{\bibinfo{person}{Anvesh Komuravelli},
  \bibinfo{person}{Nikolaj~S. Bj{\o}rner}, \bibinfo{person}{Arie Gurfinkel},
  {and} \bibinfo{person}{Kenneth~L. McMillan}.}
  \bibinfo{year}{2015}\natexlab{}.
\newblock \showarticletitle{Compositional Verification of Procedural Programs
  using Horn Clauses over Integers and Arrays}. In
  \bibinfo{booktitle}{\emph{Formal Methods in Computer-Aided Design, {FMCAD}
  2015, Austin, Texas, USA, September 27-30, 2015}},
  \bibfield{editor}{\bibinfo{person}{Roope Kaivola} {and}
  \bibinfo{person}{Thomas Wahl}} (Eds.). \bibinfo{publisher}{{IEEE}},
  \bibinfo{pages}{89--96}.
\newblock
\urldef\tempurl%
\url{https://doi.org/10.1109/FMCAD.2015.7542257}
\showDOI{\tempurl}


\bibitem[Komuravelli et~al\mbox{.}(2014)]%
        {smt-recursive-programs}
\bibfield{author}{\bibinfo{person}{Anvesh Komuravelli}, \bibinfo{person}{Arie
  Gurfinkel}, {and} \bibinfo{person}{Sagar Chaki}.}
  \bibinfo{year}{2014}\natexlab{}.
\newblock \showarticletitle{SMT-Based Model Checking for Recursive Programs}.
  In \bibinfo{booktitle}{\emph{Computer Aided Verification - 26th International
  Conference, {CAV} 2014, Held as Part of the Vienna Summer of Logic, {VSL}
  2014, Vienna, Austria, July 18-22, 2014. Proceedings}}
  \emph{(\bibinfo{series}{Lecture Notes in Computer Science},
  Vol.~\bibinfo{volume}{8559})}, \bibfield{editor}{\bibinfo{person}{Armin
  Biere} {and} \bibinfo{person}{Roderick Bloem}} (Eds.).
  \bibinfo{publisher}{Springer}, \bibinfo{pages}{17--34}.
\newblock
\urldef\tempurl%
\url{https://doi.org/10.1007/978-3-319-08867-9\_2}
\showDOI{\tempurl}


\bibitem[Lahiri and Qadeer(2006)]%
        {well-founded-lists}
\bibfield{author}{\bibinfo{person}{Shuvendu~K. Lahiri} {and}
  \bibinfo{person}{Shaz Qadeer}.} \bibinfo{year}{2006}\natexlab{}.
\newblock \showarticletitle{Verifying properties of well-founded linked lists}.
  In \bibinfo{booktitle}{\emph{Proceedings of the 33rd {ACM} {SIGPLAN-SIGACT}
  Symposium on Principles of Programming Languages, {POPL} 2006, Charleston,
  South Carolina, USA, January 11-13, 2006}},
  \bibfield{editor}{\bibinfo{person}{J.~Gregory Morrisett} {and}
  \bibinfo{person}{Simon L.~Peyton Jones}} (Eds.). \bibinfo{publisher}{{ACM}},
  \bibinfo{pages}{115--126}.
\newblock
\urldef\tempurl%
\url{https://doi.org/10.1145/1111037.1111048}
\showDOI{\tempurl}


\bibitem[Lahiri and Qadeer(2007)]%
        {well-founded-reachability}
\bibfield{author}{\bibinfo{person}{Shuvendu~K Lahiri} {and}
  \bibinfo{person}{Shaz Qadeer}.} \bibinfo{year}{2007}\natexlab{}.
\newblock \bibinfo{booktitle}{\emph{A decision procedure for well-founded
  reachability}}.
\newblock \bibinfo{type}{{T}echnical {R}eport}. \bibinfo{institution}{Technical
  Report MSR-TR-2007-43, Microsoft Research}.
\newblock


\bibitem[Lahiri and Qadeer(2008)]%
        {well-founded-reachability-revisiting}
\bibfield{author}{\bibinfo{person}{Shuvendu~K. Lahiri} {and}
  \bibinfo{person}{Shaz Qadeer}.} \bibinfo{year}{2008}\natexlab{}.
\newblock \showarticletitle{Back to the future: revisiting precise program
  verification using {SMT} solvers}. In \bibinfo{booktitle}{\emph{Proceedings
  of the 35th {ACM} {SIGPLAN-SIGACT} Symposium on Principles of Programming
  Languages, {POPL} 2008, San Francisco, California, USA, January 7-12, 2008}},
  \bibfield{editor}{\bibinfo{person}{George~C. Necula} {and}
  \bibinfo{person}{Philip Wadler}} (Eds.). \bibinfo{publisher}{{ACM}},
  \bibinfo{pages}{171--182}.
\newblock
\urldef\tempurl%
\url{https://doi.org/10.1145/1328438.1328461}
\showDOI{\tempurl}


\bibitem[Lamport(1998)]%
        {lamport-paxos}
\bibfield{author}{\bibinfo{person}{Leslie Lamport}.}
  \bibinfo{year}{1998}\natexlab{}.
\newblock \showarticletitle{The Part-Time Parliament}.
\newblock \bibinfo{journal}{\emph{{ACM} Trans. Comput. Syst.}}
  \bibinfo{volume}{16}, \bibinfo{number}{2} (\bibinfo{year}{1998}),
  \bibinfo{pages}{133--169}.
\newblock
\urldef\tempurl%
\url{https://doi.org/10.1145/279227.279229}
\showDOI{\tempurl}


\bibitem[Leino(2010)]%
        {dafny}
\bibfield{author}{\bibinfo{person}{K.~Rustan~M. Leino}.}
  \bibinfo{year}{2010}\natexlab{}.
\newblock \showarticletitle{Dafny: An Automatic Program Verifier for Functional
  Correctness}. In \bibinfo{booktitle}{\emph{Logic for Programming, Artificial
  Intelligence, and Reasoning - 16th International Conference, LPAR-16, Dakar,
  Senegal, April 25-May 1, 2010, Revised Selected Papers}}
  \emph{(\bibinfo{series}{Lecture Notes in Computer Science},
  Vol.~\bibinfo{volume}{6355})}, \bibfield{editor}{\bibinfo{person}{Edmund~M.
  Clarke} {and} \bibinfo{person}{Andrei Voronkov}} (Eds.).
  \bibinfo{publisher}{Springer}, \bibinfo{pages}{348--370}.
\newblock
\urldef\tempurl%
\url{https://doi.org/10.1007/978-3-642-17511-4\_20}
\showDOI{\tempurl}


\bibitem[Leino(2012)]%
        {induction-smt-solver}
\bibfield{author}{\bibinfo{person}{K.~Rustan~M. Leino}.}
  \bibinfo{year}{2012}\natexlab{}.
\newblock \showarticletitle{Automating Induction with an {SMT} Solver}. In
  \bibinfo{booktitle}{\emph{Verification, Model Checking, and Abstract
  Interpretation - 13th International Conference, {VMCAI} 2012, Philadelphia,
  PA, USA, January 22-24, 2012. Proceedings}} \emph{(\bibinfo{series}{Lecture
  Notes in Computer Science}, Vol.~\bibinfo{volume}{7148})},
  \bibfield{editor}{\bibinfo{person}{Viktor Kuncak} {and}
  \bibinfo{person}{Andrey Rybalchenko}} (Eds.). \bibinfo{publisher}{Springer},
  \bibinfo{pages}{315--331}.
\newblock
\urldef\tempurl%
\url{https://doi.org/10.1007/978-3-642-27940-9\_21}
\showDOI{\tempurl}


\bibitem[Lewis(1980)]%
        {epr-decidable}
\bibfield{author}{\bibinfo{person}{Harry~R. Lewis}.}
  \bibinfo{year}{1980}\natexlab{}.
\newblock \showarticletitle{Complexity Results for Classes of Quantificational
  Formulas}.
\newblock \bibinfo{journal}{\emph{J. Comput. Syst. Sci.}} \bibinfo{volume}{21},
  \bibinfo{number}{3} (\bibinfo{year}{1980}), \bibinfo{pages}{317--353}.
\newblock
\urldef\tempurl%
\url{https://doi.org/10.1016/0022-0000(80)90027-6}
\showDOI{\tempurl}


\bibitem[L{\"{o}}ding et~al\mbox{.}(2018)]%
        {natural-proofs}
\bibfield{author}{\bibinfo{person}{Christof L{\"{o}}ding}, \bibinfo{person}{P.
  Madhusudan}, {and} \bibinfo{person}{Lucas Pe{\~{n}}a}.}
  \bibinfo{year}{2018}\natexlab{}.
\newblock \showarticletitle{Foundations for natural proofs and quantifier
  instantiation}.
\newblock \bibinfo{journal}{\emph{Proc. {ACM} Program. Lang.}}
  \bibinfo{volume}{2}, \bibinfo{number}{{POPL}} (\bibinfo{year}{2018}),
  \bibinfo{pages}{10:1--10:30}.
\newblock
\urldef\tempurl%
\url{https://doi.org/10.1145/3158098}
\showDOI{\tempurl}


\bibitem[Lotan et~al\mbox{.}(2024)]%
        {raz-squeezers}
\bibfield{author}{\bibinfo{person}{Raz Lotan}, \bibinfo{person}{Eden Frenkel},
  {and} \bibinfo{person}{Sharon Shoham}.} \bibinfo{year}{2024}\natexlab{}.
\newblock \showarticletitle{Proving Cutoff Bounds for Safety Properties in
  First-Order Logic}.
\newblock \bibinfo{journal}{\emph{CoRR}}  \bibinfo{volume}{abs/2408.10685}
  (\bibinfo{year}{2024}).
\newblock
\urldef\tempurl%
\url{https://doi.org/10.48550/ARXIV.2408.10685}
\showDOI{\tempurl}
\showeprint[arXiv]{2408.10685}


\bibitem[McMillan and Padon(2018)]%
        {mcmillan-decidable-ivy}
\bibfield{author}{\bibinfo{person}{Kenneth~L. McMillan} {and}
  \bibinfo{person}{Oded Padon}.} \bibinfo{year}{2018}\natexlab{}.
\newblock \showarticletitle{Deductive Verification in Decidable Fragments with
  Ivy}. In \bibinfo{booktitle}{\emph{{SAS}}} \emph{(\bibinfo{series}{Lecture
  Notes in Computer Science}, Vol.~\bibinfo{volume}{11002})}.
  \bibinfo{publisher}{Springer}, \bibinfo{pages}{43--55}.
\newblock
\urldef\tempurl%
\url{https://doi.org/10.1007/978-3-319-99725-4\_4}
\showDOI{\tempurl}


\bibitem[Murali et~al\mbox{.}(2022)]%
        {fossil}
\bibfield{author}{\bibinfo{person}{Adithya Murali}, \bibinfo{person}{Lucas
  Pe{\~{n}}a}, \bibinfo{person}{Eion Blanchard}, \bibinfo{person}{Christof
  L{\"{o}}ding}, {and} \bibinfo{person}{P. Madhusudan}.}
  \bibinfo{year}{2022}\natexlab{}.
\newblock \showarticletitle{Model-guided synthesis of inductive lemmas for
  {FOL} with least fixpoints}.
\newblock \bibinfo{journal}{\emph{Proc. {ACM} Program. Lang.}}
  \bibinfo{volume}{6}, \bibinfo{number}{{OOPSLA2}} (\bibinfo{year}{2022}),
  \bibinfo{pages}{1873--1902}.
\newblock
\urldef\tempurl%
\url{https://doi.org/10.1145/3563354}
\showDOI{\tempurl}


\bibitem[Padon et~al\mbox{.}(2017)]%
        {paxos-made-epr}
\bibfield{author}{\bibinfo{person}{Oded Padon}, \bibinfo{person}{Giuliano
  Losa}, \bibinfo{person}{Mooly Sagiv}, {and} \bibinfo{person}{Sharon Shoham}.}
  \bibinfo{year}{2017}\natexlab{}.
\newblock \showarticletitle{Paxos made {EPR:} decidable reasoning about
  distributed protocols}.
\newblock \bibinfo{journal}{\emph{Proc. {ACM} Program. Lang.}}
  \bibinfo{volume}{1}, \bibinfo{number}{{OOPSLA}} (\bibinfo{year}{2017}),
  \bibinfo{pages}{108:1--108:31}.
\newblock
\urldef\tempurl%
\url{https://doi.org/10.1145/3140568}
\showDOI{\tempurl}


\bibitem[Paulson(1994)]%
        {isabelle}
\bibfield{author}{\bibinfo{person}{Lawrence~C. Paulson}.}
  \bibinfo{year}{1994}\natexlab{}.
\newblock \bibinfo{booktitle}{\emph{Isabelle - {A} Generic Theorem Prover (with
  a contribution by T. Nipkow)}}. \bibinfo{series}{Lecture Notes in Computer
  Science}, Vol.~\bibinfo{volume}{828}.
\newblock \bibinfo{publisher}{Springer}.
\newblock
\showISBNx{3-540-58244-4}
\urldef\tempurl%
\url{https://doi.org/10.1007/BFB0030541}
\showDOI{\tempurl}


\bibitem[Rabin(1969)]%
        {rabin-mso-s2s}
\bibfield{author}{\bibinfo{person}{Michael~O Rabin}.}
  \bibinfo{year}{1969}\natexlab{}.
\newblock \showarticletitle{Decidability of second-order theories and automata
  on infinite trees.}
\newblock \bibinfo{journal}{\emph{Transactions of the american Mathematical
  Society}}  \bibinfo{volume}{141} (\bibinfo{year}{1969}),
  \bibinfo{pages}{1--35}.
\newblock
\urldef\tempurl%
\url{https://doi.org/10.2307/1995086}
\showDOI{\tempurl}


\bibitem[Ramsey(1930)]%
        {bsr-epr}
\bibfield{author}{\bibinfo{person}{Frank~P Ramsey}.}
  \bibinfo{year}{1930}\natexlab{}.
\newblock \showarticletitle{On a Problem of Formal Logic}.
\newblock \bibinfo{journal}{\emph{Procedures of London Mathematical Society}}
  \bibinfo{volume}{30} (\bibinfo{year}{1930}), \bibinfo{pages}{264--285}.
\newblock
\urldef\tempurl%
\url{https://doi.org/10.1007/978-0-8176-4842-8\_1}
\showDOI{\tempurl}


\bibitem[Riazanov and Voronkov(1999)]%
        {vampire}
\bibfield{author}{\bibinfo{person}{Alexandre Riazanov} {and}
  \bibinfo{person}{Andrei Voronkov}.} \bibinfo{year}{1999}\natexlab{}.
\newblock \showarticletitle{Vampire}. In \bibinfo{booktitle}{\emph{{CADE}}}
  \emph{(\bibinfo{series}{Lecture Notes in Computer Science},
  Vol.~\bibinfo{volume}{1632})}. \bibinfo{publisher}{Springer},
  \bibinfo{pages}{292--296}.
\newblock
\urldef\tempurl%
\url{https://doi.org/10.1007/3-540-48660-7\_26}
\showDOI{\tempurl}


\bibitem[Schoisswohl and Kov{\'{a}}cs(2021)]%
        {vampire-reflection-induction}
\bibfield{author}{\bibinfo{person}{Johannes Schoisswohl} {and}
  \bibinfo{person}{Laura Kov{\'{a}}cs}.} \bibinfo{year}{2021}\natexlab{}.
\newblock \showarticletitle{Automating Induction by Reflection}. In
  \bibinfo{booktitle}{\emph{{LFMTP}}} \emph{(\bibinfo{series}{{EPTCS}},
  Vol.~\bibinfo{volume}{337})}. \bibinfo{pages}{39--54}.
\newblock
\urldef\tempurl%
\url{https://doi.org/10.4204/EPTCS.337.4}
\showDOI{\tempurl}


\bibitem[Segall(1983)]%
        {broadcast-echo}
\bibfield{author}{\bibinfo{person}{Adrian Segall}.}
  \bibinfo{year}{1983}\natexlab{}.
\newblock \showarticletitle{Distributed network protocols}.
\newblock \bibinfo{journal}{\emph{{IEEE} Trans. Inf. Theory}}
  \bibinfo{volume}{29}, \bibinfo{number}{1} (\bibinfo{year}{1983}),
  \bibinfo{pages}{23--34}.
\newblock
\urldef\tempurl%
\url{https://doi.org/10.1109/TIT.1983.1056620}
\showDOI{\tempurl}


\bibitem[Shelah(1977)]%
        {shelah1977decidability}
\bibfield{author}{\bibinfo{person}{Saharon Shelah}.}
  \bibinfo{year}{1977}\natexlab{}.
\newblock \showarticletitle{Decidability of a portion of the predicate
  calculus}.
\newblock \bibinfo{journal}{\emph{Israel Journal of Mathematics}}
  \bibinfo{volume}{28}, \bibinfo{number}{1} (\bibinfo{year}{1977}),
  \bibinfo{pages}{32--44}.
\newblock
\urldef\tempurl%
\url{https://doi.org/10.1007/BF02759780}
\showDOI{\tempurl}


\bibitem[Sheng et~al\mbox{.}(2022)]%
        {reasoning-about-vectors}
\bibfield{author}{\bibinfo{person}{Ying Sheng}, \bibinfo{person}{Andres
  N{\"{o}}tzli}, \bibinfo{person}{Andrew Reynolds}, \bibinfo{person}{Yoni
  Zohar}, \bibinfo{person}{David~L. Dill}, \bibinfo{person}{Wolfgang
  Grieskamp}, \bibinfo{person}{Junkil Park}, \bibinfo{person}{Shaz Qadeer},
  \bibinfo{person}{Clark~W. Barrett}, {and} \bibinfo{person}{Cesare Tinelli}.}
  \bibinfo{year}{2022}\natexlab{}.
\newblock \showarticletitle{Reasoning About Vectors Using an {SMT} Theory of
  Sequences}. In \bibinfo{booktitle}{\emph{Automated Reasoning - 11th
  International Joint Conference, {IJCAR} 2022, Haifa, Israel, August 8-10,
  2022, Proceedings}} \emph{(\bibinfo{series}{Lecture Notes in Computer
  Science}, Vol.~\bibinfo{volume}{13385})},
  \bibfield{editor}{\bibinfo{person}{Jasmin Blanchette}, \bibinfo{person}{Laura
  Kov{\'{a}}cs}, {and} \bibinfo{person}{Dirk Pattinson}} (Eds.).
  \bibinfo{publisher}{Springer}, \bibinfo{pages}{125--143}.
\newblock
\urldef\tempurl%
\url{https://doi.org/10.1007/978-3-031-10769-6\_9}
\showDOI{\tempurl}


\bibitem[Taube et~al\mbox{.}(2018)]%
        {modularity-for-decidability}
\bibfield{author}{\bibinfo{person}{Marcelo Taube}, \bibinfo{person}{Giuliano
  Losa}, \bibinfo{person}{Kenneth~L. McMillan}, \bibinfo{person}{Oded Padon},
  \bibinfo{person}{Mooly Sagiv}, \bibinfo{person}{Sharon Shoham},
  \bibinfo{person}{James~R. Wilcox}, {and} \bibinfo{person}{Doug Woos}.}
  \bibinfo{year}{2018}\natexlab{}.
\newblock \showarticletitle{Modularity for decidability of deductive
  verification with applications to distributed systems}. In
  \bibinfo{booktitle}{\emph{{PLDI}}}. \bibinfo{publisher}{{ACM}},
  \bibinfo{pages}{662--677}.
\newblock
\urldef\tempurl%
\url{https://doi.org/10.1145/3192366.3192414}
\showDOI{\tempurl}


\bibitem[Team(2024)]%
        {coq}
\bibfield{author}{\bibinfo{person}{The Coq~Development Team}.}
  \bibinfo{year}{2024}\natexlab{}.
\newblock \bibinfo{booktitle}{\emph{The Coq Proof Assistant}}.
\newblock
\urldef\tempurl%
\url{https://doi.org/10.5281/zenodo.11551307}
\showDOI{\tempurl}


\bibitem[{The Open Logic Project}({[n.\,d.]})]%
        {open-logic-definability}
\bibfield{author}{\bibinfo{person}{{The Open Logic Project}}.}
  \bibinfo{year}{[n.\,d.]}\natexlab{}.
\newblock \bibinfo{booktitle}{\emph{Frame Definability}}.
\newblock
\urldef\tempurl%
\url{https://builds.openlogicproject.org/content/normal-modal-logic/frame-definability/frame-definability.pdf}
\showURL{%
\tempurl}


\bibitem[Van~Rossum and Drake(2009)]%
        {python}
\bibfield{author}{\bibinfo{person}{Guido Van~Rossum} {and}
  \bibinfo{person}{Fred~L. Drake}.} \bibinfo{year}{2009}\natexlab{}.
\newblock \bibinfo{booktitle}{\emph{Python 3 Reference Manual}}.
\newblock \bibinfo{publisher}{CreateSpace}, \bibinfo{address}{Scotts Valley,
  CA}.
\newblock
\showISBNx{1441412697}
\urldef\tempurl%
\url{https://doi.org/10.5555/1593511}
\showDOI{\tempurl}


\bibitem[Yao et~al\mbox{.}(2022)]%
        {duoai}
\bibfield{author}{\bibinfo{person}{Jianan Yao}, \bibinfo{person}{Runzhou Tao},
  \bibinfo{person}{Ronghui Gu}, {and} \bibinfo{person}{Jason Nieh}.}
  \bibinfo{year}{2022}\natexlab{}.
\newblock \showarticletitle{DuoAI: Fast, Automated Inference of Inductive
  Invariants for Verifying Distributed Protocols}. In
  \bibinfo{booktitle}{\emph{16th {USENIX} Symposium on Operating Systems Design
  and Implementation, {OSDI} 2022, Carlsbad, CA, USA, July 11-13, 2022}},
  \bibfield{editor}{\bibinfo{person}{Marcos~K. Aguilera} {and}
  \bibinfo{person}{Hakim Weatherspoon}} (Eds.). \bibinfo{publisher}{{USENIX}
  Association}, \bibinfo{pages}{485--501}.
\newblock
\urldef\tempurl%
\url{https://www.usenix.org/conference/osdi22/presentation/yao}
\showURL{%
\tempurl}


\bibitem[Yao et~al\mbox{.}(2021)]%
        {distai}
\bibfield{author}{\bibinfo{person}{Jianan Yao}, \bibinfo{person}{Runzhou Tao},
  \bibinfo{person}{Ronghui Gu}, \bibinfo{person}{Jason Nieh},
  \bibinfo{person}{Suman Jana}, {and} \bibinfo{person}{Gabriel Ryan}.}
  \bibinfo{year}{2021}\natexlab{}.
\newblock \showarticletitle{DistAI: Data-Driven Automated Invariant Learning
  for Distributed Protocols}. In \bibinfo{booktitle}{\emph{{OSDI}}}.
  \bibinfo{publisher}{{USENIX} Association}, \bibinfo{pages}{405--421}.
\newblock


\bibitem[Zaval{\'{\i}}a et~al\mbox{.}(2023)]%
        {chc-adt}
\bibfield{author}{\bibinfo{person}{Lucas Zaval{\'{\i}}a},
  \bibinfo{person}{Lidiia Chernigovskaia}, {and} \bibinfo{person}{Grigory
  Fedyukovich}.} \bibinfo{year}{2023}\natexlab{}.
\newblock \showarticletitle{Solving Constrained Horn Clauses over Algebraic
  Data Types}. In \bibinfo{booktitle}{\emph{Verification, Model Checking, and
  Abstract Interpretation - 24th International Conference, {VMCAI} 2023,
  Boston, MA, USA, January 16-17, 2023, Proceedings}}
  \emph{(\bibinfo{series}{Lecture Notes in Computer Science},
  Vol.~\bibinfo{volume}{13881})}, \bibfield{editor}{\bibinfo{person}{Cezara
  Dragoi}, \bibinfo{person}{Michael Emmi}, {and} \bibinfo{person}{Jingbo Wang}}
  (Eds.). \bibinfo{publisher}{Springer}, \bibinfo{pages}{341--365}.
\newblock
\urldef\tempurl%
\url{https://doi.org/10.1007/978-3-031-24950-1\_16}
\showDOI{\tempurl}


\end{thebibliography}

\ifappendix
\pagebreak
\appendix
\section{Detailed Proofs}
\label{sec:proofs}

\Proofs{}

\fi
	
\end{document}